\documentclass[twocolumn,aps,prl,showpacs,superscriptaddress]{revtex4}
\usepackage[dvips]{graphicx}
\usepackage{bm}
\begin{document}
%\draft
%\twocolumn[\hsize\textwidth\columnwidth\hsize
%\csname @twocolumnfalse\endcsname
\title{Single-ion and exchange anisotropy effects and multiferroic behavior in high-symmetry tetramer single molecule magnets}
\author{Richard A. Klemm}
\email{klemm@physics.ucf.edu} \affiliation{Department of Physics,
 University of Central Florida, Orlando, FL 32816 USA}
\author{Dmitri V. Efremov}
\email{efremov@theory.phy.tu-dresden.de} \affiliation{Institut
f{\"u}r Theoretische Physik, Technische Universit{\"a}t Dresden,
01062 Dresden, Germany}
\date{\today}
\begin{abstract}
We study single-ion and  exchange  anisotropy effects in
equal-spin $s_1$ tetramer single molecule magnets exhibiting
$T_d$, $D_{4h}$, $D_{2d}$, $C_{4h}$, $C_{4v}$, or $S_4$ ionic
point group symmetry. We first write the group-invariant quadratic
single-ion and symmetric anisotropic exchange Hamiltonians in the
appropriate local coordinates. We then rewrite these local
Hamiltonians in the molecular or laboratory representation, along
with the  group-invariant Dzyaloshinskii-Moriya (DM), and
isotropic Heisenberg, biquadratic, and three-center quartic
Hamiltonians. Using our exact, compact forms for the single-ion
spin matrix elements, we evaluate the eigenstate energies
analytically to first order in the microscopic anisotropy
interactions, corresponding to the strong exchange limit, and
provide tables of simple formulas for the energies of the lowest
four eigenstate manifolds of ferromagnetic (FM) and
anitiferromagnetic (AFM) tetramers with arbitrary $s_1$. For AFM
tetramers, we illustrate the first-order level-crossing inductions
for $s_1=1/2,1,3/2$, and obtain a preliminary estimate of the
microscopic parameters in a Ni$_4$ from a fit to magnetization
data.
 Accurate analytic expressions
  for the thermodynamics, electron
paramagnetic resonance  absorption and inelastic neutron
scattering cross-section are given, allowing for a determination
of three of the microscopic anisotropy interactions from the
second excited state manifold of FM tetramers. We also predict
that tetramers with symmetries $S_4$ and $D_{2d}$ should exhibit
both DM interactions and multiferroic states, and  illustrate our
predictions for $s_1=1/2, 1$.
\end{abstract}
\vskip0pt\vskip0pt \maketitle
\section{I. Introduction}
Single molecule magnets (SMM's) have been a topic of great
interest for more than a decade,\cite{general} because of their
potential uses in quantum computing and/or magnetic
storage,\cite{moregeneral} which are possible due to magnetic
quantum tunneling  (MQT) and entangled states.  In fits to a
wealth of data,  the Hamiltonian within an SMM cluster was assumed
to be the  Heisenberg exchange interaction plus weaker total
(global, or giant) spin anisotropy interactions, with a fixed
overall total spin quantum number $s$.\cite{general} MQT and
entanglement were only studied in this simple model.

The simplest SMM clusters  are dimers.\cite{ek,ekshort,ek2}
Surprisingly, two antiferromagnetic dimers, an Fe$_2$,
[Fe(salen)Cl]$_2$, where salen is
$N,N'$-ethylenebis(salicylideneiminato), and a Ni$_2$,
Na$_2$Ni$_2$(C$_2$O$_4$)$_3$(H$_2$O)$_2$, appear to have
substantial single-ion anisotropy without any appreciable total
spin anisotropy.\cite{Shapira,Mennerich,ek2} The presence of
single-ion or exchange anisotropy actually precludes the total
spin $s$ from being a good quantum number.\cite{ekshort,ek2}
Although the most common SMM clusters have ferromagnetic (FM)
intramolecular interactions and contain $n\ge8$ magnetic
ions,\cite{Dalal,Fe8} a number of intermediate-sized FM SMM
clusters with $n=4$ and rather simple molecular structures were
recently studied.  Fits to electron paramagnetic resonance (EPR)
Ni$_4$ data assuming a fixed $s$ were also problematic, suggesting
single-ion or exchange anisotropy in that tetramer, as
well.\cite{Hill1,Hill2}  The Cu$_4$ tetramer
 Cu$_4$OCl$_6$(TPPO)$_4$, where TPPO is triphenylphosphine oxide, has four spin 1/2 ions
on the corners of a regular tetrahedron, with an $s=2$ ground
state and approximate $T_d$ symmetry.\cite{Black1,Black2,Black3}
In this case, there are no single-ion anisotropy effects, but
anisotropic symmetric exchange interactions were thought to be
responsible for the zero-field energy
splittings.\cite{Black1,Buluggiu} The Co$_4$,
Co$_4$(hmp)$_4$(MeOH)$_4$Cl$_4$, where hmp is
hydroxymethylpyridyl, and Cr$_4$,
[Cr$_4$S(O$_2$CCH$_3$)$_8$(H$_2$O)$_4$](NO$_3$)$_2\cdot$H$_2$O,
compounds have $s=6$ ground states with spin 3/2 ions on the
corners of tetrahedrons.\cite{Co4,Cr4} Those  compounds have $S_4$
and approximate $D_{2d}$ symmetry, respectively.\cite{Co4,Cr4} A
number of high symmetry $s=4$ ground state Ni$_4$ structures with
spin 1 ions were
reported.\cite{Ni4,Ni4Maria,Edwards,Ni4S4,Hendrickson} Two of
these, [Ni(hmp)(ROH)Cl]$_4$, where R is an alkyl group, such as
methyl, ethyl, or 3,3-dimethyl-1-butyl and hmp is
2-hydroxymethylpyridyl, form tetramers with precise $S_4$ group
symmetry.\cite{Hendrickson,Ni4S4} Two others, Ni$_4$(ROH)L$_4$,
where R is methyl or ethyl and H$_2$L is
salicylidene-2-ethanolamine, had approximate $S_4$ symmetry,
although the precise symmetry was only $C_1$.\cite{Ni4}  Several
planar Mn$_4$ compounds with the Mn$^{+3}$ spin 2 ions on the
corners of squares were made, with overall $s=8$ tetramer ground
states.\cite{Boskovic} Although two of these complexes had only
approximate $S_4$ symmetry, one of these complexes,
Mn$_4$Cl$_4$(L')$_4$, where H$_2$L' is
4-$t$-butyl-salicylidene-2-ethanolamine, had perfect $S_4$
symmetry.\cite{Boskovic} Inelastic neutron scattering (INS)
experiments provided strong evidence for single-ion anisotropy in
a Co$_4$ and a Ni$_4$ with approximate $S_4$
symmetry.\cite{Co4,Ni4}

We note that {\it ab initio} calculations of the intramolecular
spin-spin interactions in SMM clusters have not yet been always
successful in calculating even the strongest, intramolecular
isotropic Heisenberg interactions accurately, and have been
incapable of calculating any of the local anisotropic spin-spin
interactions within an SMM
cluster.\cite{Parkpublished,Pedersonpublished,Pedersonreview} Even
to obtain the Heisenberg interactions accurately, it seems one
needs to extend the local spin-density approximation (LSDA) to
include on-site repulsions with strength $U$ (the LSDA+U model),
which would have to be introduced phenomenologically to fit the
lowest two energy level manifolds in zero applied magnetic
field.\cite{Pedersonreview,Stolbov,Park,NRL,Harmon} We therefore
define a microscopic model to be a model constructed in terms of
the individual spins and from the local interactions between them,
with parameters describing the strengths of the various types of
local spin-spin interactions and interactions between the local
spins and the magnetic field. This is distinct from a model
constructed solely from the anisotropies of the total spin of an
SMM cluster, which we denote as a phenomenological model.  Our
definition of a microscopic model is analogous to the standard
model of the interactions of quarks and gluons within a hadron.

Recently there have been microscopic treatments of
dimers,\cite{ekshort,ek2} trimers, and tetramers, including Zeeman
$g$-tensor anisotropy, single-ion anisotropy, and anisotropic
exchange interactions.\cite{Bocabook} Most of those treatments and
their recent extensions to more general systems expressed the
single-spin matrix elements only  in terms of Wigner $3j$, $6j$,
and $9j$ symbols.\cite{Bocabook,WG} While such treatments are very
helpful in fitting experimental data, more compact analytic forms
are desirable to study microscopic models of FM SMM clusters in
which the MQT and entanglement issues crucial for quantum
computing can be understood. We constructed the quadratic
single-ion and anisotropic near-neighbor (NN) and
next-nearest-neighbor (NNN) exchange SMM cluster
 Hamiltonians from the respective local
axial and azimuthal vector groups for equal-spin tetramer SMM
clusters with point group symmetries $g=T_d$, $D_{4h}$, $D_{2d}$,
$C_{4h}$, $C_{4v}$, and $S_4$, and found  compact analytic
expressions for the single-spin matrix elements of four general
spins. Each local vector group generates site-dependent molecular
single-ion and exchange anisotropy. We then show that for $D_{2d}$
and $S_4$ symmetries, the antisymmetric exchange interactions lead
to non-vanishing spin currents that may be accompanied by electric
polarizations, leading to multiferroic effects.  We evaluate
 the magnetization, specific heat,
EPR and INS transitions  in the Hartree approximation, and provide
a procedure for extracting three of the effective site-independent
microscopic parameters using EPR.  We also show analytically how
to include the effects of weak biquadratic exchange.
\begin{figure}
\includegraphics[width=0.23\textwidth]{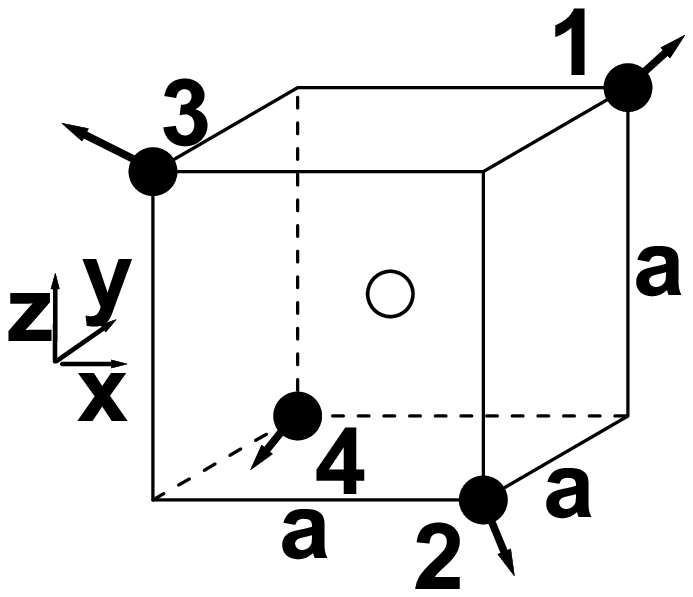}\hskip5pt\includegraphics[width=0.23\textwidth]{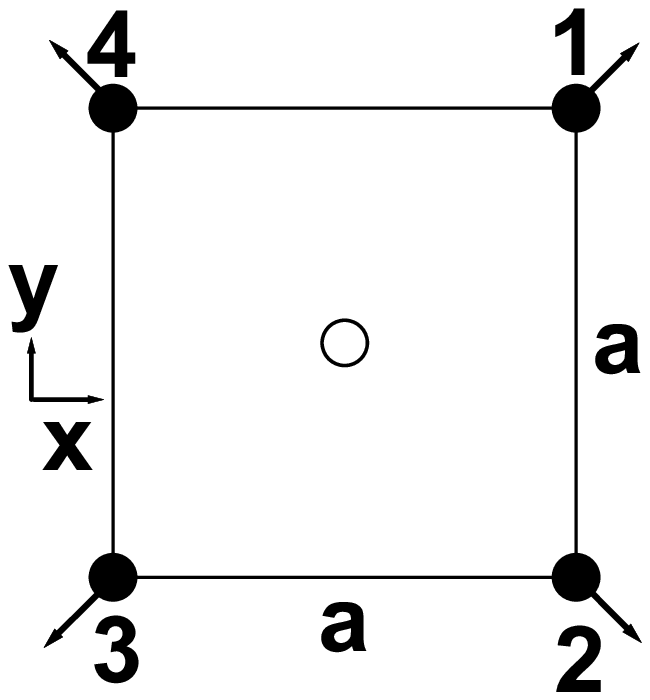}
\caption{$T_{d}$ (left) and $D_{4h}$ (right)  ion sites (filled).
Circle: origin. Arrows: local axial $\hat{\bm z}_n^{T_d}$ (left),
azimuthal $\hat{\bm x}_n^{D_{4h}}$ (right) single-ion vectors. The
axial vectors $\hat{\bm z}_n^{D_{4h}}=\hat{\bm z}$, normal to the
ionic plane.}\label{fig1}
\end{figure}

\begin{figure}
\includegraphics[width=0.23\textwidth]{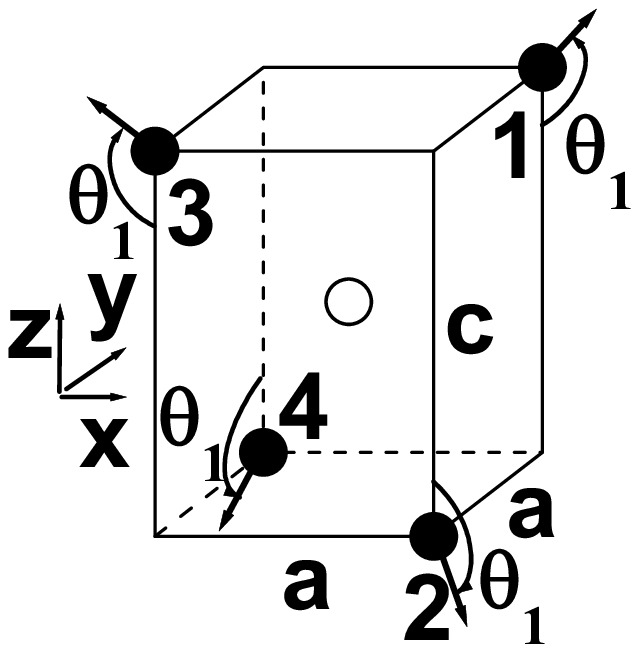}\hskip2pt\includegraphics[width=0.235\textwidth]{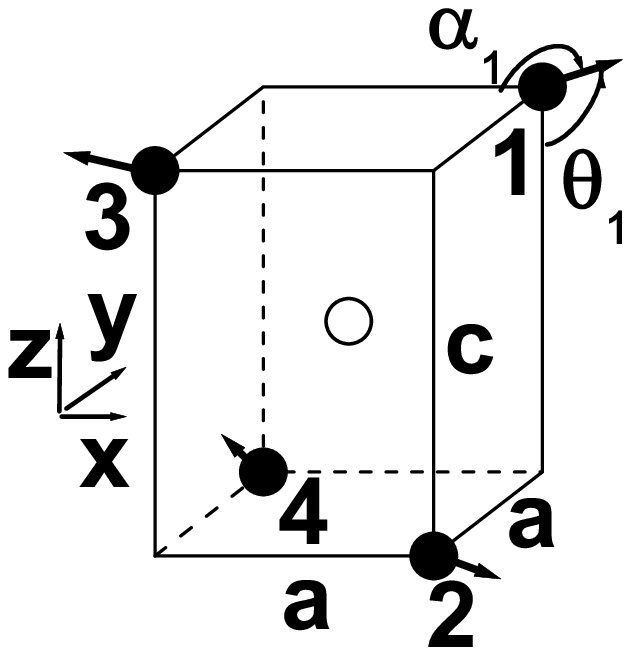}
\caption{$D_{2d}$ (left) and $S_4$ (right) ion sites (filled).
Circle: origin.  Arrows:  local axial single-ion vectors.  The
$g=D_{2d},S_4$ axial vectors $\hat{\bm z}^g_1$ make the angles
$\theta_1^g$ with the $z$ axis, and the $S_4$ axial vector
$\hat{\bm z}^{S_4}_1$ also makes the angle $\alpha_1$ with the $x$
axis, where
$\cos\alpha_1=\sin\theta^{S_4}_1\cos\phi^{S_4}_1$.}\label{fig2}
\end{figure}

\begin{figure}
\includegraphics[width=0.225\textwidth]{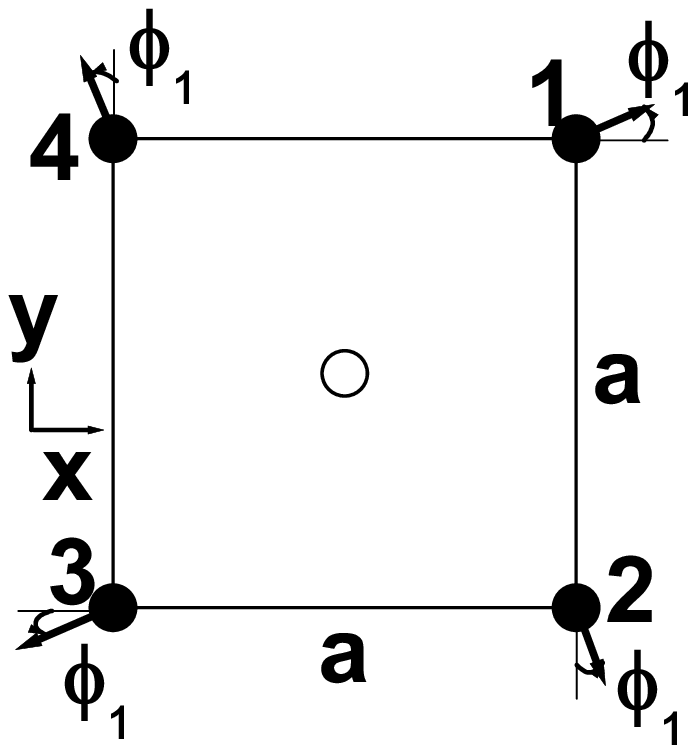}\hskip5pt\includegraphics[width=0.22\textwidth]{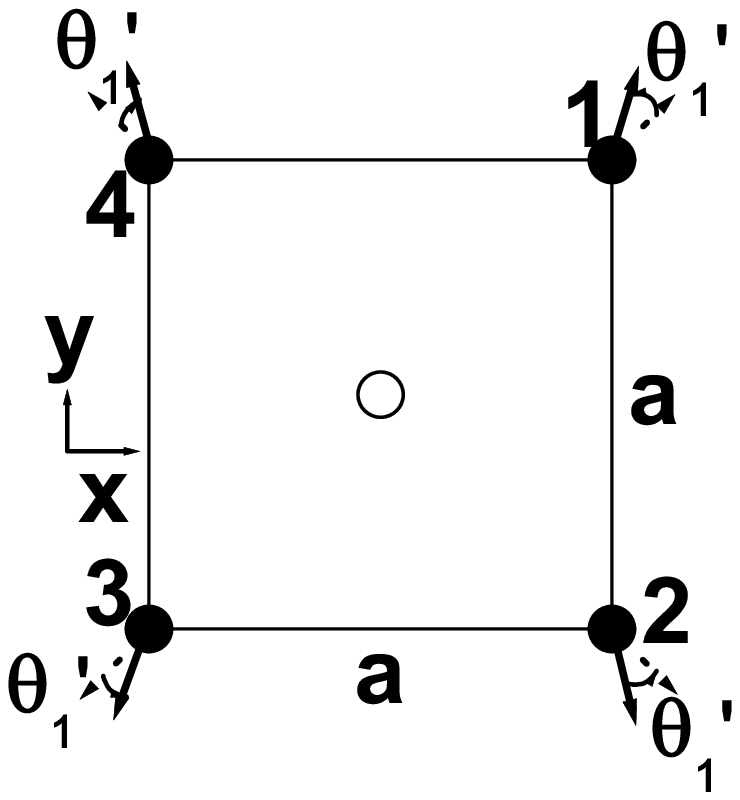}
\caption{$C_{4h}$ (left) and $C_{4v}$ (right) ion sites (filled).
Circle: origin.  Arrows:  local azimuthal $\hat{\bm x}_n^{C_{4h}}$
(left) axial $\hat{\bm z}_n^{C_{4v}}$ (right) single-ion vectors.
The axial vectors $\hat{\bm z}_n^{C_{4h}}=\hat{\bm z}$. The
$C_{4v}$ axial vectors each make the angle
$\theta_1=\pi/2-\theta_1'$ with the $z$ axis. The dotted arrows
(equivalent to the $D_{4h}$ azimuthal single-ion vectors $\hat{\bm
x}_n^{D_{4h}}$) are their projections in the $xy$
plane.}\label{fig3}
\end{figure}

An outline of the paper is as follows.  In Sec. II, we discuss the
six structures and the general quadratic spin Hamiltonian.  In
Sec. III, we write the single-ion and symmetric anisotropic
exchange Hamiltonians in terms of the local coordinates,  and the
antisymmetric exchange Hamiltonian in the molecular coordinates.
 In Sec. IV, we impose the operations
of the six group symmetries, and discuss the effects of
antisymmetric anisotropic exchange interactions and the related
electric polarizations
 in lower
symmetry systems. In Sec. V, the resulting group-symmetric
Hamiltonians are written in the molecular representation, and the
isotropic biquadratic exchange interactions are introduced.
Section VI contains the eigenstates of the full Hamiltonian to
first order in the anisotropy and NN biquadratic exchange
interactions.  These eiqenstates are used to obtain the
 level-crossing inductions for  AFM tetramers,
 and particular examples with $s_1=1/2, 1, 3/2$ are presented. In
 Sec. VI, we also evaluate quantitatively some effects of antisymmetric
 anisotropic exchange and provide our related predictions for
 multiferroic behavior.
 In Sec. VII, the self-consistent Hartree approximation (or
 strong-exchange limit) is used to provide simple but
 accurate results for the thermodynamics, EPR resonant inductions,
 and INS cross-sections, and describe how EPR experiments in the
 excited states of FM tetramers can provide a measure of some of the
  microscopic anisotropy interations strengths.  Finally, in Sec. VIII, we
 discuss the significance of our results, and provide a preliminary fit to
 magnetization data on an AFM Ni$_4$ tetramer, and in Sec. IX, we present our conclusions.

\section{II. Structures and Bare Hamiltonian}

For SMM clusters with ionic
 site point groups $g=T_d, D_{4h}$, we assume the four equal-spin $s_1$ ions sit  on opposite
corners of a cube or square of side $a$ centered at the origin, as
pictured in Fig. 1. For clusters with $g=D_{2d},S_4$, we take the
ions to sit on opposite corners of a tetragonal prism with sides
$(a,a,c)$ centered at the origin, as in Fig. 2.   The ions for
$g=C_{4h}, C_{4v}$ also sit on the corners of a square of side $a$
centered at the origin, as pictured in Fig. 3, but the ligand
groups have different symmetries than for the simpler $D_{4h}$
case pictured in Fig. 1.\cite{Tinkham} In each case, we take the
origin to be at the geometric center, so that $\sum_{n=1}^4{\bm
r}_n=0$, where the relative ion site vectors are
\begin{eqnarray}
{\bm
r}_{n}&=&\frac{a}{\sqrt{2}}\Bigl[\sin\Bigl(\frac{(2n-1)\pi}{4}\Bigr)\hat{\bm
x}+\cos\Bigl(\frac{(2n-1)\pi}{4}\Bigr)\hat{\bm
y}\Bigr]\nonumber\\
& &\qquad-\frac{c}{2}(-1)^n\hat{\bm z}.\label{rn}
\end{eqnarray}
 Tetrahedrons with
$g=T_d$, $c/a=1$, approximately as in Cu$_4$,\cite{Black1} are a
four-spin example of the equivalent-neighbor model.\cite{ka}  In
squares with $g=D_{4h}$, $C_{4h}$, or $C_{4v}$, $c=0$. The high
$D_{4h}$ symmetry is approximately exhibited by the square Nd$_4$
compound, Nd$_4$(OR)$_{12}$, where R is 2,2-dimethyl-1-propyl, in
which the Nd$^{+3}$ ions have equal total angular momentum
$j=9/2$.\cite{Nd4,KlemmLuban} We note that the Mn$_4$ clusters
with approximate or exact $S_4$ symmetry also have
$c=0$.\cite{Boskovic}  In tetragonal prisms with $g=D_{2d}$ or
$S_4$, $c/a>1$, approximately as in a Co$_4$,\cite{Co4} or
$c/a<1$, as in some Mn$_4$ and a Ni$_4$.\cite{Boskovic,Ni4Maria}
For comparison with the planar symmetries $g=C_{4h}, D_{4h},$ and
$D_{4v}$, we assume for $g=D_{2d}, S_4$ that $c/a<1$, so that
there are four NN sites and two NNN sites. For each $g$, $\hat{\bm
x}, \hat{\bm y}, \hat{\bm z}$ are the molecular (or laboratory)
unit coordinate axis vectors.

The most general Hamiltonian quadratic in the four spin operators
${\bm S}_n$ may be written for group $g$ as
\begin{eqnarray}
{\cal H}^g=-\mu_B \sum_{n=1}^4{\bm B}\cdot\tensor{\bm
g}^g_n\cdot{\bm S}_n+\sum_{n,n'=1}^4{\bm S}_n\cdot\tensor{\bm
D}_{n,n'}^g\cdot{\bm S}_{n'},\label{generalH}
\end{eqnarray}
where $\mu_B$ is the Bohr magneton and  ${\bm
B}=B(\sin\theta\cos\phi,\sin\theta\sin\phi,\cos\theta)$ is the
magnetic induction at an arbitrary direction $(\theta,\phi)$
relative to the molecular (or cluster) coordinates $(\hat{\bm
x},\hat{\bm y},\hat{\bm z})$.\cite{Bocabook,BenciniGatteschi}

For simplicity, we take $\tensor{\bm g}^g_n$ to be diagonal,
isotropic, and site-independent, so that the Zeeman interaction
may be written in terms of a single gyromagnetic ratio
$\gamma\approx2\mu_B$. Thus in the following, $g$ only refers to
the molecular group.    We separate $\tensor{\bm D}^g_{n,n'}$ into
its symmetric and antisymmetric parts, $\tensor{\bm
D}^g_{n,n'}=\tensor{\bm D}^{g,s}_{n,n'}+\tensor{\bm
D}^{g,a}_{n,n'}$, respectively.  For $n'=n$, the single-ion
$\tensor{\bm D}^g_{n,n}$ is necessarily symmetric, so $\tensor{\bm
D}^{g,a}_{n,n}=0$.  For each $g$, the four  $\tensor{\bm
D}^{g,s}_{n,n}$ contain the local single-ion structural
information, and the six distinct symmetric $\tensor{\bm
D}^{g,s}_{n,n'}$ contain the local symmetric exchange structural
information, which lead to the isotropic, or Heisenberg, exchange
interactions, and the remaining symmetric anisotropic exchange
interactions. The six distinct antisymmetric
 $\tensor{\bm
D}^{g,a}_{n,n'}$  contain additional local structural information
which lead to the Dzyaloshinskii-Moriya (DM)
interactions.\cite{Moriya,Dzyaloshinskii} Physically, the
symmetric anisotropic exchange interactions also contain the
intramolecular dipole-dipole interactions, which can be even
larger in magnitude than the terms originating from actual
anisotropic exchange.\cite{BenciniGatteschi,Jackson}

As is well known, each of the symmetric rank-three tensors (or
matrices) $\tensor{\bm D}_{n,n'}^g$ can be diagonalized by three
rotations: a rotation by the angle $\phi^g_{n,n'}$ about the
molecular $z$ axis, then a rotation by the angle $\theta^g_{n,n'}$
about the rotated $\tilde{x}$ axis, followed by a rotation by the
angle $\psi^g_{n,n'}$ about the rotated $\tilde{z}$
axis.\cite{Goldstein} This necessarily leads to the three
principal axes $\hat{\tilde{\bm x}}^g_{n,n'}$, $\hat{\tilde{\bm
y}}^g_{n,n'}$, and $\hat{\tilde{\bm z}}^g_{n,n'}$.   For the
single-ion axes with $n'=n$, we denote these principal axes to be
$\hat{\tilde{\bm x}}^g_n$, $\hat{\tilde{\bm y}}^g_n$, and
$\hat{\tilde{\bm z}}^g_n$, respectively,  which are written
explicitly in Sec. III.  The non-vanishing matrix elements in
these locally-diagonalized symmetric matrix coordinates are
$\tilde{D}^{g,s}_{n,n',xx}$, $\tilde{D}^{g,s}_{n,n',yy}$ and
$\tilde{D}^{g,s}_{n,n',zz}$. Since the structural information in
each of the $\tensor{\tilde{\bm D}}^{g,s}_{n,n'}$ depends upon the
local environment, in the absence of molecular group $g$ symmetry,
each of these angles would in principle be different from one
another.

Although an antisymmetric exchange matrix $\tensor{\bm
D}_{n,n'}^{g,a}$ can generally be diagonalized by a unitary
transformation, it contains at most three independent, real
parameters, which can be incorporated into the components of a
three-vector, ${\bm d}_{n,n'}^g$, with an effective spin-spin
interaction of the form ${\bm d}_{n,n'}^g\cdot({\bm S}_n\times{\bm
S}_{n'})$,\cite{Moriya,Dzyaloshinskii} which is easiest to write
in the molecular representation.

For the six high-symmetry groups under study, we analyze the
effects of molecular group symmetry upon the single-ion and
anisotropic exchange parts of ${\cal H}^g$.  The  group
symmetries further restrict the number of independent parameters.

In the absence of any anisotropy interactions, the bare
Hamiltonian ${\cal H}^g_0$ is given by the Zeeman and Heisenberg
interactions,
\begin{eqnarray}
{\cal H}_0^g&=&-\gamma {\bm B}\cdot{\bm S}-J_g'({\bm S}_1\cdot{\bm
S}_3+{\bm S}_2\cdot{\bm S}_4)\nonumber\\
& &-J_g({\bm S}_1\cdot{\bm S}_2+{\bm S}_2\cdot{\bm S}_3+{\bm
S}_3\cdot{\bm S}_4+{\bm S}_4\cdot{\bm S}_1),\label{H0bare}
\end{eqnarray}
which can be rewritten as
\begin{eqnarray}
 {\cal
H}_0^g&=&-\frac{J_g}{2}{\bm S}^2-\gamma{\bm B}\cdot{\bm
S}-\frac{(J_g'-J_g)}{2}({\bm S}_{13}^2+{\bm S}_{24}^2),\label{H0}
\end{eqnarray}
where  ${\bm S}_{13}={\bm S}_1+{\bm S}_3$, ${\bm S}_{24}={\bm
S}_2+{\bm S}_4$, and ${\bm S}={\bm S}_{13}+{\bm S}_{24}$ is the
total spin operator,\cite{KlemmLuban} and we dropped an
irrelevant, overall constant. In Eq. (\ref{H0}),
 \begin{eqnarray} J_{T_d}'&=&J_{T_d},\label{JTdp}\\
J_{g}'&\ne&J_{g}\label{Jgp} \end{eqnarray} for
$g=D_{2d},S_4,D_{4h}, C_{4h}$, and $C_{4v}$.  In terms of the
diagonalized matrix elements,
$-2J_g=\tilde{D}^{g,s}_{1,2,xx}+\tilde{D}^{g,s}_{1,2,yy}$ and
$-2J_g'=\tilde{D}^{g,s}_{1,3,xx}+\tilde{D}^{g,s}_{1,3,yy}$, for
instance. For our $c/a<1$ convention, $-J_g$ and $-J_g'$ are the
NN and NNN Heisenberg interactions for $g=D_{2d},S_4$, $C_{4v},
C_{4h}$ and $D_{4h}$.

\section{III. The single-ion and anisotropic exchange Hamiltonians}

To  take account of the molecular group $g$ symmetries, it is
useful to write the single-ion and symmetric anisotropic exchange
interactions in terms of the  local coordinates.  In this section,
we write the local Hamiltonian for these interactions, and the
molecular Hamiltonian for the antisymmetric exchange interactions.
In Sec. IV, we then impose the group symmetries on these
interactions for $C_{4h}, D_{4h}, C_{4v}, S_4, D_{2d}$, and $T_d$
molecular group symmetries, respectively.

\subsection{A. Local single-ion Hamiltonian}

For the single-ion anisotropy, we  define the local  vector basis
for the $n$th site to be $\{\hat{\tilde{\bm x}}_n^g,
\hat{\tilde{\bm y}}_n^g, \hat{\tilde{\bm z}}_n^g\}$ for each $g$.
These basis elements are the vectors that diagonalize the single
ion matrix from $\tensor{\bm D}_{n,n}^g$ to $\tensor{\tilde{\bm
D}}_{n,n}^g$.\cite{Goldstein} Since we employ these vectors
repeatedly, we write them here for simplicity of presentation.
 The diagonalized vector set elements  may be written in the molecular
 $(\hat{\bm x},\hat{\bm y},\hat{\bm z})$ representation as
\begin{eqnarray}
\hat{\tilde{\bm x}}_{n}^g&=&\left(\begin{array}{c}
\cos\phi_n^g\cos\psi_n^g-\cos\theta_n^g\sin\phi_n^g\sin\psi_n^g\\
\sin\phi_n^g\cos\psi_n^g+\cos\theta_n^g\cos\phi_n^g\sin\psi_n^g\\
\sin\theta_n^g\sin\psi_n^g\end{array}\right),\label{xng}\\
\hat{\tilde{\bm y}}_{n}^g&=&\left(\begin{array}{c}
-\cos\phi_n^g\sin\psi_n^g-\cos\theta_n^g\sin\phi_n^g\cos\psi_n^g\\
-\sin\phi_n^g\sin\psi_n^g+\cos\theta_n^g\cos\phi_n^g\cos\psi_n^g\\
\sin\theta_n^g\cos\psi_n^g\end{array}\right),\label{yng}\\
\hat{\tilde{\bm z}}_{n}^g&=&\left(\begin{array}{c}
\sin\theta_n^g\sin\phi_n^g\\
-\sin\theta_n^g\cos\phi_n^g\\
\cos\theta_n^g\end{array}\right),\label{zng}
\end{eqnarray}
which satisfy $\hat{\tilde{\bm x}}^g_n\times\hat{\tilde{\bm
y}}_n^g=\hat{\tilde{\bm z}}_n^g$. We then write the most general
quadratic single-ion anisotropy interaction as
\begin{eqnarray}
{\cal H}^{g,\ell}_{si}&=&-\sum_{n=1}^4\Bigl(J_{a,n}^g({\bm
S}_n\cdot\hat{\tilde{\bm
z}}^g_n)^2\nonumber\\
& &+J_{e,n}^g[({\bm S}_n\cdot\hat{\tilde{\bm x}}^g_n)^2-({\bm
S}_n\cdot\hat{\tilde{\bm y}}^g_n)^2]\Bigr),\label{Hsi}
\end{eqnarray}
in terms of the site-dependent axial and azimuthal interactions
$J_{a,n}^g, J_{e,n}^g$, analogous in notation  to that for
homoionic dimers.\cite{ekshort,ek2} In terms of the diagonalized
matrix elements,
$-J_{a,n}^g=\tilde{D}^{g,s}_{n,n,zz}-(\tilde{D}^{g,s}_{n,n,xx}+\tilde{D}^{g,s}_{n,n,yy})/2$
and
$-J_{e,n}^g=(\tilde{D}^{g,s}_{n,n,xx}-\tilde{D}^{g,s}_{n,n,yy})/2$.

\subsection{B. Local symmetric anisotropic exchange Hamiltonian}

In addition to the single-ion interactions, the other microscopic
anisotropic interactions are the anisotropic exchange
interactions, which include the intracluster dipole-dipole
interactions.\cite{Jackson} The intercluster dipole-dipole
interactions can lead to low-$T$ hysteresis in the
phenomenological total spin model,\cite{Marisol} but in the
microscopic individual spin model, are generally much weaker than
the intracluster ones due to the larger distances involved. Hence,
we neglect those and all other intercluster interactions, such as
those mediated by phonons. As for the single-ion interactions, we
first construct the symmetric anisotropic exchange Hamiltonian
${\cal H}_{ae}^g$ in the local group coordinates.   In this case,
there are distinct local vector sets for the NN and NNN exchange
interactions. Diagonalization of the symmetric anisotropic
exchange matrix $\tensor{\bm D}_{n,n'}^{g,s}$ leads to
$\tensor{\tilde{\bm D}}_{n,n'}^{g,s}$ and the vector basis
$\{\hat{\tilde{\bm x}}_{n,n'}^g, \hat{\tilde{\bm y}}_{n,n'}^g,
\hat{\tilde{\bm z}}_{n,n'}^g\}$, given by Eqs.
(\ref{xng})-(\ref{zng}) with the subscript $n$ replaced by $n,n'$.

The local symmetric anisotropic exchange Hamiltonian ${\cal
H}_{ae}^{g,\ell}$ is then generally given by
\begin{eqnarray}
 {\cal
H}_{ae}^{g,\ell}&=&-\sum_{q=1}^2\sum_{n=1}^{6-2q}\Bigl[J^{f,g}_{n,n+q}({\bm
S}_n\cdot\hat{\tilde{\bm z}}_{n,n+q}^g)({\bm
S}_{n+q}\cdot\hat{\tilde{\bm
z}}_{n,n+q}^g)\nonumber\\
& &+J^{c,g}_{n,n+q}\Bigl(({\bm S}_n\cdot\hat{\tilde{\bm
x}}_{n,n+q}^g)({\bm
S}_{n+q}\cdot\hat{\tilde{\bm x}}_{n,n+q}^g)\nonumber\\
& &\qquad-({\bm S}_n\cdot\hat{\tilde{\bm y}}_{n,n+q}^g)({\bm
S}_{n+q}\cdot\hat{\tilde{\bm
y}}_{n,n+q}^g)\Bigr)\Bigr],\label{Hae}
\end{eqnarray}
where we define ${\bm S}_5\equiv{\bm S}_1$, as if the four NN
spins were on a ring.  In Eq. (\ref{Hae}), the axial and azimuthal
interaction strengths
$-J_{n,n'}^{f,g}=\tilde{D}^{g,s}_{n,n',zz}-(\tilde{D}^{g,s}_{n,n',xx}+\tilde{D}^{g,s}_{n,n',yy})/2$
and
$-J_{n,n'}^{c,g}=(\tilde{D}^{g,s}_{n,n',xx}-\tilde{D}^{g,s}_{n,n',yy})/2$,
as for the single-ion interaction strengths.  The subscripts $a,e$
and superscripts $f,c$ correspond to our dimer notation.\cite{ek2}

\subsection{C. Antisymmetric anisotropic exchange Hamiltonian}
As noted above, we  write the antisymmetric anisotropic exchange,
or Dzyaloshinskii-Moriya (DM),\cite{Moriya,Dzyaloshinskii}
Hamiltonian ${\cal H}_{DM}^g$ in the molecular
representation,\cite{BenciniGatteschi}
\begin{eqnarray}
{\cal H}_{DM}^g&=&\sum_{q=1}^2\sum_{n=1}^{6-2q}{\bm
d}^{g}_{n,n+q}\cdot\Bigl({\bm S}_n\times{\bm
S}_{n+q}\Bigr).\label{HDM}
\end{eqnarray}
We note that in these molecular coordinates, the DM interaction
three-vectors ${\bm d}^{g}_{n,n+q}$ depend explicitly upon the
exchange bond indices $n,n+q$ for each group $g$. We then employ
the local group symmetries to relate them to one another.

The rules for the directions of the ${\bm d}_{n,n+q}^g$ were given
by Moriya,\cite{Moriya} and were employed for a dimer example by
Bencini and Gatteschi.\cite{BenciniGatteschi}  The Moriya rules
are: (1) ${\bm d}^g_{n,n'}$ vanishes if a center of inversion
connects ${\bm r}_n$ and ${\bm r}_{n'}$. (2) When a mirror plane
contains ${\bm r}_n$ and ${\bm r}_{n'}$, ${\bm d}^g_{n,n'}$ is
normal to the mirror plane.  (3) When a mirror plane is the
perpendicular bisector of ${\bm r}_n-{\bm r}_{n'}$, ${\bm
d}_{n,n'}^g$ lies in the mirror plane. (4) When a two-fold
rotation axis is the perpendicular bisector of ${\bm r}_n-{\bm
r}_{n'}$, then ${\bm d}_{n,n'}^g$ is orthogonal to the rotation
axis. (5) When ${\bm r}_n-{\bm r}_{n'}$ is an $r$-fold rotation
axis with $r>2$, then ${\bm d}_{n,n'}^g$ is parallel to ${\bm
r}_n-{\bm r}_{n'}$. As noted above, we shall incorporate these
rules  in the molecular representation.  For example, in
NaV$_2$O$_5$, the lack of inversion symmetry between interacting
spins has been shown to lead to a DM interaction.\cite{NaV2O5}

\section{IV. Group symmetry invariance}

\subsection{A. General considerations}

In this section, we impose the set of allowed group $g$ symmetry
operations upon the full Hamiltonian ${\cal H}^g$. These symmetry
operations are represented by the matrices ${\cal O}_{\lambda}$
for $\lambda=1,\ldots,26$ listed in Subsection A of the Appendix.
For each $g$, we require ${\cal H}^g$ to be invariant under each
symmetry operation ${\cal O}_{\lambda}^g$ for each allowed
$\lambda$.

For the six $g$ cases under study, the set $\{{\cal
O}_{\lambda}^g\}$ of group operations greatly reduces the number
of single-ion and symmetric anisotropic exchange parameters.  As
we shall see, in each group $g$, these reduce  the single-ion and
symmetric anisotropic exchange interaction strength set to
\begin{eqnarray}
\{J_j^g\}&\equiv&\{J_a^g,J_e^g,J_{f,q}^g,J_{c,q}^g\},
\end{eqnarray}
for $q=1,2$, which are independent of the site index $n$.   That
is, for each $g$, there are at most two single-ion, two NN and two
NNN symmetric anisotropic exchange interaction strengths. In
addition, for these six $g$ cases, the group operations further
limit the number of vector set parameters to
\begin{eqnarray}
\mu_1^g&=&\{\theta_1^g,\phi_1^g,\psi_1^g\},\\
\mu_{1q}^g&=&\{\theta_{1p}^g,\phi_{1p}^g,\psi_{1p}^g\},
\end{eqnarray}
where $p=q+1=2,3$, and we used the notation
$\theta_1^g=\theta_{1,1}^g$, $\theta_{1p}^g=\theta_{1,p}^g$, etc.
Some of these parameters may be further restricted.  In addition,
however, the  molecular single-ion and anisotropic exchange
Hamiltonians contain both site-independent and site-dependent
terms.

For ${\cal H}_{DM}^{g}$, we first impose the Moriya rules on each
anisotropic exchange pair,\cite{Moriya,BenciniGatteschi} and then
impose the required group symmetries on the six pairs.  For the
six groups under study, the group symmetries place restrictions
upon the ${\bm d}_{n,n+q}^g$, leading to the anisotropic exchange
parameter set
\begin{eqnarray}
d^g&=&\{d_{z}^g,d_{x1}^g,d_{y1}^g,d_{x2}^g,d_{y2}^g\}.
\end{eqnarray}
For each of the six $g$ cases, the NNN DM parameter set has at
least one more restriction than does the NN DM parameter set. Some
$g$ symmetries lead to site-dependent signs of the components of
$d^g$.

\begin{table}
\begin{tabular}{cccc}
$g$&$\theta_1^g$&$\phi_1^g$&$\psi_1^g$\\
\noalign{\vskip3pt}
\hline\\
\noalign{\vskip-7pt}
$C_{4h}$&0&$\phi_1^{g}$&0\\
\noalign{\vskip5pt}
$D_{4h}$&0&$\frac{\pi}{4}$&0\\
\noalign{\vskip5pt}
$C_{4v},D_{2d}$&$\theta_1^{g}$&$\frac{3\pi}{4}$&$-\frac{\pi}{2}$\\
\noalign{\vskip5pt}
$S_4$&$\theta_1^{S_4}$&$\phi_1^{S_4}$&$\psi_1^{S_4}$\\
\noalign{\vskip5pt} $T_d$&$\tan^{-1}\sqrt{2}$&$\frac{3\pi}{2}$&0
\end{tabular}
\caption{Lists of the single-ion parameter sets
$\mu_1^g$.}\label{tab1}
\end{table}
\begin{table}
\begin{tabular}{cccc}
$g$&$\theta_{1p}^g$&$\phi_{1p}^g$&$\psi_{1p}^g$\\
\noalign{\vskip3pt}
\hline\\
\noalign{\vskip-7pt}
$C_{4h}$&0&$\phi_{1p}^{C_{4h}}$&0\\
$D_{4h}$&0&$\frac{\pi}{4}$&0\\
$C_{4v}$&0&$\frac{(p-2)\pi}{4}$&0\\
\noalign{\vskip5pt}
$S_4$&$\theta_{1p}^{S_4}$&$\phi_{1p}^{S_4}$&$\psi_{1p}^{S_4}$\\
\noalign{\vskip5pt}
$D_{2d}$&$\theta_{12}^{D_{2d}}\delta_{p,2}+\frac{\pi}{2}\delta_{p,3}$&$\frac{(-1)^p\pi}{2(p-1)}$&$\frac{(p-2)\pi}{2}$\\
\end{tabular}
\caption{Lists of the relevant NN ($p=2$) and NNN ($p=3$)
parameter sets $\mu_{1p}^g$.}\label{tab2}
\end{table}
\begin{table}
\begin{tabular}{cccccc}
$g$&$d_z^g$&$d_{x1}^g$&$d_{y1}^g$&$d_{x2}^g$&$d_{y2}^g$\\
\noalign{\vskip3pt}
\hline\\
\noalign{\vskip-7pt} $C_{4h},D_{4h}$&$d_z^{g}$&0&0&0&0\\
 \noalign{\vskip5pt}
$S_4$&$d_z^{S_4}$&$d_{x1}^{S_4}$&$d_{y1}^{S_4}$&$d_{x2}^{S_4}$&$d_{y2}^{S_4}$\\
\noalign{\vskip5pt}
$D_{2d}$&$d_z^{D_{2d}}$&0&$d_{y1}^{D_{2d}}$&$d_{x2}^{D_{2d}}$&0\\
\noalign{\vskip5pt}
 $T_d, C_{4v}$&0&0&0&0&0
\end{tabular}
\caption{Lists of the DM parameter sets $d^g$.}\label{tab3}
\end{table}

\subsection{C. Imposing the group symmetries}

In Subsection A of the Appendix, we describe the matrices ${\cal
O}_{\lambda}$ for $\lambda=1,\ldots,26$ representing the group
symmetry operations for $g = C_{4h}, D_{4h}, C_{4v}, S_4, D_{2d}$,
and $T_d$. For each molecular group $g$, the allowed symmetry
operations ${\cal O}_{\lambda}$ commute with the Hamiltonian.  For
a particular $\lambda$, ${\cal O}_{\lambda}{\bm r}_n={\bm
r}_{n'(\lambda)}$.   We therefore take ${\bm S}_n={\bm S}({\bm
r}_n)$, so that ${\cal O}_{\lambda}{\bm S}_n={\bm S}({\cal
O}_{\lambda}{\bm r}_n)={\bm S}({\bm r}_{n'(\lambda)})={\bm
S}_{n'(\lambda)}$.

For $C_{4h}$ symmetry,  besides the trivial identity operation,
the allowed group operations  are clockwise and counterclockwise
rotations by $\pi/2$ about the $z$ axis, and reflections in the
$xy$ plane.\cite{Tinkham}  These operations are represented
respectively by the matrices ${\cal O}_{1,2,6}$.  We use this
simple case to illustrate how the symmetries are imposed.  We
first consider the axial part of ${\cal H}_{si}^{C_{4h},\ell}$,
and set
\begin{eqnarray}
\sum_{n=1}^4J_{a,n}^{C_{4h}}({\bm S}_n\cdot\hat{\tilde{\bm
z}}_n^{C_{4h}})^2&=&\sum_{n=1}^4J_{a,n}^{C_{4h}}{\cal O}_1({\bm
S}_n\cdot\hat{\tilde{z}}_n^{C_{4h}})^2{\cal O}_1^{T}.\label{symmetry}\nonumber\\
\end{eqnarray}
We interpret $\hat{\tilde{\bm z}}_n^{C_{4h}}$ as its vector
transpose, $(\hat{\tilde{\bm z}}_n^{C_{4h}})^{T}$ in Eq.
(\ref{symmetry}), and obtain
\begin{eqnarray}
{\cal O}_1{\bm S}_n&=&{\bm S}_{n+1},\label{O1Sn}\\
 (\hat{\tilde{\bm
z}}_n^{C_{4h}})^{T}{\cal O}_1^{T}&=&\Bigl(-\sin\theta_n^{C_{4h}}\cos\phi_n^{C_{4h}},\nonumber\\
& &\qquad-\sin\theta_n^{C_{4h}}\sin\phi_n^{C_{4h}},\cos\theta_n^{C_{4h}}\Bigr).\label{O1zn}\nonumber\\
\end{eqnarray}
Substituting these into the right-hand side of  Eq.
(\ref{symmetry}),  setting $n\rightarrow n+1$ in the left-hand
side, and equating coefficients of $S_{n+1,\alpha}S_{n+1,\beta}$
for $\alpha,\beta=x,y,z$ leads to
\begin{eqnarray}
J_{a,n}^{C_{4h}}&=&J_{a,n+1}^{C_{4h}}=J_a^{C_{4h}},\\
\theta_{n}^{C_{4h}}&=&\theta_{n+1}^{C_{4h}}=\theta_1^{C_{4h}},\\
\phi_{n}^{C_{4h}}&=&\phi_{n+1}^{C_{4h}}+\frac{\pi}{2}.
\end{eqnarray}
Then, imposing ${\cal O}_6$ symmetry, we have ${\cal O}_6{\bm
S}_n={\bm S}_n$, and either $\theta_1^{C_{4h}}=\pi/2$ or
$\theta_1^{C_{4h}}=0$, both of which lead to invariance of this
part of the Hamiltonian under ${\cal O}_6$  We therefore take the
easy-axis case, $\theta_1^{C_{4h}}=0$.  Carrying out similar
transformations on the azimuthal single-ion Hamiltonian leads to
\begin{eqnarray}
J_{e,n}^{C_{4h}}&=&J_e^{C_{4h}},\\
\chi_n^{C_{4h}}&=&\phi_{n}^{C_{4h}}+\psi_n^{C_{4h}}=\chi_{n+1}^{C_{4h}}+\frac{\pi}{2}.
\end{eqnarray}
We could then choose $\psi_1^{C_{4h}}=0$, leaving one free angle
parameter $\phi_1^{C_{4h}}$, as listed in Table I, plus the two
interaction strengths $J_a^{C_{4h}}, J_e^{C_{4h}}$.

The symmetric anisotropic exchange Hamiltonian,  ${\cal
H}_{ae}^{C_{4h},\ell}$ can be made invariant under ${\cal
O}_1,{\cal O}_2,{\cal O}_6$ in a very similar fashion. The
operations for the other five $g$ symmetries are listed in
Subsection A of the Appendix, along with the associated matrices.
Our results for the single-ion, symmetric anisotropic exchange,
and DM interaction parameters are compiled in Tables I-III.

\subsection{D. Induced electric polarizations}

  As shown by Katsura {\it et
al.},\cite{Katsura} the spin-orbit interactions  between spins at
sites $n$ and $n'$ can induce an electric polarization
\begin{eqnarray}{\bm P}_{n,n'}\sim\hat{\bm r}_{n,n'}\times({\bm S}_n\times{\bm
S}_{n'}),
\end{eqnarray}
where $\hat{\bf r}_{n,n'}$ is a unit vector directed from site $n$
to site $n'$.
 In our model, the  thermodynamic averages of such polarizations vanish in the absence
 of  DM interactions, but non-vanishing in-plane vector components ${\bm d}_q^g$ of the ${d}^g$ DM
 interaction parameter sets allow them to become
 finite.
 Tetramers with the rather low molecular group
symmetries $S_4$ and $D_{2d}$ have no overall center of inversion
symmetry, and contain a complex set of DM interactions.  Depending
upon the polarizability of the attached ligand groups, this  may
lead to a combined spin-induced electric polarization
\begin{eqnarray} {\bm
P}_s&\propto&\frac{1}{2}\sum_{n,n'=1}^4\hat{\bm
r}_{n,n'}\times\langle{\bm S}_n\times{\bm
S}_{n'}\rangle,\end{eqnarray} where $\langle\ldots\rangle$
represents the thermodynamic average in the presence of the full
Hamiltonian, including the relevant DM interactions.

Besides the direct DM interactions, we predict the possibility of
dual, or induced, DM interactions.  Although  ${\bm d}_q^g$ DM
interactions between individual spin pairs are allowed in
tetramers with the lowest $C_{4h}$ and $D_{4h}$ symmetries
studied, the group symmetry causes the dipole moments on opposite
sides of their square geometries to cancel one another.
 Although we have not studied this point in detail, tetramers with these symmetries can in principle be made to exhibit
additional effective DM interactions by application of an electric
field ${\bm E}\ne0$.\cite{Katsura,Mostovoy} Thus,  in tetramers
with $S_4$, $D_{2d}$, or lower symmetry, a multiferroic effect can
occur,\cite{Mostovoy} in which both DM interactions and ${\bm
P}_s\ne0$.  More generally, multiferroic effects arise in systems
such as some dimers, trimers, and tetramers that generally do not
have a center of inversion at the midpoints of the ${\bm
r}_{n,n'}$.\cite{ek2}

\section{V. The Hamiltonian in the molecular
representation}
\subsection{A.  The molecular single-ion Hamiltonian}

To make contact with experiment, we use the group symmetries to
rewrite ${\cal H}^{g,\ell}_{si}$ in the molecular $(\hat{\bm
x},\hat{\bm y},\hat{\bm z})$ representation,
\begin{eqnarray}{\cal
H}^g_{si}&=&-\sum_{n}\Bigl(J_z^gS_{n,z}^2+(-1)^nJ_{xy}^g(S_{n,x}^2-S_{n,y}^2)\nonumber\\
&
&+\sum_{\alpha\ne\beta}K_{\alpha\beta}^g(n)S_{n,\alpha}S_{n,\beta}\Bigr),\label{Hsimolecular}
\end{eqnarray}
where $\alpha,\beta=x,y,z$, and we subtracted an irrelevant
constant.  ${\cal H}_{si}^g$ contains the site-independent
interactions $J_z^g$ and the site-dependent interactions
$(-1)^nJ_{xy}^g$ and $K_{\alpha\beta}^g(n)$, which are written in
terms of the parameter sets $\mu_1^g$ in Subsection B of the
Appendix. Most important is the result that for $T_d$ symmetry,
\begin{eqnarray}
J_z^{T_d}&=&0. \end{eqnarray}

The first-order contributions to the eigenstate energies from the
site-dependent interactions $(-1)^nJ_{xy}^g$ and
$K_{\alpha\beta}^g(n)$ vanish.  Hence, these interactions only
contribute to the eigenstate energies to second and higher orders
in the interactions $J_a^g$ and $J_e^g$. For $C_{4v}, D_{2d}$ and
$S_4$, the effective axial site-independent interactions $J_z^g$
arise from a combination of the local axial and azimuthal
interactions $J_a^g$ and $J_e^g$.
 For $g\ne T_d$, $J_z^g$ can be large, even if the molecular structure is
 nearly $T_d$.

\subsection{B. Symmetric anisotropic exchange in the molecular
representation}
 We then
construct the group-invariant symmetric anisotropic exchange
Hamiltonian in the molecular coordinates for the six $g$
symmetries. For $g=D_{2d},S_4$, there are renormalizations of the
isotropic exchange interactions, modifying ${\cal H}_0^g$ to
\begin{eqnarray}
{\cal H}_0^{g,r}&=&-\frac{\tilde{J}_g}{2}{\bm S}^2-\gamma{\bm
B}\cdot{\bm S}-\frac{(\tilde{J}_g'-\tilde{J}_g)}{2}({\bm
S}_{13}^2+{\bm S}_{24}^2),\nonumber\\
\end{eqnarray}
where
\begin{eqnarray}
\tilde{J}_g&=&J_g+\delta J_{g},\\
 \tilde{J}_{g}'&=&J_g'+\delta J_g',
\end{eqnarray}
where $J_g, J_g'$ are given by Eqs. (\ref{JTdp}) and (\ref{Jgp}),
and the $\delta J_g$ and $\delta J_g^{'}$ are given in in terms of
the parameters sets $\mu_{1p}^g$ in Subsection C of the Appendix.
For the three planar symmetries, $g=C_{4h}, D_{4h}$, and $C_{4v}$,
$\delta J_g=\delta J^{'}_g=0$. For $T_d$ symmetry, there are no
group-satisfying azimuthal symmetric exchange vectors, so
$J_{c,q}^{T_d}=0$ for $q=1,2$. However, the axial vectors parallel
to ${\bm r}_{n,n'}$ satisfy all of the group symmetries, so that
the $J_{f,q}^{T_d}$ could exist, provided that
$J_{f,1}^{T_d}=J_{f,2}^{T_d}$. However, the requirement
$\tilde{J}_{T_d}=\tilde{J}_{T_d}'$ to preserve the $T_d$ symmetry
of the renormalized Heisenberg interactions forces
$J_{f,2}^{T_d}=J_{f,1}^{T_d}/2$.  Hence, we must conclude that
$J_{f,q}^{T_d}=0$ for $q=1,2$ and $\delta J_{T_d}=\delta
J_{T_d}'=0$.

 ${\cal H}_{ae}^{g,\ell}$ also leads to additional
interactions $\delta{\cal H}_{ae}^g$ in the molecular frame,
\begin{eqnarray}
\delta{\cal
H}_{ae}^{g}&=&\sum_{q=1}^2\sum_{n=1}^{6-2q}\biggl[J_{q,z}^gS_{n,z}S_{n+q,z}\nonumber\\
&&+(-1)^{n+1}\biggl(J_{q,xy}^g\nonumber\\
& &\times\Bigl[S_{n,x}S_{n+q,x}-S_{n,y}S_{n+q,y}\Bigr]\nonumber\\
&
&+\sum_{\alpha\ne\beta}K_{q,\alpha\beta}^g(n)S_{n,\alpha}S_{n+q,\beta}\biggr)\biggr],
\label{Haemolecular}
\end{eqnarray}
where $\alpha,\beta=x,y,z$. As for the single-ion interactions in
the molecular representation, the site-independent symmetric
exchange interactions $J_{q,z}^g$ contribute to the eigenstate
energies to first order, but the first-order contributions to the
eigenstate energies from the site-dependent interactions vanish.
Both the site-independent and site-dependent symmetric exchange
interactions are given in terms of the parameter sets $\mu_{1p}^g$
in Subsection C of the Appendix.

\subsection{C. Antisymmetric exchange Hamiltonian}

In Secs. IV and V, we already evaluated the antisymmetric exchange
Hamiltonians in the molecular representation, and the parameter
sets are listed in Table III. These six ${\cal H}_{DM}^g$ may be
combined  as
\begin{eqnarray}
{\cal H}_{DM}^g&=&\sum_{q=1}^2\sum_{n=1}^{6-2q}\bigl({\bm
S}_n\times{\bm
S}_{n+q}\bigr)\cdot\Bigl(d^g_z(n)\delta_{q,1}\hat{\bm z}\nonumber\\
& &+{\bm d}_q^g\sin(n\pi/2) +(\hat{\bm z}\times{\bm
d}^g_q)\cos(n\pi/2)\Bigr),\nonumber\\\label{HDM}
\end{eqnarray}
where the scalar $d_z^g(n)$ and the two two-vectors ${\bm d}_q^g$
all vanish for $g=C_{4v},T_d$, but for the other four symmetries
are given in Subsection C of the Appendix.

We  note that both site-dependent and site-independent DM
interactions give rise to second-order eigenstate energy
corrections, and can only be neglected in fits to experiment for
tetramers with symmetries very close to $T_d, C_{4v}$, or higher.
In Subsection D of the Appendix, we give ${\cal H}_{DM}^g$ for the
lower symmetry $C_{2v}^{13}$ tetramers.

\subsection{D. Biquadratic and three-center quartic isotropic exchange interactions}
In the previous subsections, we listed the quadratic single-ion
and anisotropic exchange interactions for the six high-symmetry
tetramer  groups under study.  However, in the lower-symmetry AFM
tetramers $\{$Ni$_4$Mo$_{12}\}$, with $C_{1v}$
symmetry,\cite{Schnack} and  Ni$_4$ and Co$_4$ [2$\times2$] grids
(or rhombuses), with approximate $C_{2v}^{13}$
symmetry,\cite{WaldmannNi4,Waldmannreview,KRS} fits to
magnetization data were facilitated by the inclusion of
biquadratic interactions.\cite{Schnack,WaldmannNi4,Waldmannreview}
In the former case, the powder fits  assumed field-dependent
interaction parameters, but the single-ion interactions were
assumed to have $T_d$ symmetry, which vanish to first order in
their strength, and the anisotropic exchange interactions were
neglected. In the latter $C_{2v}^{13}$ case, the authors neglected
the NN DM interactions given in the Appendix.  Subsequently,
Kostyuchenko showed that three-center isotropic quartic
interactions should be comparable in magnitude to the biquadratic
interactions, and provided a fit to the midpoints of the
level-crossing magnetization behavior on $\{$Ni$_4$Mo$_{12}\}$,
without making the assumption of strong field dependence to the
Heisenberg interactions.\cite{Kostyuchenko}  Here we provide
preliminary fits to the AFM $\{$Ni$_4$Mo$_{12}\}$ magnetization
data, extending the treatment of Kostyuchenko to include the first
order anisotropy interactions, which can fit the widths of the
transitions, as well as the midpoints. More complete fits to those
experiments and to experiments on the grid SMM's will be presented
elsewhere.\cite{ekfuture} Such fits are greatly aided by an
analytic treatment of biquadratic and three-center isotropic
quartic exchange.

For tetramers with the six $g$ symmetries under study, the
biquadratic interactions may be written as
\begin{eqnarray}
{\cal H}_{\rm b}^g&=&\sum_{q=1}^2{\cal H}_{{\rm b},q}^g\\
{\cal H}_{{\rm b},q}^g&=&-J_{{\rm
b},q}^g\sum_{n=1}^{6-2q}\bigl({\bm S}_n\cdot{\bm S}_{n+q}\bigr)^2.
\end{eqnarray}
 For $g=T_d$, we
take $J_{\rm b,1}^{T_d}=J_{\rm b,2}^{T_d}$, but otherwise $J_{\rm
b,2}^{g}\ne J_{\rm b,1}^{g}$. For the six $g$ symmetries, ${\cal
H}_{\rm b}^g$ is  invariant under all of the appropriate
symmetries.

The three-center quartic interactions for systems with the six $g$
symmetries may be written as
\begin{eqnarray}
{\cal H}_{\rm t}^g&=&\sum_{q=1}^2{\cal H}_{{\rm t},q}^g\\
{\cal H}_{{\rm t},1}^g&=&-J_{{\rm t},1}^g\sum_{{n=1}\atop{\rm
odd}}^4\sum_{{n'=1}\atop{\rm even}}^4\bigl({\bm S}_n\cdot{\bm
S}_{n+1}\bigr)\bigl({\bm S}_{n'}\cdot{\bm S}_{n'+1}\bigr),\\
{\cal H}^g_{{\rm t},2}&=&-J^g_{{\rm
t},2}\sum_{n=1}^4\sum_{n'=1}^2\bigl({\bm S}_n\cdot{\bm
S}_{n+1}\bigr)\bigl({\bm S}_{n'}\cdot{\bm S}_{n'+2}\bigr).
\end{eqnarray}

\section{VI. Eigenstates of the full Hamiltonian}

\subsection{A. Induction representation}
We assume a molecular Hamiltonian
\begin{eqnarray}{\cal
H}^g&=&{\cal H}_0^{g,r}+{\cal H}^g_{si}+\delta{\cal
H}_{ae}^g+{\cal H}_{DM}^g+{\cal H}_{\rm b}^g.
\end{eqnarray}
 To
take proper account of ${\bm B}$ in ${\cal H}_0^{g,r}$, we
construct our SMM eigenstates in the induction representation by
\begin{eqnarray}
\left(\begin{array}{c}\hat{\bm x}\\
\hat{\bm y}\\
\hat{\bm z}\end{array}\right)=\left(\begin{array}{ccc}\cos\theta\cos\phi &-\sin\phi &\sin\theta\cos\phi\\
\cos\theta\sin\phi &\cos\phi &\sin\theta\sin\phi\\
-\sin\theta
&0&\cos\theta\end{array}\right)\left(\begin{array}{c}\hat{\bm
x}'\\
\hat{\bm y}'\\
\hat{\bm z}'\end{array}\right),\label{rotation}\nonumber\\
\end{eqnarray}
so that ${\bm B}=B\hat{\bm z}'$. A subsequent arbitrary rotation
about $\hat{\bm z}'$ does not affect the eigenstates.\cite{ek2} We
then set $\hbar=1$ and write
\begin{eqnarray}
{\bm S}^2|\psi_{s,m}^{s_{13},s_{24}}\rangle&=&s(s+1)|\psi_{s,m}^{s_{13},s_{24}}\rangle,\label{S}\\
{\bm S}_{13}^2|\psi_{s,m}^{s_{13},s_{24}}\rangle&=&s_{13}(s_{13}+1)|\psi_{s,m}^{s_{13},s_{24}}\rangle,\\
{\bm S}_{24}^2|\psi_{s,m}^{s_{13},s_{24}}\rangle&=&s_{24}(s_{24}+1)|\psi_{s,m}^{s_{13},s_{24}}\rangle,\\
S_{\tilde{z}}|\psi_{s,m}^{s_{13},s_{24}}\rangle&=&m|\psi_{s,m}^{s_{13},s_{24}}\rangle,\label{Sz}\\
 S_{\tilde{\sigma}}|\psi_{s,m}^{s_{13},s_{24}}\rangle&=&A_s^{\tilde{\sigma}
m}|\psi_{s,m+\tilde{\sigma}}^{s_{13},s_{24}}\rangle,\label{SxSy}\\
A_s^{m}&=&\sqrt{(s-m)(s+m+1)},\label{Asm}
\end{eqnarray}
 where $S_{\tilde{\sigma}}=S_{\tilde{x}}+i\tilde{\sigma} S_{\tilde{y}}$ with $\tilde{\sigma}=\pm$.
For brevity, we define
\begin{eqnarray}
\nu&\equiv&\{s,m,s_{13},s_{24},s_1\},\label{nu}\\
\overline{\nu}&\equiv&\{s,s_{13},s_{24},s_1\},
\label{overlinenu}\end{eqnarray} where $\overline{\nu}$ excludes
$m$, and write
$|\nu\rangle\equiv|\psi_{s,m}^{s_{13},s_{24}}\rangle$. ${\cal
H}_{\rm b,2}^g$ and ${\cal H}_0^{g,r}$
 are both diagonal in this representation, but ${\cal H}_{si}^g,
 \delta{\cal H}_{ae}^g$, and ${\cal H}_{\rm b,1}^g$ are
not, and we  therefore  assume their interaction strengths to be
small, relative to $\tilde{J}_g$ and $\tilde{J}_g'$. From Eqs.
(\ref{S}) to (\ref{Sz}), $\langle\nu'|{\cal H}^{g,r}_0+{\cal
H}_{\rm b,2}^g|\nu\rangle=E_{\nu,0}^g\delta_{\nu',\nu}$, where
\begin{eqnarray}
E_{\nu,0}^g&=&-\frac{\tilde{J}_g}{2}s(s+1)-\gamma
Bm+\delta E_{\overline{\nu},0}^g\label{E0}\\
\delta E_{\overline{\nu},0}^g&=&-\frac{1}{2}\sum_{n=1}^2\Bigl[(\tilde{J}_g'-\tilde{J}_g)s_{n,n+2}(s_{n,n+2}+1)\nonumber\\
& &+\frac{J_{\rm
b,2}^{g}}{2}[-2s_1(s_1+1)+s_{n,n+2}(s_{n,n+2}+1)]^2\nonumber\\
& &+\frac{J^g_{{\rm t},2}}{2}[-2s_1(s_1+1)+s_{n,n+2}(s_{n,n+2}+1)]\nonumber\\
& &\times\Bigl(s(s+1)-\sum_{n'=1}^2s_{n',n'+2}(s_{n',n'+2}+1)\Bigr)\Bigr],\label{deltaE0} \nonumber\\
 \label{deltaE0}
\end{eqnarray}
where the $\tilde{J}_g$ and $\tilde{J}_g'$ are given by Eqs.
(\ref{JTdp}), (\ref{Jgp}), and (\ref{deltaJC4h})-(\ref{deltaJS4}).
Since ${\cal H}_0^{g,r}$ and ${\cal H}_{\rm b,2}^g$ are invariant
under all rotations, $E_{\nu,0}^g$ is independent of
$\theta,\phi$.

\subsection{B. First-order eigenstates}

 In the induction representation, we write ${\cal H}_{si}^g+\delta{\cal H}_{ae}^g+{\cal H}_{DM}^g$
as ${\cal H}_{si}'^{,g}+{\cal H}_{ae}'^{,g}+{\cal H}_{DM}'^{,g}$.
${\cal H}_{\rm b,1}^g$ is a scalar independent of the direction of
${\bm B}$.
 We then make a standard perturbation expansion for the seven
remaining microscopic anisotropy energies
$\{J_j^g\}\equiv\{J_a^g,J_e^g,J_{f,q}^g,J_{c,q}^g,J_{\rm b,1}^g\}$
for $q=1,2$ small relative to $|\tilde{J}_g|,
|\tilde{J}_g'|$.\cite{ek2} To do so, it is necessary to evaluate
the single-ion matrix elements analytically, as they contain much
of the interesting physics. Compact expressions for these matrix
elements for general $(s_1,s_2,s_3,s_4)$ are given in Subsection E
of the Appendix.

 At arbitrary ${\bm B}$ angles $(\theta,\phi)$, the first order corrections $E_{\nu,1}^g=\langle\nu|{\cal H}_{si}^{',g}+{\cal H}_{ae}^{',g}+{\cal H}_{DM}'^{,g}+{\cal H}_{{\rm b},1}^g+{\cal H}^g_{{\rm t},1}|\nu\rangle$ to the eigenstate energies for
$g=C_{4h}$, $D_{4h}$, $C_{4v}$, $S_4$, $D_{2d}$, and $T_d$
symmetries are
\begin{eqnarray}
E^{g}_{\nu,1}&=&\frac{\tilde{J}^{g,\overline{\nu}}_{z}}{2}[m^2-s(s+1)]-\delta \tilde{J}_z^{g,\overline{\nu}}\nonumber\\
&
&-\frac{[3m^2-s(s+1)]}{2}\tilde{J}^{g,\overline{\nu}}_{z}\cos^2\theta, \label{E1}\\
\tilde{J}_{z}^{g,\overline{\nu}}&=&J_z^ga^{+}_{\overline{\nu}}-J_{1,z}^gc_{\overline{\nu}}^{-}
-\frac{1}{2}J_{2,z}^ga_{\overline{\nu}}^{-},\label{Jztildemu}\\
\delta\tilde{J}_z^{g,\overline{\nu}}&=&J_z^gb_{\overline{\nu}}^{+}
-\frac{1}{4}J_{1,z}^g(b_{\overline{\nu}}^{+}+b_{\overline{\nu}}^{-})-\frac{1}{4}J_{2,z}^gb_{\overline{\nu}}^{-}\nonumber\\
& &+J_{\rm b,1}^g{\cal B}_{\overline{\nu}}+J_{{\rm t},1}^g{\cal
T}_{\overline{\nu}},\label{deltaJztildemu}
\end{eqnarray}
where analytic expressions for the $a_{\overline{\nu}}^{\pm}$,
$b_{\overline{\nu}}^{\pm}$,
 $c_{\overline{\nu}}^{\pm},$ ${\cal B}_{\overline{\nu}}$ and ${\cal T}_{\overline{\nu}}$ for general $\overline{\nu}$ are given
 in Subsection F of the Appendix, along with Tables IV-VII and VIII-XI of their simple analytic forms for the lowest
 four eigenstate manifolds of FM and AFM tetramers, respectively. We note that all of these interaction
 coefficients are invariant under $s_{13}\leftrightarrow s_{24}$, as expected.
 The DM   and all site-dependent interactions vanish in this first-order perturbation.
   Second order corrections
to the eigenstate energies will be presented
elsewhere.\cite{ekfuture}

For all six $g$ symmetries, $E_{\nu,1}^{g}$ has a form analogous
to that of the equal-spin dimer in the absence of azimuthal
single-ion and symmetric anisotropic exchange
interactions.\cite{ek2} For these high-symmetry tetramers, to
first order in the anisotropy interactions, the azimuthal
single-ion and anisotropic exchange interactions merely
renormalize the respective effective  site-independent axial
interactions. Thus, to first order, we only have two effective
isotropic exchange, two biquadratic isotropic exchange, and three
effective anisotropy interactions, $\tilde{J}_g$, $\tilde{J}_g'$,
$J_{\rm b,1}^g$, $J_{\rm b,2}^g$, $J_z^g$, $J_{1,z}^g$, and
$J_{2,z}^g$, which are fixed for a particular SMM. Nevertheless,
the first-order eigenstate energies
$E_{\nu}^g=E_{\nu,0}^g+E_{\nu,1}^{g}$, given by Eqs. (\ref{E0})
and (\ref{E1}), contain these seven effective interaction
strengths in ways that depend strongly upon the quantum number set
${\nu}$ and upon $\theta$. These different ${\nu},\theta$
dependencies can be employed to provide definitive measures of at
least some of the seven $\overline{\nu}$-independent effective
isotropic exchange and anisotropy interactions.

 \begin{figure}
\includegraphics[width=0.45\textwidth]{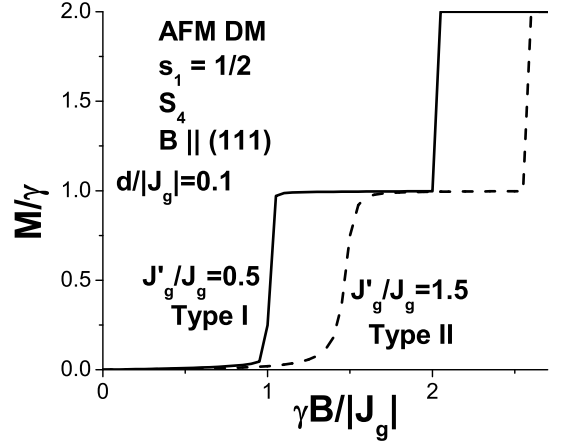}
\caption{Plots of the magnetization $M/\gamma$ versus $\gamma
B/|\tilde{J}_g|$ of an $s_1=1/2$ tetramer at $T=0$ with $c=a$,
$g=S_4$, $d_z^g=d_{1y}^g=d_{2x}^g=d_{2y}^g=d$, $d_{1x}^g=0$,
$d/|\tilde{J}_g|=0.1$, and ${\bm B}||(111)$.  The solid and dashed
curves are for the Type I ($\tilde{J}_g'/\tilde{J}_g=1.5$) and
Type II ($\tilde{J}_g'/\tilde{J}_g=0.5$) tetramers,
respectively.}\label{fig4}
\end{figure}

\begin{figure}
\includegraphics[width=0.45\textwidth]{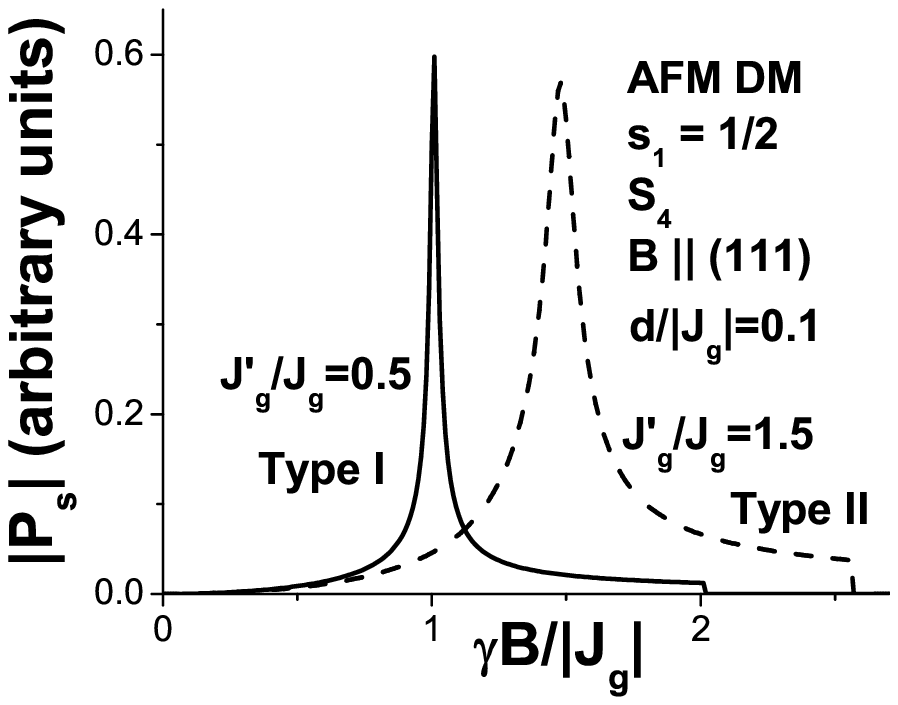}
\caption{Plots of the magnitude of the spin-derived polarization
$|{\bf P}_s|$ in arbitrary units versus $\gamma B/|\tilde{J}_g|$
at $T=0$ of an $s_1=1/2$ tetramer with $c=a$, $g=S_4$,
$d_z^g=d_{1y}^g=d_{2x}^g=d_{2y}^g=d$, $d_{1x}^g=0$,
$d/|\tilde{J}_g|=0.1$, and ${\bm B}||(111)$.  The solid and dashed
curves are for the Type I ($\tilde{J}_g'/\tilde{J}_g=1.5$) and
Type II ($\tilde{J}_g'/\tilde{J}_g=0.5$) tetramers,
respectively.}\label{fig5}
\end{figure}

\begin{figure}
\includegraphics[width=0.45\textwidth]{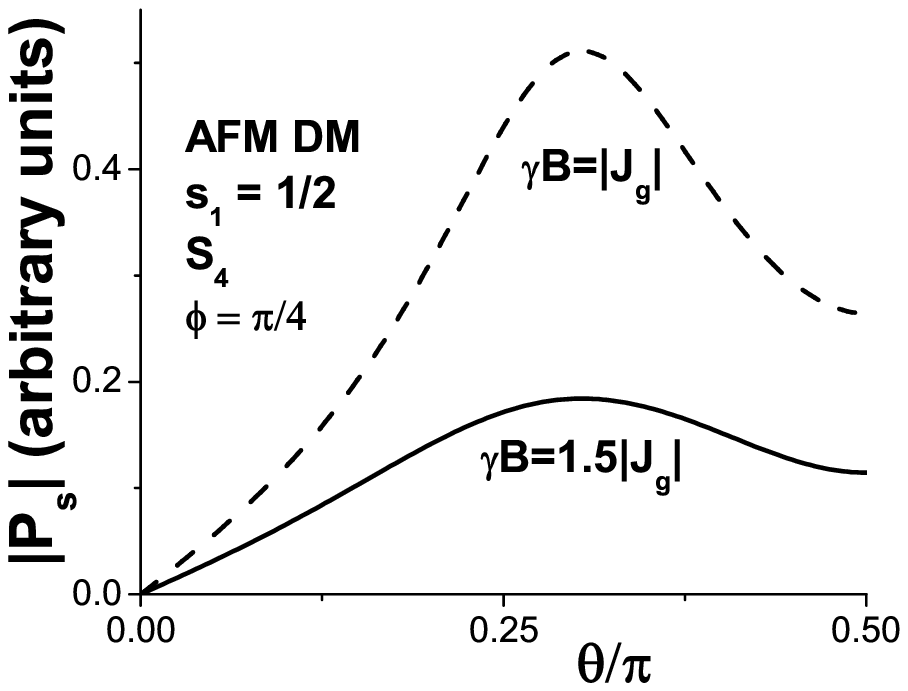}
\caption{Plots of the magnitude of the spin-derived polarization
$|{\bf P}_s|$ in arbitrary units versus $\theta/\pi$ at $T=0$ of
an $s_1=1/2$ tetramer with $c=a$, $g=S_4$,
$\tilde{J}_g'/\tilde{J}_g=1$,
$d_z^g=d_{1y}^g=d_{2x}^g=d_{2y}^g=d$, $d_{1x}^g=0$, $d/J_g= 0.1$,
and $\phi=\pi/4$. The  solid and dashed curves are for $\gamma
B/|\tilde{J}_g|=1.5,1$, respectively.}\label{fig6}
\end{figure}

 \begin{figure}
\includegraphics[width=0.45\textwidth]{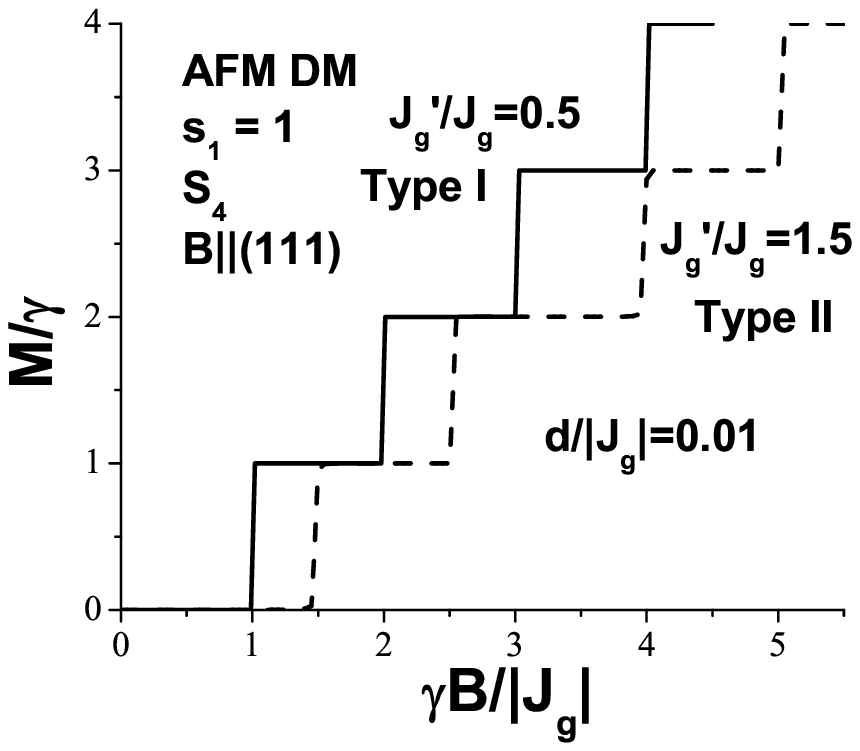}
\caption{Plots of the magnetization $M/\gamma$ versus $\gamma
B/|\tilde{J}_g|$ of an $s_1=1$ tetramer at $T=0$ with $c=a$,
$g=S_4$, $d_z^g=d_{1y}^g=d_{2x}^g=d_{2y}^g=d$, $d_{1x}^g=0$,
$d/|\tilde{J}_g|=0.01$, and ${\bm B}||(111)$.  The solid and
dashed curves are for the Type I ($\tilde{J}_g'/\tilde{J}_g=1.5$)
and Type II ($\tilde{J}_g'/\tilde{J}_g=0.5$) tetramers,
respectively.}\label{fig7}
\end{figure}

\begin{figure}
\includegraphics[width=0.45\textwidth]{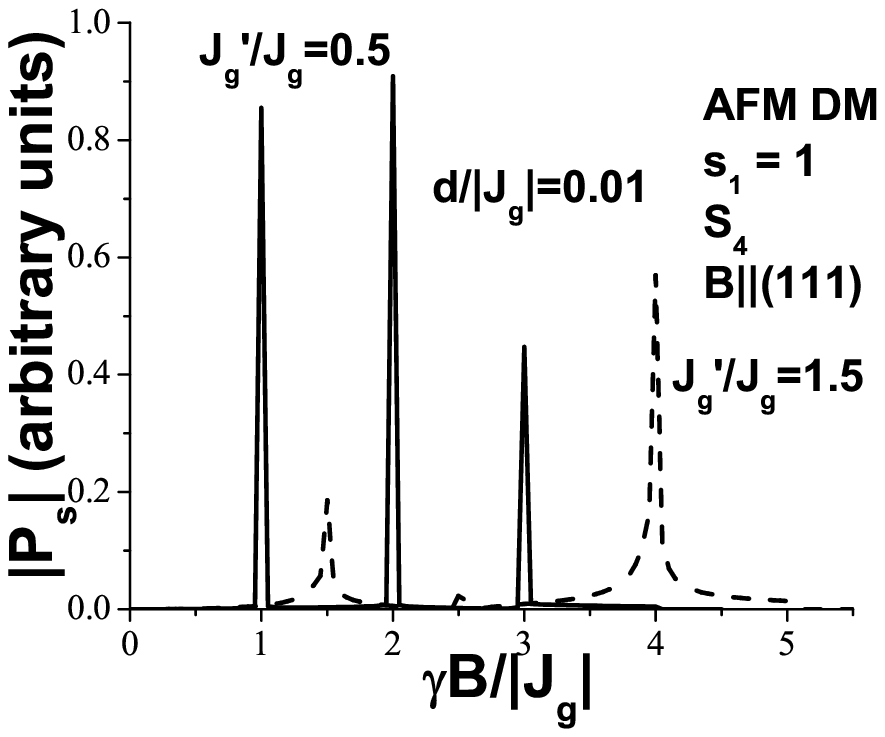}
\caption{Plots of the magnitude of the spin-derived polarization
$|{\bf P}_s|$ in arbitrary units versus $\gamma B/|\tilde{J}_g|$
at $T=0$ of an $s_1=1$ tetramer with $c=a$, $g=S_4$,
$d_z^g=d_{1y}^g=d_{2x}^g=d_{2y}^g=d$, $d_{1x}^g=0$,
$d/|\tilde{J}_g|=0.01$, and ${\bm B}||(111)$.  The solid and
dashed curves are for the Type I ($\tilde{J}_g'/\tilde{J}_g=1.5$)
and Type II ($\tilde{J}_g'/\tilde{J}_g=0.5$) tetramers,
respectively.}\label{fig8}
\end{figure}

\begin{figure}
\includegraphics[width=0.45\textwidth]{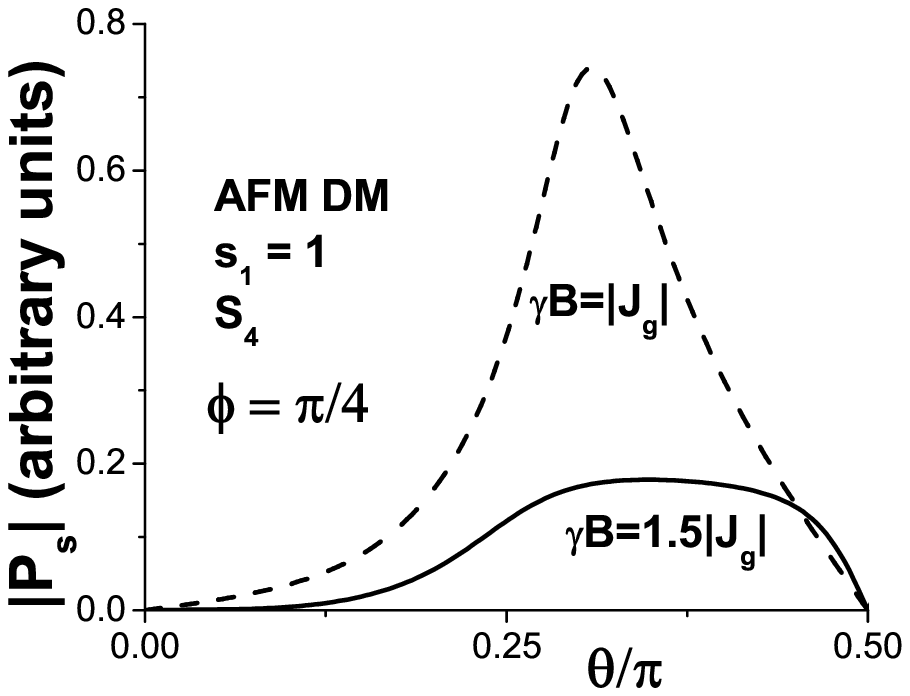}
\caption{Plots of the magnitude of the spin-derived polarization
$|{\bf P}_s|$ in arbitrary units versus $\theta/\pi$ at
 $T=0$ of an $s_1=1$ tetramer with $c=a$,
$g=S_4$, $\tilde{J}_g'/\tilde{J}_g=1$,
$d_z^g=d_{1y}^g=d_{2x}^g=d_{2y}^g=d$, $d_{1x}^g=0$, $d/J_g= 0.01$,
and $\phi=\pi/4$. The  solid and dashed curves are for $\gamma
B/|\tilde{J}_g|=1.5,1$, respectively.}\label{fig9}
\end{figure}

\subsection{C. Type I and Type II tetramers}
There are at least two types of FM and AFM tetramers.  To the
extent that single-ion, symmetric anisotropic exchange, and
biquadratic exchange interactions are small relative to the
Heisenberg interactions, there are just two types of tetramers.
The criterion is simply based upon $\tilde{J}_g'-\tilde{J}_g$ in
$\delta E_{\overline{\nu},0}^g$, which we assume to be larger in
magnitude than all anisotropy and biquadratic interaction
strengths. Type I tetramers have $\tilde{J}_g'-\tilde{J}_g>0$,
which can occur for either sign of $\tilde{J}_g$, provided $g\ne
T_d$.  For Type I, the lowest energy state in each $s$ manifold
occurs for the maximum values $s_{13},s_{24}=2s_1$.  Thus, at low
$T$, Type I tetramers behave as pairs of spin $2s_1$ dimers.  Type
II tetramers  with $\tilde{J}_g'-\tilde{J}_g<0$ are frustrated,
with the lowest energy state in each $s$ manifold occurring for
the minimal $s_{13},s_{24}$ values. For even $s$, these minima
occur for $s_{13},s_{24}=s/2$, but for odd $s$, the energy minimum
is doubly degenerate, occurring at
$s_{13},s_{24}=(s\pm1)/2,(s\mp1)/2$. Hence, explicit formulas for
first-order eigenstate energy parameters
$a^{\pm}_{\overline{\nu}}$, $b^{\pm}_{\overline{\nu}}$,
$c^{\pm}_{\overline{\nu}}$, and ${\cal B}_{\overline{\nu}}$ with
arbitrary $s$ in the three special
 cases of $(s_{13},s_{24})=(2s_1,2s_1)$,
$(s_{13},s_{24})=(s/2,s/2)$ for even $s$, and
$(s_{13},s_{24})=[(s\pm1)/2,(s\mp1)/2)]$ for odd $s$ are
  given in Subsections G and H of the Appendix. For sufficiently strong
$\tilde{J}_g'-\tilde{J}_g$, these
 formulas can apply to the lowest energy eigenstate  in each $s$ manifold.
When $\tilde{J}_g'-\tilde{J}_g$ is small relative to the other
interactions, the situation becomes more complicated, as the
lowest energy eigenstate for a particular ${\bm B}$ can depend
upon more of the effective interaction values.  Tables IV-XI in
the Appendix are sufficient for full analyses of such cases for
$s_1\le3/2$, but such studies will be made
subsequently.\cite{ekfuture}

\subsection{D. DM interactions and spin-induced polarizations}

 Although the first order correction to the  energy arising from the DM interactions vanishes,
  for $s_1=1/2, 1$, it is not too difficult to diagonalize the
Hamiltonian matrix exactly, and hence to take precise account of
the effects of the DM interactions. To focus on the effects of DM
interactions, in Figs. 4-9, we omit the single-ion, symmetric
anisotropic exchange interactions, and biquadratic interactions,
keeping only the AFM Heisenberg, Zeeman, and weak DM interactions.
In Fig. 4, we plotted $M/\gamma$ versus $\gamma B/|\tilde{J}_g|$
for $s_1=1/2$ AFM tetramers with $S_4$ symmetry with ${\bm B}||
(111)$, $T=0$, and $d/|\tilde{J}_g|=0.1$, where
$d_z^g=d_{1y}^g=d_{2x}^g=d_{2y}^g=d$, $d_{1x}^g=0$, in the limit
$c=a$.  We note that for $\tilde{J}_g<0$ and
$\tilde{J}_g'/\tilde{J}_g=0.5$ the magnetization exhibits sharp
steps close to the integral values $s=1,2$ of $\gamma
B/|\tilde{J}_g|$, characteristic of Type I AFM tetramers.  For
$\tilde{J}_g<0$ and $\tilde{J}_g'/\tilde{J}_g=1.5$, the two steps
are shifted to higher $B$ values, and the first step is broadened.
This is Type II AFM tetramer behavior for the DM interaction.  In
Fig. 5, the corresponding curves for the spin-induced polarization
$|{\bm P}_s|$ are shown.  In both cases, there is a sharp peak at
the inflection point of the first magnetization step, at which the
total spin value $s$ changes from 0 to 1.  In addition, there is a
discontinuity in slope at the positions of the second
magnetization steps, at which  ${\bm P}_s$ begins its rapid
decrease to zero, which it reaches at the $B$ value for which $M$
just reaches saturation. Note that ${\bm P}_s\rightarrow0$ at
large $B$, since the large ${\bm B}$ aligns the spins at all
sites, causing their vector products to vanish. Some results for
the intermediate case $\tilde{J}_g'/\tilde{J}_g=1$ are shown in
Fig. 6.  In this figure, we plotted $|{\bm P}_s|$ in arbitrary
units versus $\theta/\pi$ for $s_1=1/2$ AFM tetramers with $S_4$
symmetry at $\phi=\pi/4$ with $d/|\tilde{J}_g|=0.1$ with $\gamma
B/|\tilde{J}_g|=1.5,1$, respectively.  The curves are symmetric
about $\theta=\pi/2$. We note that the curves exhibit broad maxima
at $\theta/\pi\approx 0.3$ for both $B$ values.  These broad
curves reflect the strong frustration of the spins with
$\tilde{J}_g'=\tilde{J}_g$, for which no preferred $s_{13},
s_{24}$ values exist.

In Figs. 7-9, we plotted analogous curves for $s_1=1$ tetramers,
except that $d/|\tilde{J}_g|=0.01$.  Curves with
$d/|\tilde{J}_g|=0.1$ for $s_1=1$ exhibit steps or peaks which are
much more broadened than the corresponding ones for $s_1=1/2$
pictured in Figs. 4-7.  We note that in Fig. 7, the dashed curve
for the Type II case $\tilde{J}'_g/\tilde{J}_g=1.5$ has a wider
steps at $M/\gamma=0,2$ than at 1,3.  This low value of the DM
strength $d/|\tilde{J}_g|=0.01$ leads to very sharp peaks in
$|{\bm P}_s|$ for both Type I and II tetramers, as shown in Fig.
8. However, the three peak tails are much broader for Type II
tetramers with $\tilde{J}'_g/\tilde{J}_g=1.5$ than for Type I
tetramers with $\tilde{J}'_g/\tilde{J}_g=0.5$.  In each case, the
peak positions correspond to the first three magnetization step
$\gamma B/|\tilde{J}_g|$ values, and the polarization also
vanishes at the $\gamma B/|\tilde{J}_g|$ value at which the fourth
magnetization step is completed.  Finally, in Fig. 9 we plotted
$|{\bm P}_s|$ in arbitary units versus $\theta/\pi$ with the same
parameters as in Fig. 6, except that $d/|\tilde{J}_g|=0.01$.  The
curve with $\gamma B/|\tilde{J}_g|=1$ has a peak at
$\theta/\pi\approx 0.3$, as for the corresponding curve with
$s_1=1/2$, but the solid curve for $\gamma B/|\tilde{J}_g|=1.5$
has a flat region for $\theta/\pi$ between 0.3 and 0.45, and both
curves become vanishingly small at $\theta=\pi/2$, which differs
strongly from the behavior shown in Fig. 6 for $s_1=1/2$.  This
difference suggests an interesting  parity effect at
$\theta=\pi/2$, with ${\bm P}_s(\theta)\propto
(\theta-\pi/2)^{2s_1+1}$.

\subsection{E. First-order AFM level crossing inductions}

For AFM tetramers, $\tilde{J}_g<0$.    There will be $2s_1+1$
level crossings, as exhibited by the magnetization steps in Figs.
4 and 7 for $s_1=1/2, 1$, respectively, provided that the lowest
energy state in each $s$ manifold does not exhibit level
repulsion. In order to specify the level-crossing inductions, we
first write
$E_{s,m}^{g}(s_{13},s_{24},s_1)=E^g_{\nu,0}+E^{g}_{\nu,1}$.  We
then note that the $s_{13},s_{24}$ values involved in level
crossings are those corresponding to the lowest energies for a
particular $s$ value.  These are different for Type I and II
tetramers.  For Type I tetramers, the level crossing-inductions
occur for
\begin{eqnarray}
E_{s,s}^{g}(2s_1,2s_1,s_1)&=&E_{s-1,s-1}^{g}(2s_1,2s_1,s_1).\label{TypeIlevelcrossings}
\end{eqnarray}
and for Type II tetramers, they occur for
\begin{eqnarray}
E_{s,s}^{g}(s/2,s/2,s_1)&=&E_{s-1,s-1}^{g}[(s-1\pm1)/2,\nonumber\\
& &(s-1\mp1)/2,s_1]
\end{eqnarray}
for even $s$, and
\begin{eqnarray}
E_{s,s}^{g}[(s\pm1)/2,(s\mp1)/2,s_1]&=&E_{s-1,s-1}^{g}[(s-1)/2,\nonumber\\
& &(s-1)/2,s_1]
\end{eqnarray}
for odd $s$.
 In Subsections G and H of the Appendix, we  presented the formulas
for the level-crossing induction parameters for both Types I and
II tetramers.

 For Type I tetramers, the first-order level-crossing inductions $B_{s_1,s}^{g,{\rm lc}(1)}$ obtained from Eq.
(\ref{TypeIlevelcrossings}) have the remarkably simple form,
\begin{eqnarray}
\gamma B_{s_1,s}^{g,{\rm
lc}(1)}(\theta)&=&-s\tilde{J}_g-J_z^g\frac{b^{+}}{2}-J_{{\rm
b},1}^gd-J_{{\rm t},1}^ge-2J_{{\rm t},2}^gss_1^2\nonumber\\& &
-\frac{c_1^{-}}{3}\Bigl(J_{\rm
eff}^g-\frac{(2s_1-1)}{(4s_1-1)}J_z^g\Bigr)(1-3\cos^2\theta),\label{Bqz}\nonumber\\
& &\\
J_{\rm
eff}^g&=&\frac{J_{1,z}^g}{2}+\frac{s_1J_{2,z}^g}{4s_1-1},\label{Jeff}
\end{eqnarray}
 where
$b^{+}=b_I^{s_1,+}(s), c_1^{-}=c_{I,1}^{s_1,-}(s)$,
$d=d_{I}^{s_1}(s)$, and $e=e_{I}^{s_1}(s)$ are given in Subsection
G of the Appendix. We note that this is independent of
$\tilde{J}_g'$ and $J_{{\rm b},2}^g$, and that the NN and NNN
symmetric anisotropic exchange interactions combine to yield the
universal Type-I level crossing form $J_{\rm eff}^g$. Furthermore,
the $\theta$-dependencies of the single-ion and symmetric
anisotropic exchange contributions have the same $s$-dependencies
for fixed $s_1$.  However the $\theta$-independent contributions
from $J_z^g$, $J_{\rm eff}^g$, and $J_{{\rm b},1}^g$  depend
separately upon $s$ for fixed $s_1$.

For the Type II AFM tetramer level-crossing inductions,   the
contributions from the near-neighbor anisotropic exchange
interaction $J_{1,z}^g$ has a rather simple form. As shown in
Subsection H of the Appendix, these contributions $\gamma
B_{1,z}^{g,{\rm lc}(1)}=J_{1,z}^gf_{1,z}(s,\theta)$ to $\gamma
B_{s_1,s}^{g,{\rm lc}(1)}(\theta)$  are independent of $s_1$,
where
\begin{eqnarray}
f_{1,z}(s,\theta)&=&\left\{\begin{array}{ll}
\frac{(s-1)}{4s}\Bigl(1+(2s-1)\cos^2\theta\Bigr),&
s\>\>{\rm odd}\\
\frac{s}{4(s-1)}\Bigl(1+(2s-3)\cos^2\theta\Bigr),&s\>\>{\rm
even}.\end{array}\right.
\end{eqnarray}  Note in particular that for $s=1$, $f_{1,z}(1,\theta)=0$.  However, the single-ion and NNN symmetric anisotropic exchange
contributions to the level crossing inductions depend upon both
$s$ and $s_1$ in different ways.

\subsubsection{$s_1=1/2$ first-order AFM level crossings}

 For the simplest case
$s_1=1/2$, as in AFM Cu$_4$ tetramers, the single-ion interaction
$J_z^g$ does not contribute to the level-crossing inductions, as
for the dimer of equal $s_1=1/2$ spins.\cite{ek2} Using the
results given in Subsections G and H of the Appendix, the
expressions for the $\gamma B_{1/2,s}^{g,{\rm lc}(1)}(\theta)$
functions are particularly simple.  For $s_1=1/2$ effective-dimer
Type I tetramers, $\tilde{J}_g'-\tilde{J}_g>0$,
\begin{eqnarray}
\gamma B_{1/2,1}^{g,{\rm
lc}(1)}(\theta)&=&-\tilde{J}_g+\frac{1}{2}(J_{{\rm b},1}^g-J^g_{{\rm t},2})\nonumber\\
& &+\frac{1}{6}J^g_{\rm eff}(1-3\cos^2\theta),\label{Cu4firstTypeIlevelcrossing}\\
\gamma B_{1/2,2}^{g,{\rm lc}(1)}(\theta)&=&-2\tilde{J}_g+J^g_{{\rm
b},1}-J^g_{{\rm t},2}\nonumber\\
& &-\frac{1}{2}J^g_{\rm eff}(1-3\cos^2\theta),
\end{eqnarray}
where $J^g_{\rm eff}$ is given by Eq. (\ref{Jeff}) with $s_1=1/2$.
For  frustrated Type II tetramers with $s_1=1/2$,
$\tilde{J}_g-\tilde{J}_g'>0$,
\begin{eqnarray}
\gamma B_{1/2,1}^{g,{\rm
lc}(1)}(\theta)&=&-\tilde{J}_g'-\frac{1}{4}(3J_{{\rm b},1}^g-J^g_{{\rm t},1})+\frac{J_{\rm b,2}^g}{2}\nonumber\\
& &+\frac{1}{4}J^g_{2,z}(1+\cos^2\theta),\\
\gamma B_{1/2,2}^{g,{\rm
lc}(1)}(\theta)&=&-\tilde{J}_g-\tilde{J}_g'+\frac{5}{4}(J_{{\rm b},1}^g-J^g_{{\rm t},1})\nonumber\\
& &+\frac{1}{2}(J_{\rm b,2}^g-J^g_{{\rm t},2})\nonumber\\
&
&+\frac{1}{4}\Bigl(2J^g_{1,z}+J_{2,z}^g\Bigr)(1+\cos^2\theta).\label{Cu4TypeIIlevelcrossings}
\end{eqnarray}
Even in this simplest of all tetramer cases,  there is still a
qualitative difference between the level crossing inductions of
Type I and Type II AFM $s_1=1/2$ tetramers.  For Type I, there is
only one effective anisotropic exchange interaction, $J_{\rm
eff}^g=(J_{1,z}^g+J_{2,z}^g)/2$ that affects the level crossing.
However, for Type II $s_1=1/2$ tetramers, the level crossing is
different for the NN and NNN anisotropic exchange interactions.
The only effect of the group symmetry is to provide restrictions
upon the values of the interactions. These expressions also show
that Type II AFM tetramers have a more complex level-crossing
induction variation than do Type I AFM tetramers, as the
first-order Type I level-crossing behavior is fully described by
three parameters, whereas the first-order Type II level-crossing
behavior depends upon four independent parameters. On the other
hand, for this special $s_1=1/2$ example, the $\theta$
dependencies of the first and second $\gamma B_{1/2,s}^{g,{\rm
lc}(1)}$ are opposite in sign for Type I, but can have the same
sign for Type II. In Fig. 10, we illustrate these $s_1=1/2$
behaviors for the  Type I with $J^g_{\rm eff}/\tilde{J}_g=0.2$ and
for Type II  with $J_{2,z}^g/\tilde{J}_g=0.2$ and
$\tilde{J}_g-\tilde{J}_g'=0.5|\tilde{J}_g|$.

In the special case of $T_d$ symmetry, we have
$\tilde{J}_{T_d}=\tilde{J}_{T_d}'=J$, $J_{{\rm b},1}^{T_d}=J_{{\rm
b},2}^{T_d}=J_b$, $J_{{\rm t},1}^{T_d}=J_{{\rm t},2}^{T_d}=J_bt$,
and $J_{q,z}^{T_d}=0$.  Since for $s_1=1/2$, ${\cal H}_b^g$ and
${\cal H}_t^g$ are diagonal, the $J_b$ and $J_t$ dependencies of
the eigenstate energies are exact.  Hence, the only difference
between  Type I and Type II $s_1=1/2$ tetramers with $T_d$
symmetry is determined by the sign of $J_b-J_t$, as is evident by
comparing Eqs.
 (\ref{Cu4firstTypeIlevelcrossing})-(\ref{Cu4TypeIIlevelcrossings}). In any event, there is no $\theta$ dependence to
the first-order level-crossing inductions for $s_1=1/2$ AFM
tetramers with $T_d$ symmetry.

\begin{figure}
\includegraphics[width=0.45\textwidth]{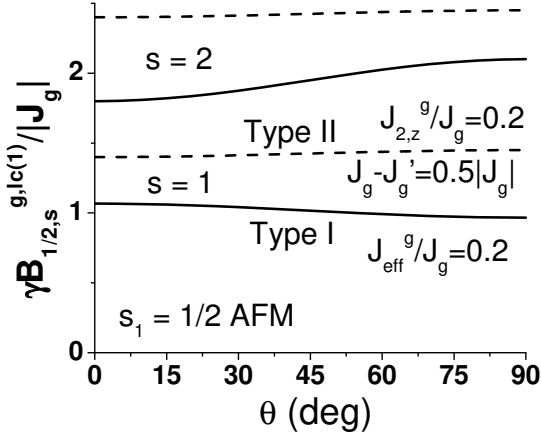}
\caption{Plots of the $s_1=1/2$ first-order level-crossing $\gamma
B_{1/2,s}^{g,\rm lc (1)}(\theta)/|\tilde{J}_g|$, where $\theta$ is
the angle between ${\bm B}$ and $\hat{\bm z}$,  with $J_{\rm
b,q}^g=0$, $J^g_{\rm eff}/\tilde{J}_g=0.2$,
$\tilde{J}_g'-\tilde{J}_g>0$ (solid, Type I) and
$J_{2,z}^g/\tilde{J}_g=0.2$,
 $\tilde{J}_g-\tilde{J}_g'=0.5|\tilde{J}_g|$ (dashed, Type II).}\label{fig10}
\end{figure}

\subsubsection{$s_1\ge1$ first-order AFM level crossings}

 For $s_1>1/2$,  the AFM level-crossings
 become much more complex than for the $s_1=1/2$ case, as single-ion anisotropies $J_z^g$ are
  allowed, and the biquadratic interactions $J_{{\rm b},q}^g$ affect
  the various level-crossings differently.  We first consider the simplest
 $s_1>1/2$ case, $s_1=1$, appropriate for AFM Ni$_4$ tetramers.
 Exact expressions for the $s=1,2,3,4$ first-order level-crossing
 inductions $B_{1,s}^{g,{\rm lc}(1)}(\theta)$
 for Type I and II $s_1=1$ tetramers are given  in Subsection F of the Appendix.
 In Figs. 11 and 12, we  plotted the $\theta$-dependence of the
first-order level crossing induction $\gamma B_{1,s}^{g,\rm lc
(1)}(\theta)/|\tilde{J}_g|$ for $g=C_{4h}, D_{4h}, C_{4v}, S_4$,
and $D_{2d}$, for two AFM Type I examples and for three Type II
examples with $\tilde{J}_g-\tilde{J}_g'=0.5|\tilde{J}_g|$,
respectively. In each curve, we allow only one of the anisotropy
interactions (or effective interactions) to be non-vanishing.  In
Fig. 11, the solid and dashed curves are for
$J_z^g/\tilde{J}_g=0.2$ and $J^g_{\rm eff}/\tilde{J}_g=0.2$,
respectively, where $J_{\rm eff}^g=J_{1,z}^g/2+J_{2,z}^g/3$ for
$s_1=1$. We note from Eq. (\ref{Bqz}) and from Fig. 11 that for
Type I, the single-ion and symmetric exchange anisotropies lead to
opposite $\theta$-dependencies, both having a change in sign just
before the second level crossing, and the dependence of $\gamma
B_{1,s}^{g, {\rm lc}(1)}$ upon $J_{{\rm b},1}^g$ decreases with
increasing $s$.

In Fig. 12, we illustrate the $s_1=1$ Type II level crossings,
setting $\tilde{J}_g-\tilde{J}_g'=0.5|\tilde{J}_g|$.  The solid
curves are for $J_z^g/\tilde{J}_g=0.2$ for $g=C_{4h}, D_{4h},
C_{4v}, S_4$, and $D_{2d}$, as in Fig. 11. The dashed and dotted
curves are for $J_{1,z}^g/\tilde{J}_g=0.4$ and
$J_{2,z}^g/\tilde{J}_g=0.4$, respectively.    For each curve, the
Type II isotropic exchange parameters lead to a larger gap between
the $s=2$ and $s=3$ level crossings.  The sign of the
$\theta$-dependencies of the single-ion (solid) curves changes
between $s=2$ and $s=3$.  The effects of the NN symmetric
anisotropic exchange interactions vanish for $s=1$, but increase
in magnitude with increasing $s$ for $s=2,3,4$.  The sign of the
$\theta$-dependence of the level crossing due to the NNN symmetric
anisotropic exchange interactions does not change, but its
magnitude increases monotonically.    Thus, Type II AFM $s_1=1$
tetramers have a richer set of first-order level-crossing
behaviors than do Type I AFM $s_1=1$ tetramers, and the unpictured
biquadratic interactions increase this richness.

\begin{figure}
\includegraphics[width=0.45\textwidth]{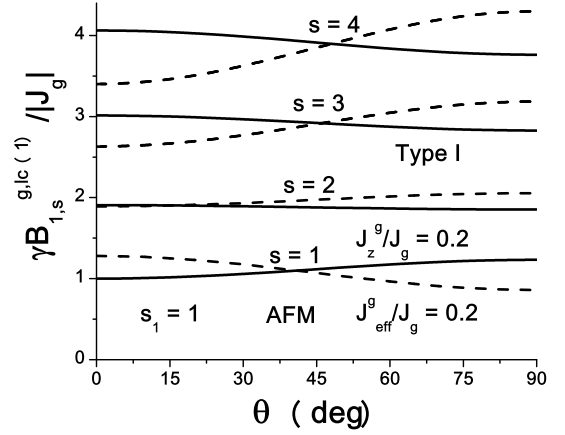}
\caption{Plots of the first-order level-crossing $B_{1,s}^{g,\rm
lc (1)}(\theta)/|\tilde{J}_g|$ for Type I,
$\tilde{J}_g-\tilde{J}_g'<0$, $J_{\rm b,q}^g=0$, and $s_1=1$.
Solid curves: $J_z^g/\tilde{J}_g=0.2$. Dashed curves: $J^g_{\rm
eff}/\tilde{J}_g=0.2$.}\label{fig11}
\end{figure}

\begin{figure}
\includegraphics[width=0.45\textwidth]{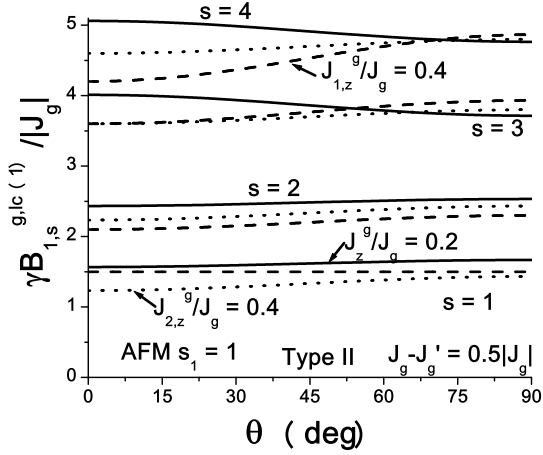}
\caption{Plots of the first-order level-crossing $\gamma
B_{1,s}^{g,\rm lc (1)}(\theta)/|\tilde{J}_g|$ for Type II with
$g=C_{4h}, D_{4h}, C_{4v}, S_4,D_{2d}$,
$\tilde{J}_g-\tilde{J}_g'=0.5|\tilde{J}_g|$, $J_{\rm b,q}^g=0$,
and $s_1=1$. Solid curves: $J_z^g/\tilde{J}_g=0.2$. Dashed curves:
$J_{1,z}^g/\tilde{J}_g=0.4$. Dotted curves:
$J_{2,z}^g/\tilde{J}_g=0.4$.}\label{fig12}
\end{figure}

In Figs. 13 and 14, the analogous Type I and Type II first-order
AFM level crossing inductions are plotted versus $\theta$ for
$s_1=3/2$ equal-spin tetramers, such as Cr$_4$.  The notation is
the same as in Figs. 11 and 12.  For Type I $s_1=3/2$ AFM
tetramers, the single-ion and symmetric anisotropic exchange
interactions lead to different $\theta$-dependencies of the
level-crossing inductions, each with a change in sign  in the
$\theta$ dependence at about the second level crossings, as seen
in Fig. 13.  For Type II $s_1=3/2$ AFM tetramers, the sign changes
appear between the second and third level crossings, as shown in
Fig. 14.  Although not pictured, the contributions for $s_1=3/2$
to the level-crossing inductions from  the $J_{\rm b,q}^g$ can be
easily calculated from Eq. (\ref{deltaE0}) and the expressions for
$d_{IIo}^{3/2}(s)$ and $d_{IIe}^{3/2}(s)$ in Subsection H of the
Appendix. They contribute to a more complex level-crossing pattern
than for $s_1=1$.

\begin{figure}
\includegraphics[width=0.45\textwidth]{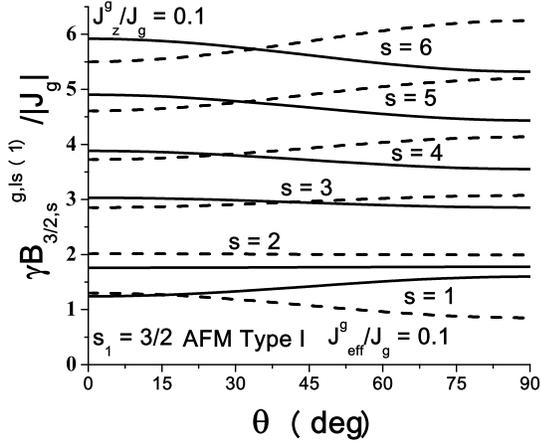}
\caption{Plots of the first-order level-crossing $\gamma
B_{3/2,s}^{g,\rm lc (1)}(\theta)/|\tilde{J}_g|$  for Type I,
$\tilde{J}_g-\tilde{J}_g'<0$, $J_{\rm b,q}^g=0$, and $s_1=3/2$.
Solid curves: $J_z^g/\tilde{J}_g=0.1$. Dashed curves: $J^g_{\rm
eff}/\tilde{J}_g=0.1$.}\label{fig13}
\end{figure}

\begin{figure}
\includegraphics[width=0.45\textwidth]{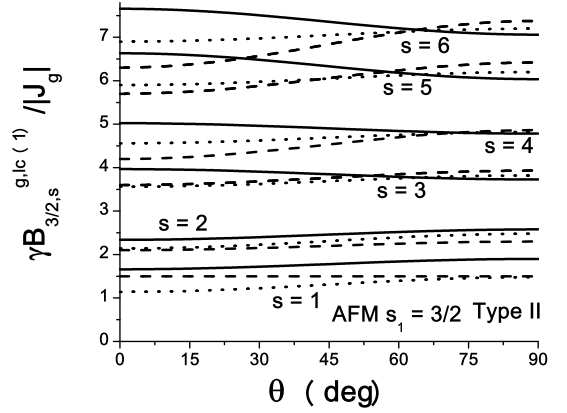}
\caption{Plots of the first-order level-crossing $\gamma
B_{3/2,s}^{g,\rm lc (1)}(\theta)/|\tilde{J}_g|$ for Type II with
$g=C_{4h}, D_{4h}, C_{4v}, S_4, D_{2d}$,
$\tilde{J}_g-\tilde{J}_g'=0.5|\tilde{J}_g|$, $J_{\rm b,q}^g=0$,
and $s_1=3/2$. Solid curves: $J_z^g/\tilde{J}_g=0.2$. Dashed
curves: $J_{1,z}^g/\tilde{J}_g=0.4$. Dotted curves:
$J_{2,z}^g/\tilde{J}_g=0.4$.}\label{fig14}
\end{figure}

For $g=T_d$ with $s_1>1/2$, we still have $J_{q,z}^g=J_z^g=0$, so
that the first-order level-crossing inductions $\gamma
B_{s_1,s}^{T_d,{\rm lc}(1)}(\theta)/|\tilde{J}_{T_d}|$ vary from
integral values only by $\theta$-independent constants due to the
biquadratic interaction $J_{{\rm b},1}^{T_d}=J_{{\rm b},2}^{T_d}$.

\section{VII. The self-consistent Hartree approximation}
\subsection{A. Partition function and thermodynamics}

The self-consistent Hartree approximation, or strong exchange
limit,\cite{BenciniGatteschi} provides  accurate results for the
${\bm B}$ dependence of the specific heat and magnetization at low
$k_BT/|\tilde{J}_g|$ and $\gamma B/|\tilde{J}_g|$ not too
small,\cite{ek2} where $k_B$ is Boltzmann's constant. In this
approximation, $E_{\nu}^g=E_{\nu,0}^g+E_{\nu,1}^g$ is given by
Eqs. (\ref{E0}) and (\ref{E1}), respectively. We shall present the
self-consistent Hartree approximation  of four measurable
quantities in the induction representation. We first define the
trace valid for our eiqenstate representation,
\begin{eqnarray}
{\rm
Tr}_{\nu}&\equiv&\sum_{\nu}=\sum_{s_{13},s_{24}=0}^{2s_1}\sum_{m=-s}^s\sum_{s=|s_{13}-s_{24}|}^{s_{13}+s_{24}},\label{trace}
\end{eqnarray}
  The partition
function in the self-consistent Hartree approximation may then be
written
 \begin{eqnarray} Z_{g}^{(1)}&=&{\rm Tr}_{\nu}e^{-\beta
E_{\nu}^{g}}, \end{eqnarray} where $\beta=1/(k_BT)$.
 In this compact notation, the self-consistent Hartree
magnetization $M_g^{(1)}({\bm B},T)$ and specific heat
$C_{g,V}^{(1)}({\bm B},T)$ are  given by
\begin{eqnarray}
M_g^{(1)}({\bm B},T)&=&\gamma{\rm
Tr}_{\nu}\Bigl(me^{-\beta E_{\nu}^g}\Bigr)/Z_g^{(1)},\\
\frac{C_{g,V}^{(1)}({\bm B},T)}{k_B\beta^2}&=&\frac{{\rm
Tr}_{\nu}\Bigl(\bigl(E_{\nu}^g\bigr)^2e^{-\beta
E_{\nu}^g}\Bigr)}{Z_g^{(1)}}\nonumber\\
& &-\biggl[\frac{{\rm Tr}_{\nu}\Bigl(E_{\nu}^{g} e^{-\beta
E_{\nu}^g}\Bigr)}{Z_g^{(1)}}\biggr]^2.
\end{eqnarray}

We note that there are strong differences between the low-$T$
behavior of FM and AFM tetramers.  We assume
$|\tilde{J}_g|>|\tilde{J}_g'-\tilde{J}_g|$.  For FM tetramers with
$\tilde{J}_g>0$, the low-$T$ thermodynamic behavior is dominated
by the $s=4s_1$, $m=-4s_1$ state, leading to
\begin{eqnarray}M_g^{(1)}({\bm
B},T)&{{\approx}\atop{T\rightarrow0}}&\gamma\hat{\bm B} {\cal
B}_{4s_1}(\beta\gamma B),
\end{eqnarray} where ${\cal B}_S(x)$ is the Brillouin function.  The universality
of this function renders thermodynamic studies useless for the
determination of the microscopic parameters.  For AFM tetramers with
$\tilde{J}_g<0$, however, there will be interesting level-crossing
effects, which can be employed to measure the microscopic
interaction parameters, as discussed in detail in Sec. V.  As for
dimers, $C_V(B,T)$ for AFM tetramers at sufficiently low $T$ exhibit
$2s_1$ central minima at the level-crossing inductions
$B_{s_1,s}^{g,{\rm lc}}(\theta)\approx B_{s_1,s}^{g,{\rm
lc}(1)}(\theta)$ that vanish as $T\rightarrow0$, equally surrounded
by peaks of equal height.\cite{ek2}  As for the magnetization,
$C_V(B,T)$ for FM tetramers at low $T$ reduces to that of a monomer
with spin $4s_1$, yielding a rather uninteresting Schottky anomaly.

\subsection{B. Electron paramagnetic resonance}

However, the microscopic nature of FM tetramers can be better probed
either by EPR or INS techniques. The self-consistent Hartree EPR
absorption ${\Im}\chi_{-\sigma,\sigma}^{g,(1)}({\bm B},\omega)$ for
clockwise ($\sigma=1$) or counterclockwise ($\sigma=-1$) circularly
polarized oscillatory fields normal to ${\bm B}$  is
\begin{eqnarray}
{\Im}\chi^{g,(1)}_{-\sigma,\sigma}&=&\frac{\gamma^2}{Z_g^{(1)}}{\rm
Tr}_{\nu}{\rm Tr}_{\nu'} e^{-\beta E_{\nu}^g}\bigl|M_{\nu,\nu'}\bigr|^2\nonumber\\
&
&\times\bigl[\delta(E_{\nu}^g-E_{\nu'}^g+\omega)-\delta(E_{\nu'}^g-E_{\nu}^g+\omega)\bigr],\nonumber\\
\end{eqnarray}
where  $M_{\nu,\nu'}=A_s^{\sigma
m}\delta_{m',m+\sigma}\delta_{s',s}\delta_{s_{13}',s^{}_{13}}\delta_{s_{24}',s^{}_{24}}$
and ${\rm Tr}_{\nu'}=\sum_{\nu'}$. The strong resonant inductions
 appear at
\begin{eqnarray}
\gamma B^{g,(1)}_{\rm
res}&=&\pm\omega+\frac{(2m+\sigma)}{2}(1-3\cos^2\theta)\tilde{J}_{z}^{g,\overline{\nu}},\label{Bres}\nonumber\\
\end{eqnarray}
where $\tilde{J}_{z}^{g,\overline{\nu}}$ is given by Eq.
(\ref{Jztildemu}) We note that $\tilde{J}_{z}^{g,\overline{\nu}}$
contains the three effective microscopic interactions, $J_z^g$,
$J_{1,z}^g$, and $J_{2,z}^g$, multiplied by the constants
$a_{\overline{\nu}}^{+}$, $-c_{\overline{\nu}}^{-}$, and
$-a_{\overline{\nu}}^{-}/2$, respectively.  In Tables IV and VI of
Subsection F of the Appendix, the values of these parameters
 for the FM ground state and the first three
excited state manifolds for arbitrary $s_1$ are given.  We note
that EPR measurements are insensitive to the Heisenberg and
biquadratic exchange interactions, which preserve $m$.

  For either  FM or AFM $s_1=1/2$ tetramers, EPR measurements can only probe the two microscopic
  symmetric anisotropic exchange interaction
parameters $J_{1,z}^g$ and $J_{2,z}^g$, and measurements of the
 two $s=1$ excited states are
sufficient to determine them.  For  FM tetramers with $s_1\ge1$,
it is a bit more difficult. From Tables IV and VI and from Eq.
(\ref{apm}) in Subsection G of the Appendix, it is easily seen
that the $(s,s_{13},s_{24})=(s,2s_1,2s_1)$ states all provide
measurements of the same combination of these three microscopic
interactions. Hence, for FM tetramers with $s_1>1/2$, measurements
of the ground $s=4s_1$ and the  first excited state manifold  with
$s=4s_1-1$ are insufficient to completely determine the three
microscopic interactions.
 In order to stay within a
single $s$ value for FM tetramers, one would need to study the
second (or higher) excited state manifold with  $s=4s_1-2$ (or
lower), in order to obtain sufficient information to determine the
three microscopic interaction strengths.  For AFM tetramers with
$s_1\ge1$, EPR transitions in the ground state are not allowed,
but  measurements of the first excited $s=1$ state manifold  would
suffice to determine $J_z^g$ and the $J_{q,z}^g$, as seen from the
formulas in Tables VIII and X in the Appendix.

\subsection{C. Inelastic neutron scattering}

The Hartree INS cross-section $S_g^{(1)}({\bm B},{\bm q},\omega)$
is
\begin{eqnarray}
S_g^{(1)}&=&{\rm Tr}_{\nu}{\rm Tr}_{\nu'}e^{-\beta E^g_{\nu}}\sum_{\tilde{\alpha},\tilde{\beta}}\bigl(\delta_{\tilde{\alpha},\tilde{\beta}}-\hat{q}_{\tilde{\alpha}}\hat{q}_{\tilde{\beta}}\bigr)\sum_{n,n'}\nonumber\\
& &\times e^{i{\bm q}\cdot({\bm r}_n-{\bm
r}_{n'})}\langle\nu|S_{n',\tilde{\alpha}}^{\dag}|\nu'\rangle\langle\nu'|S_{n,\tilde{\beta}}|\nu\rangle\nonumber\\
& &\qquad\times\delta(\omega+E^g_{\nu}-E^g_{\nu'}),
\end{eqnarray}
where
$\tilde{\alpha},\tilde{\beta}=\tilde{x},\tilde{y},\tilde{z}$,
$\hat{q}_{\tilde{x}}=\sin\theta_{b,q}\cos\phi_{b,q}$,
$\hat{q}_{\tilde{y}}=\sin\theta_{b,q}\sin\phi_{b,q}$, and
$\hat{q}_{\tilde{z}}=\cos\theta_{b,q}$,  $\theta_{b,q}$ and
$\phi_{b,q}$ describe the  relative orientations of ${\bm B}$ and
${\bm q}$,\cite{ek2}  the ${\bm r}_n$ are given by Eq. (\ref{rn}),
and the $\langle\nu'|S_{n,\tilde{\alpha}}|\nu\rangle$ are given by
Eqs.
 (\ref{Mz}) and (\ref{Msigma}) in  Subsection E of the Appendix. The
scalar ${\bm q}\cdot({\bm r}_n-{\bm r}_{n'})$ is invariant under
the rotation, Eq. (\ref{rotation}). After some algebra, we rewrite
$S_g^{(1)}({\bm q},\omega)$ as
\begin{eqnarray}
S_g^{(1)}&=&{\rm Tr}_{\nu}e^{-\beta
E^g_{\nu}}\sum_{\nu'}\delta(\omega+E_{\nu}^g-E_{\nu'}^g)\nonumber\\
& &\times\Bigl(\sin^2\theta_{b,q}L_{\nu,\nu'}({\bm
q})+\frac{(2-\sin^2\theta_{b,q})}{4}M_{\nu,\nu'}({\bm
q})\Bigr),\label{Sofqomega}\nonumber\\
\end{eqnarray}
where the Hartree functions $L_{\nu,\nu'}({\bm q})$ and
$M_{\nu,\nu'}({\bm q})$ are given in Subsection I of  the
Appendix. They are independent of  ${\bm B}$.  Since $E_{\nu}^g$
is well-behaved as ${\bm B}\rightarrow0$, Eq. (\ref{Sofqomega}) is
accurate for all ${\bm B}$.

 As for the dimer,\cite{ek2} additional EPR and INS transitions
with amplitudes higher order in the anisotropy parameters $J_z^g$,
$J_{1,z}^g$, and $J_{2,z}^g$ relative to $\tilde{J}_g$ are
obtained in the extended Hartree approximation, but will be
presented elsewhere for brevity.\cite{ekfuture}

\section{VIII. Discussion}

The quadratic phenomenological total spin anisotropy model widely
used in fitting experimental data on SMM's is
\begin{eqnarray}
{\cal H}_p&=&{\cal A}-{\cal D}S_z^2-{\cal
E}(S_x^2-S_y^2),\label{Hph}
\end{eqnarray}
where ${\cal A}$ represents the isotropic total spin interactions,
and ${\cal D}$ and ${\cal E}$ are measures of the axial and
azimuthal total spin anisotropy, respectively.\cite{general}
Often, additional quartic terms are added.\cite{Hendrickson,Fe4}
The anisotropy is defined relative to the total spin principal
axes, which for equal spin, high-symmetry systems are the
molecular axis vectors. It is easy to evaluate
$E^p_{\nu}=\langle\nu|\tilde{\cal H}_p|\nu\rangle$ in the
induction representation. One obtains Eqs. (\ref{E0}) and
(\ref{E1}), provided that
\begin{eqnarray}
{\cal A}&=&{\cal A}_0+\delta {\cal A},\\
{\cal A}_0&=&-\tilde{J}_gs(s+1)/2-\gamma Bm,\\
\delta{\cal A}&=&\delta E_{\overline{\nu},0}^g-\delta\tilde{J}_z^{g,\overline{\nu}},\\
 {\cal D}&=&\tilde{J}_z^{g,\overline{\nu}},\\
{\cal E}&=&0,
\end{eqnarray}
where $\delta E_{\overline{\nu},0}^g$,
$\tilde{J}_z^{g,\overline{\nu}}$ and
$\delta\tilde{J}_z^{g,\overline{\nu}}$ are given by Eqs.
(\ref{deltaE0}), (\ref{Jztildemu}) and (\ref{deltaJztildemu}),
respectively, which contain the constants
$a_{\overline{\nu}}^{\pm}$, $b_{\overline{\nu}}^{\pm}$,
$c_{\overline{\nu}}^{\pm}$, and ${\cal B}_{\overline{\nu}}$.
Precise formulas for arbitrary $\overline{\nu}$ appear in
Subsection F of the Appendix, along with Tables IV-XI of their
values for arbitrary $s_1$ in the ground and lowest three excited
state manifolds for FM and AFM tetramers, respectively. Usually,
one assumes the strong exchange limit, so that the isotropic
${\cal A}_0$ is sufficiently large that it remains constant for
$B=0$, and can be neglected. A non-vanishing ${\cal E}$ would lead
to a term in Eq. (\ref{E1}) proportional to
$\sin^2\theta\cos(2\phi)$, as for the dimer,\cite{ek2} which does
not arise in the first-order calculation for the high-spin
tetramers under consideration, based upon the microscopic
parameters alone. Hence, the ${\cal D}$ term
 in ${\cal H}_p$ alone describes the $\theta$ and $m$ dependencies
of $E_{\nu,0}^g+E_{\nu,1}^{g}$ correctly, provided that the
quantum numbers $s_{13}$ and $s_{24}$ remain constant.

More important, the additional constant term $\delta{\cal A}$ has
generally been neglected.  Even to zeroth order, the sign of
$\tilde{J}_g-\tilde{J}_g^{'}$ in $\delta E_{\overline\nu,0}^g$
distinguishes between Type I and Type II tetramers, which
distinction is absent in the phenomenological model. Moreover the
different first-order dependencies of ${\cal A}$ and ${\cal D}$
upon the $\overline{\nu}$  are important in determining the
level-crossing inductions for AFM tetramers, each of which
involves two values of $s$ and $m$, and hence different $s_{13}$
and $s_{24}$ values, as well. The zero-field energy spectrum is
thus more complicated than that given by the usual
phenomenological model, which could lead to substantially
different  fits to experiment.

For simplicity, the only higher order interactions we have
considered are the isotropic NN and NNN biquadratic exchange
interactions.  These isotropic interactions are rotationally
invariant, so they are independent of $\theta$ in the induction
representation. Hence, they only contribute to $\delta A$.  Thus
they modify the positions but not the $\theta$-dependencies of the
AFM level-crossing inductions, and do not modify any EPR
transitions.

In the ground state of FM tetramers, $\overline{\nu}$ is
restricted to
 the
single set of values, $(s,s_{13},s_{24})=(4s_1,2s_1,2s_1)$. In
this high-spin case, ${\cal H}_p$ can provide a correct
phenomenology of the  ground state energy. However, if applied to
the two low-lying excited states with $s=4s_1-1$, for instance,
one would infer two different ${\cal A}$ and ${\cal D}$ values
from those obtained in the ground state. However, as noted above,
since all states with $(s_{13},s_{24})=(2s_1,2s_1)$ contain the
same combination of $a_{\overline{\nu}}^{\pm}$ and
$c^{-}_{\overline{\nu}}$, in order to exploit the $\overline{\nu}$
dependence of ${\cal D}$ to obtain an unambiguous EPR measurement
of the three microscopic parameters $J_z^g, J_{1,z}^g$, and
$J_{2,z}^g$, for $s_1>1/2$, one needs to examine higher state
manifolds, such as the manifold with $s=4s_1-2$. For AFM
tetramers, ${\cal H}_p$ also correctly provides a vanishing
first-order correction to the $s=0$ manifold of states with
 $s_{13}=s_{24}=0,1,\ldots, 2s_1$. However, ${\cal H}_p$ also
has  problems describing the first excited  manifold of AFM states
with $s=1$, because it leads to the  choice of $\delta {\cal A}$
independent of the quantum numbers $\overline{\nu}$, which is
unphysical except for tetramers with $T_d$ symmetry. Hence, the
phenomenological model works best in describing only a single
 state with fixed $(s,s_{13},s_{24})$. This is more
restrictive than the usual assumption of its applicability to all
states with fixed $s$.\cite{Ni4,Fe4,CorniaFe4,Rastelli,Fe4spin5}

We note that the FM Cu$_4$ tetramer Cu$_4$OCl$_6$(TPPO)$_6$ was
claimed to have $T_d$ symmetry and an $s=2$ ground
state.\cite{Black1,Black2,Black3} It is noteworthy that those
authors thought that anisotropic exchange interactions might be
responsible for their observed zero-field energy
splittings.\cite{Black1,Buluggiu}  Since tetramers with $T_d$
symmetry do not have either symmetric or antisymmetric anisotropic
exchange interactions, another explanation must be considered.
From Tables IV-VII in the Appendix, it is evident that the FM
ground state is non-degenerate for all $s_1$, even for those SMM's
with lower symmetries allowing anisotropic exchange interactions.
It therefore appears that the sample may not have been
single-phase,\cite{Black1,Black2,Black3} as in a nominally $S_4$
Ni$_4$ tetramer,\cite{Hendrickson} allowing for an apparent ground
state splitting.

We note that for the FM Fe$_4$ SMM, Fe$_4$(thme)$_2$(dpm)$_6$,
where H$_3$thme is 1,1,1-tris(hydroxymethyl)ethane and Hdpm is
dipivaloylmethane,\cite{Fe4,CorniaFe4,Rastelli} the high $D_3$
symmetry also precludes the ${\cal E}$ term in ${\cal H}_p$.
Nevertheless, in fits to INS data, it was  assumed that ${\cal
E}\ne0$,\cite{Fe4,Fe4spin5} in order to obtain the appropriate
anticrossing gaps, so that either the powdered sample did not have
pure $D_3$ symmetry, or the phenomenological model they used, Eq.
(\ref{Hph}) plus two quartic terms obeying $D_3$ symmetry, was not
appropriate.  Since single-ion interactions appeared to be
important,\cite{Fe4} the total spin might not have been a
well-defined quantum number, as in at least one Fe$_2$ dimer and
in Fe$_8$.\cite{ek2,Shapira,Dalal,Fe8}  However, we note that it
might be interesting to investigate whether second-order DM
effects might yield an effective finite ${\cal E}$ value.

Using a microscopic Hamiltonian, detailed fits to the four
magnetization step data obtained in large pulsed fields on powder
samples of the AFM Ni$_4$ tetramer
[Mo$_{12}^V$O$_{30}$($\mu_2$-OH)$_{10}$H$_2\{$Ni$^{II}$(H$_2$O)$_3\}_4$],
or $\{$Ni$_4$Mo$_{12}\}$ were presented.\cite{Schnack}  Although
the molecule has $C_{1v}$ symmetry, it is close to exhibiting
$C_{3v}$ symmetry.  Since the steps were unevenly spaced, the
authors assumed the molecule to have weak, but important
biquadratic interactions.  To limit the number of fitting
parameters, they assumed $C_{3v}$ symmetry for the Heisenberg and
biquadratic interactions,  and $T_d$ symmetry for the single-ion
and anisotropic exchange interactions.  In addition, they allowed
the two Heisenberg interaction strengths to have strong magnetic
field dependencies. Subsequently, magnetoinfrared studies of that
compound were made, revealing only very small differences in the
responses at $B=0$ and 14 T,\cite{Janprivate} providing little, if
any justification for such strong magnetic field dependencies of
the Heisenberg interaction strengths.

More recently, a remarkably simple fit to the four level-crossing
midpoints was made by Kostyuchenko.\cite{Kostyuchenko}  In this
fit, the $\{$Ni$_4$Mo$_{12}\}$ molecule was assumed to have $T_d$
symmetry, so that $\tilde{J}_g=\tilde{J}_g'=-J$, and a convincing
argument was presented that the strength $-J_3$ of the isotropic
three-center quartic interactions ought to be comparable in
magnitude to that ($-J_2$) of the biquadratic interactions.  Since
each of these terms preserves the $s$ quantum number, the
Hamiltonian matrix in the absence of single-ion and anisotropic
exchange interactions is block diagonal, and for $s_1=1$, it is
possible to obtain the exact eigenvalues in terms of the three
parameters $J,J_2,J_3$.  However, in his fit to the magnetization
level crossing midpoint data on $\{$Ni$_4$Mo$_{12}\}$, he found
$J_2=J_3$, which implies that he claimed to fit the four linear
equations for the four level crossing midpoints with two
parameters. Although two of those equations were nearly
degenerate, three were clearly non-degenerate, rendering his
two-parameter fit inappropriate.

 In Subsection F of the Appendix, we extended
the calculation of Kostyuchenko to the five lower symmetries, so
that there are six isotropic interactions $\tilde{J}_g,
\tilde{J}_g', J^g_{{\rm b},q}$, and $J^g_{{\rm t},1q}$ for
$q=1,2$. In the limit investigated by Kostyuchenko,
$\tilde{J}_g=\tilde{J}_g'=-J$, $J^g_{{\rm b},1}=J^g_{{\rm
b},2}=-J_2$, and $J^g_{{\rm t},1}=J^g_{{\rm t},2}=-J_3$, our
results agree with those of Kostyuchenko for the $s=4$ and $s=0$
states, and one each of the $s=2$ and $s=3$ states. However, our
results do not agree with his for the other $s=1,2,3$ states,
except for the special case  $J_2=J_3$, which is what he claimed
to have obtained in his inappropriate fit.\cite{Kostyuchenko}
Although Kostyuchenko presented no details of his calculations,
and provided no reference to the matrix elements, in Subsection F
of the Appendix, we provided the explicit details of our
calculations.  We note that his fit assumed no widths to each
level crossings. Although this is essentially correct for the
first two level crossings, it is certainly not true for the third
and fourth level crossings.\cite{Schnack}

Even if one takes the correct forms for the eigenstate energies
with $T_d$ symmetry (neglecting the second order single-ion
anisotropy contributions) given in the appendix, one still has to
solve four equations with the three parameters $J, J_2$, and
$J_3$.  In the case that there might be an accidental remaining
degeneracy, we have tried to do this. We first assumed $J_2\ge
J_3$, but no consistent solution to the four level-crossing
equations could be found.  We then assumed $J_3>J_2$. In this
case, the minimum energies are $E_0=16J_2-8J_3$, $E_{1-}$,
$E_2=3J+\frac{39}{4}J_2-\frac{31}{4}J_3$,
$E_3=6J+\frac{61}{9}J_2-\frac{7}{9}J_3$, and $E_4=10J+6J_2+12J_3$,
where
$E_{1-}=J+\frac{87}{8}J_2-\frac{31}{8}J_3-\frac{1}{24}\sqrt{(91J_3-75J_2)^2+320(3J_2-4J_3)^2}$.
In this case, the square root appears in the equations for the
first and second level crossings, so that we add those two
equations to obtain three linear equations in the three unknowns.
Solving these three unknowns, we then find
\begin{eqnarray}
J/k_B&=&7.511{\rm K},\\
J_2/k_B&=&0.7637{\rm K},\\
J_3/k_B&=&0.9665{\rm K}.
\end{eqnarray}
These results necessarily fit the midpoints of the third and
fourth level crossings precisely.
 Now, substituting these values into the equations for the first
 and second level crossings, we obtain 4.35 T and 9.05 T, which
 are
 in remarkably good agreement with the experimental values of 4.5
 T and 8.9T, respectively.  Hence, although Kostyuchenko didn't obtain a correct
 set of formulas or a correct fit, his idea that the three-center
 terms could be important is valid.\cite{Kostyuchenko} It is remarkable that one is
 able to a good fit to four level-crossings with only three parameters.
 Nevertheless, these parameters do not give rise to any widths to
 the transitions, unlike the experiments.\cite{Schnack}

In Subsection F of the Appendix, we listed the Type-I and Type-II
first-order level crossing inductions for $s_1=1$ AFM tetramers
quantization scheme is appropriate for $C_{1v}$ symmetry, the
number of independent parameters for that low symmetry is very
large. Nevertheless, one can quantitatively fit the four
experimental level-crossing induction midpoints at 4.5 T, 8.9 T,
20.1 T, and 32 T,
 by
assuming some approximate symmetry such as $D_{2d}$ or $S_4$, for
which $J_z^g$ is non-vanishing.  The midpoint of the level
crossings occur at $\theta=\pi/4$, and the level-crossing widths
 are obtained from the differences between
the values at $\theta=0$ and $\theta=\pi/2$. We first tried to fit
the data assuming a Type-I tetramer. In this case, the widths of
the four level crossings are determined by the single parameter
$|J^g_z-3J^g_{\rm eff}|$, and are in the proportions 49 : 19 : 65
: 105, respectively. Taking $|J^g_z-3J^g_{\rm eff}|=2.0$T$\gamma$,
we obtain the widths to be 1.4T, 0.54T, 1.85T, and 3.0T, which
overestimates the width of the first level crossing, and
underestimates the widths of the third and fourth level crossings.
Nevertheless, it is interesting to try to fit the midpoint data
with this assumption.  We note that for Type-I, the level
crossings do not depend upon $J^g_{{\rm b},2}$ and $\tilde{J}_g'$,
and $J^g_{{\rm t},2}$ only enters the level-crossing equations via
the combination $\tilde{J}_g+2J^g_{{\rm t},2}$.  We first assumed
$J^g_z-3J^g_{\rm eff}=+2.0$T$\gamma$.  In this case we could fit
the four equations with the remaining four parameters, and
obtained $(\tilde{J}_g-2J^g_{{\rm t},2})/k_B=-11.02$K,
$J^g_z/k_B=-3.522$K, $J^g_{\rm eff}/k_B=-2.069$K, $J^g_{{\rm
b},1}/k_B=-2.752$K, and $J^g_{{\rm t},1}/k_B=-0.503$K.  This fit
gives a large value to the symmetric anisotropic exchange
$J^g_{\rm eff}$.  We then assumed $J^g_z-3J^g_{\rm
eff}=-2.0$T$\gamma$.  This led to
\begin{eqnarray}
(\tilde{J}_g-2J^g_{{\rm t},2})/k_B&=&-10.24{\rm K},\\
J^g_z/k_B&=&-2.570{\rm K},\\
J^g_{\rm eff}/k_B&=&+0.039{\rm K},\\
J^g_{{\rm b},1}/k_B&=&-1.861{\rm K},\\
J^g_{{\rm t},1}/k_B&=&-0.784{\rm K},
\end{eqnarray}
which is a much smaller value of $J^g_{\rm
eff}=J_{1,z}^g/2+J_{2,z}^g/3$, and a smaller ratio of $J^g_{{\rm
b},1}$ to $J^g_{{\rm t},1}$, which is reasonable. However, the fit
to the  four level-crossing widths is mediocre, at best.

It therefore seems that with the nine parameters in the four
Type-II tetramer level-crossing inductions listed in the Appendix,
 one might do better by fitting not only the midpoints but also
the widths of at least three of the level-crossing inductions. The
four widths are governed by the three parameters $J_{q,z}^g$ and
$J_z^g$ for $q=1,2$.   Setting $J_{1,z}^g=J_{2,z}^g=-J_z^g/2$, for
instance, leaves the widths in the proportions 3 : 0 : 26 : 31,
and their magnitudes are then set by  $|J_z^g|$.  We then fit
$J_z^g$ to the half-width of the fourth transition, which is
roughly 8.0T. The  fit with more reasonable parameter values is
obtained for $J_z^g<0$.  To limit the remaining parameters, we
arbitrarily take $J_{{\rm b},2}^g=J_{{\rm t},2}^g$, and to insure
Type-II behavior, choose $\tilde{J}_g'=1.5\tilde{J}_g$. Then, we
fit the midpoints of the four transitions with the remaining four
parameters, and we find
\begin{eqnarray}
\tilde{J}_g/k_B&=&-6.799\>{\rm K},\\
\tilde{J}_g'/k_B&=&-10.198\>{\rm K},\\
J_z^g/k_B&=&-4.17\>{\rm K},\\
J_{1,z}^g/k_B&=&J_{2,z}^g=2.08\>{\rm K},\\
J_{{\rm b},1}^g/k_B&=&4.602\>{\rm K},\\
J_{{\rm t},1}^g/k_B&=&-2.529\>{\rm K},\\
J_{{\rm b},2}^g/k_B&=&J^g_{{\rm t},2}/k_B=0.614\>{\rm K}.
\end{eqnarray}
The resulting widths of the transitions are 0.8 T, 0, 6.7T, and
8.0T, respectively.  These widths are in good agreement with
experiment,\cite{Schnack} and the magnitudes of both Heisenberg
interactions are larger than those of the other interactions,
justifying the first-order perturbation fit.

While this is certainly not the best fit, it is quantitatively in
agreement with experiment, and does not involve the assumption of
strongly field-dependent Heisenberg interaction
strengths.\cite{Schnack} We emphasize that this fit is not
optimized, as we made the arbitrary choices
 $J^g_{{\rm t},2}=J^g_{{\rm b},2}$,
and $J_g'=1.5J_g$, although the only restrictions on those
parameters were $\tilde{J}_g<0$ and $\tilde{J}_g'-\tilde{J}_g<0$.
In addition, non-vanishing DM interactions (which vanish in first
order) do give some additional widths to the level crossings, and
these might provide an additional contribution to the broad third
and fourth level crossings observed in experiment.\cite{Schnack}
The best fit to experiment may not be either Type-I or Type-II,
but may involve a more complicated analysis involving other states
within some of the constant $s$ manifolds.

However, with only polycrystalline data available, it is difficult
to distinguish the different possible interactions uniquely.  When
single crystals of sufficient size for low-temperature
magnetization measurements are made, we intend to fit the data
using a more consistent set of parameters, neglecting any field
dependencies, if possible.\cite{ekfuture} To limit the number of
parameters, we intend to make the assumption of $C_{3v}$ symmetry,
which will require a new quantization scheme,
$|\nu\rangle=|s,m,s_{123},s_{12},s_1\rangle$ and the appropriate
single-ion matrix elements, which are not yet in the
literature.\cite{Bocabook}

We note that our formulation of the single-ion matrix elements in
terms of a pair of dimers is applicable to low-symmetry systems
such as
Mo$_{12}$O$_{30}(\mu_2$-OH)$_{10}$H$_2$\{Ni(H$_2$O)$_3\}_4$,
abbreviated as $\{$Ni$_4$Mo$_{12}\}$,\cite{Schnack}
 systems such as  Ni$_4$
tetramers obtained from salts of
[Ni$_4$(H$_2$O)$_2$(PW$_9$O$_{34}$)$_2]^{10-}$ with $C_{2v}$
symmetry,\cite{Ni4C2v} and the unequal-spin systems
Mn$_2^{II}$Mn$_2^{III}$ and
Ni$_2^{II}$Mn$_2^{III}$.\cite{Lecren,Ni2Mn2} In the first system
with $C_{1v}$ symmetry, one would expect many more single-ion,
symmetric anisotropic exchange, and DM interactions, making
definitive fits to the existing powder magnetization data
problematic.\cite{Schnack} However, to improve the fits to
$C_{3v}$ or $D_3$ symmetry systems, such as the Fe$_4$ compound
Fe$_4$(thme)$_2$(dpm)$_6$ and the Cr$^{III}$Ni$_3^{II}$ tetramer
with an $s=9/2$ ground state,\cite{Fe4,CrNi3} would require a
reformulation of the single-ion matrix elements as a trimer plus a
monomer.\cite{ekfuture}  Even classical Heisenberg models of such
systems show strongly different dynamics than of systems with
$T_d$ symmetry.\cite{ka,ak}

Unless they vanish identically, as for $T_d$ and $C_{4v}$
symmetries, DM interactions will appear in the second-order
eigenstate energies. This is true even when the DM interactions
are site-dependent and average to zero, just as for the cases of
site-dependent single-ion and symmetric anisotropic exchange
interactions. Hence, although they have been neglected in many
fits to experimental data, they should be included in subsequent
fits.  They are most prominent for systems with lower symmetry,
such as $S_4$ or $D_{2d}$, and the effects become increasingly
strong with increasing $s_1$ value. As an example, in the
Appendix, we derived the symmetry-allowed DM interactions for the
lower-symmetry $C_{2v}^{13}$, appropriate for the [2$\times2$]
grid tetramers.\cite{WaldmannNi4,Waldmannreview}

Finally, the DM interactions also can  give rise to an electric
polarization, and hence to multiferroic behavior.  Our results
suggest that this behavior should apply for tetramers with all
possible individual spin values, as long as there is no center of
inversion symmetry connecting the interacting spin pairs.  For
tetramers with $S_4$ and $D_{2d}$ or lower symmetry, this should
be observable.  Our results indicate that these effects should
also occur for quantum spins.  In addition, there is an
interesting parity effect present in the systems we studied, with
$|{\bm P}_s(\theta)|\propto(\theta-\pi/2)^{2s_1+1}$.  This
deserves further study to elucidate its generality.

\section{IX. Conclusions}

We presented a theory of high-symmetry single molecule magnets,
including a compact form for the exact single-spin matrix elements
for four general spins.  We used the local axial and azimuthal
vector groups to construct the  invariant single-ion and symmetric
anisotropic exchange  Hamiltonians, and the molecular
representation to obtain the Dzyaloshinskii-Moriya interactions,
for equal-spin $s_1$ tetramers with site point group symmetries
$T_d$, $D_{4h}$, $D_{2d}$, $S_4$, $C_{4h}$,  or $C_{4v}$. Each
vector group introduces site-dependent molecular single-ion and
anisotropic exchange interactions. Assuming weak effective
site-independent single-ion, symmetric exchange anisotropy, and
isotropic biquadratic exchange interactions, we evaluated the
first-order corrections to the eigenstate energies.  Depending
upon the relative strengths of the near-neighbor and
next-nearest-neighbor Heisenberg exchange interactions, there are
generally two types of high-symmetry tetramers. For the single-ion
and symmetric exchange anisotropy interactions, we provided
analytic results and illustrations of the antiferromagnetic
level-crossing inductions. We also provided Hartree expressions
for the magnetization, specific heat, EPR absorption, and INS
cross-section, which are accurate at low temperatures and
arbitrary magnetic fields. For ferromagnetic tetramers, we
provided a procedure for a precise EPR determination of three of
the microscopic anisotropy parameters.  We predict that
geometrically frustrated tetramers with symmetries $S_4$ and
$D_{2d}$, as well as $C_{2v}^{13}$, are likely candidate materials
for multiferroic states. Our procedure is extendable to more
general systems.

We thank N. S. Dalal, D. Khomskii, and J. van den Brink for
helpful comments and discussions. This work was supported in part
by the NSF under contract NER-0304665.

\section{Appendix}

\subsection{A. Symmetry operation matrices}
Rotations by $\pm\pi/2$ about the $z$ axis are represented by
\begin{eqnarray}
{\cal O}_{1,2}&=&\left(\begin{array}{ccc} 0&\pm1&0\\
\mp1&0&0\\
0&0&1\end{array}\right).\label{O12}
\end{eqnarray}
Rotations by $\pi$ about the $x$ and $y$ axes are represented by
\begin{eqnarray}
{\cal O}_{3,4}&=&\left(\begin{array}{ccc} \pm1&0&0\\
0&\mp1&0\\
0&0&-1\end{array}\right).\label{O34}
\end{eqnarray}
Rotations by $\pi$ about the $z$ axis and reflections in the $xy$
plane are respectively represented by
\begin{eqnarray}
{\cal O}_{5,6}&=&\left(\begin{array}{ccc} \mp1&0&0\\
0&\mp1&0\\
0&0&\pm1\end{array}\right).\label{O56}
\end{eqnarray}
Rotations by $\pi$ about the $y=\pm x$ diagonal axes are
represented by
\begin{eqnarray}
{\cal O}_{7,8}&=&\left(\begin{array}{ccc} 0&\pm1&0\\
\pm1&0&0\\
0&0&-1\end{array}\right).\label{O78}
\end{eqnarray}
Reflections in the $xz$ and $yz$ mirror planes are represented by
\begin{eqnarray}
{\cal O}_{9,10}&=&\left(\begin{array}{ccc} \pm1&0&0\\
0&\mp1&0\\
0&0&1\end{array}\right).\label{O910}
\end{eqnarray}
Reflections in the mirror planes containing the $z$ axis and the
diagonals $y=\pm x$ are represented by
\begin{eqnarray}
{\cal O}_{11,12}&=&\left(\begin{array}{ccc} 0&\pm1&0\\
\pm1&0&0\\
0&0&1\end{array}\right).\label{O1112}
\end{eqnarray}
Reflections in the mirror planes containing the $y$ axis and the
lines $z=\pm x$ are represented by
\begin{eqnarray}
{\cal O}_{13,14}&=&\left(\begin{array}{ccc} 0&0&\pm1\\
0&1&0\\
\pm1&0&0\end{array}\right).\label{O1314}
\end{eqnarray}
Reflections in the mirror planes containing the $x$ axis and the
lines $y=\pm z$ are represented by
\begin{eqnarray}
{\cal O}_{15,16}&=&\left(\begin{array}{ccc} 1&0&0\\
0&0&\pm1\\
0&\pm1&0\end{array}\right).\label{O1516}
\end{eqnarray}
For $T_d$, clockwise rotations by $2\pi/3$ about the cube
diagonals are represented by
\begin{eqnarray}
{\cal O}_{17,18}&=&\left(\begin{array}{ccc} 0&\pm1&0\\
0&0&1\\
\pm1&0&0\end{array}\right)\label{O1718}
\end{eqnarray}
and
\begin{eqnarray}
{\cal O}_{19,20}&=&\left(\begin{array}{ccc} 0&\pm1&0\\
0&0&-1\\
\mp1&0&0\end{array}\right).\label{O1920}
\end{eqnarray}
Counterclockwise rotations by $2\pi/3$ about the cube diagonals
are represented by ${\cal O}_{\lambda}^{T}={\cal
O}_{\lambda}^{-1}$ for $\lambda=17,\ldots,20$.  Finally, there are
the six $S_4$ improper rotations consisting of rotations about a
high-symmetry axis by $\pm\pi/2$ followed by a reflection in the
plane perpendicular to the rotation axis.  For $S_4$ symmetry, $z$
is the high symmetry axis, and the operations are represented by
\begin{eqnarray}
{\cal O}_{21,22}&=&\left(\begin{array}{ccc} 0&\pm1&0\\
\mp1&0&0\\
0&0&-1\end{array}\right).\label{O2122}
\end{eqnarray}
For $T_d$ symmetry, we also have
\begin{eqnarray}
{\cal O}_{23,24}&=&\left(\begin{array}{ccc} -1&0&0\\
0&0&\pm1\\
0&\mp1&0\end{array}\right)\label{O2324}
\end{eqnarray}
and
\begin{eqnarray}
{\cal O}_{25,26}&=&\left(\begin{array}{ccc} 0&0&\pm1\\
0&-1&0\\
\mp1&0&0\end{array}\right).\label{O2526}
\end{eqnarray}

As described in the text, $C_{4h}$ symmetry involves ${\cal
O}_1=({\cal O}_2)^{T}$ and ${\cal O}_6$.

$D_{4h}$ symmetry contains the same three symmetry operations of
$C_{4h}$ symmetry, rotations by $\pm\pi/2$ about the $z$ axis,
${\cal O}_{1,2}$, and reflections in the $xy$ plane, ${\cal O}_6$.
In addition, it is also symmetric under the four  rotations by
$\pi$ about the $x$ and  $y$ axes, represented by ${\cal
O}_{3,4}$, and about the $y=\pm x$ diagonals, represented by
${\cal O}_{7,8}$.\cite{Tinkham}

For $C_{4v}$ symmetry, there are six group operations.  These are
rotations by $\pm\pi/2$ about the $z$ axis, ${\cal O}_{1,2}$,
reflections in the $xz$ and  $yz$ planes, represented respectively
by ${\cal O}_{9,10}$, and reflections in the diagonal mirror
planes containing the $z$ axis and the lines $y=\pm x$,
represented by ${\cal O}_{11,12}$.

For the lowest group symmetry under study, $S_{4}$, the only two
group operations are clockwise and counterclockwise
 rotations by $\pi/4$ about the $z$ axis, followed by a reflection in the
   $xy$ plane.\cite{Tinkham}   These improper rotations
are represented by ${\cal O}_{21,22}$.

Besides the identity operation, $D_{2d}$ group symmetry has five
operations. The first three are rotations by $\pi$ about the $x$,
$y$, and $z$ axes, respectively represented by ${\cal O}_{3,4,6}$.
In addition, there are two diagonal mirror planes associated with
the principal axis, $z$. These are represented by ${\cal
O}_{11,12}$.

Finally, the highest symmetry under study, $T_d$, has 23
operations besides the identity.  The first five are the same as
for $D_{2d}$ symmetry:  rotations by $\pi$ about the $x$, $y$, and
$z$ axes, and reflections in the two diagonal mirror planes
associated with the $z$ axis.  Then there are the four diagonal
mirror planes associated with $x$ and $y$ axes, represented
respectively by ${\cal O}_{13,14,15,16}$.  Next, there are the
four clockwise and four counterclockwise rotations by $2\pi/3$
about the cube diagonals.  The four clockwise rotations are
represented by ${\cal O}_{17,18,19,20}$, and the four
counterclockwise rotations are represented respectively by ${\cal
O}^{T}_{17,18,19,20}={\cal O}^{-1}_{17,18,19,20}$. Finally, there
are the six improper rotations by $\pm\pi/2$ about the $x$, $y$,
and $z$ axes, represented by ${\cal O}_{21,22,23,24,25,26}$.

\subsection{B. Molecular single-ion interactions}

The site-independent interactions in the molecular representation
are
\begin{eqnarray}
J_z^{g}&=&J_a^g\qquad\qquad{\rm for}\> g=C_{4h}, D_{4h},\label{JzC4h}\\
J_z^{g}&=&\frac{1}{2}\Bigl(J_a^g(3\cos^2\theta_1^{g}-1)+3J_e^g\sin^2\theta_1^{g}\Bigr)\nonumber\\
&&\qquad\qquad\>\>{\rm for}\> g=C_{4v}, D_{2d},\label{JzC4v}\nonumber\\
& &\\
J_z^{S_4}&=&\frac{J_a^{S_4}}{2}(3\cos^2\theta_1^{S_4}-1)\nonumber\\
& &-\frac{3}{2}J_e^{S_4}\sin^2\theta_1^{S_4}\cos(2\psi_1^{S_4}),\label{JzS4}\\
 J_z^{T_d}&=&0.\label{JzTd}
\end{eqnarray}

For $g=D_{4h}$, the only non-vanishing site-dependent single-ion
interaction is
\begin{eqnarray}
K_{xy}^{D_{4h}}(n)&=&(-1)^{n+1}J_e^{D_{4h}}.
\end{eqnarray}
For $g=C_{4h}$, the two non-vanishing site-dependent single-ion
interactions are
\begin{eqnarray}
J_{xy}^{C_{4h}}&=&J_e^{C_{4h}}\cos(2\chi_1^{C_{4h}}),\\
K_{xy}^{C_{4h}}(n)&=&(-1)^{n+1}J_e^{C_{4h}}\sin(2\chi_1^{C_{4h}}).
\end{eqnarray}
For $g=C_{4v}, D_{2d}$,  the three non-vanishing site-dependent
single-ion interactions in Eq. (\ref{Hsimolecular}) are
\begin{eqnarray}
K_{xy}^{g}(n)&=&\frac{(-1)^{n+1}}{2}\Bigl(J_a^g\sin^2\theta_1^{g}\nonumber\\
& &+J_e^g(1+\cos^2\theta_1^{g})\Bigr),\label{KxyD2d}\\
K_{xz}^{g}(n)&=&\frac{1}{2}\cos[(2n-1)\pi/4]\sin(2\theta_1^{g})(J_a^g-J_e^g),\label{KxzD2d}\nonumber\\
& &\\
K_{yz}^g(n)&=&\frac{1}{2}\sin[(2n-1)\pi/4]\sin(2\theta_1^{g})(J_a^g-J_e^g).\label{KyzD2d}\nonumber\\
\end{eqnarray}
For $S_4$, the single-ion site-dependent interactions are
\begin{eqnarray}
J_{xy}^{S_4}&=&-J_1\cos(2\phi_1^{S_4})\nonumber\\
& &\qquad+J_2\sin(2\phi_1^{S_4}) ,\label{JxyofmuS4}\\
K_{xy}^{S_4}(n)&=&(-1)^n\Bigl(J_1\sin(2\phi_1^{S_4})\nonumber\\
& &\qquad+J_2\cos(2\phi_1^{S_4})\Bigr),\\
K_{xz}^{S_4}(n)&=&(-1)^{n+1}\Bigl[J_3\cos\Big(\phi_1^{S_4}-\frac{n\pi}{2}\Bigr)\nonumber\\
& &\qquad-J_4\sin\Bigl(\phi_1^{S_4}-\frac{n\pi}{2}\Bigr)\Bigr],\\
K_{yz}^{S_4}(n)&=&(-1)^{n+1}\Bigl[J_3(\sin\Bigl(\phi_1^{S_4}-\frac{n\pi}{2}\Bigr)\nonumber\\
& &\qquad+J_4\cos\Bigl(\phi_1^{S_4}-\frac{n\pi}{2}\Bigr)\Bigr],\\
J_1&=&\frac{1}{2}\Bigl(J_a^{S_4}\sin^2\theta_1^{S_4}\nonumber\\
& &-J_e^{S_4}(1+\cos^2\theta_1^{S_4})\cos(2\psi_1^{S_4})\Bigr),\\
J_2&=&J_e^{S_4}\cos\theta_1^{S_4}\sin(2\psi_1^{S_4}),\\
J_3&=&\frac{1}{2}\sin(2\theta_1^{S_4})[J_a^{S_4}+J_e^{S_4}\cos(2\psi_1^{S_4})],\nonumber\\
& &\\
J_4&=&J_e^{S_4}\sin\theta_1^{S_4}\sin(2\psi_1^{S_4}).
\end{eqnarray}

The site-dependent single-ion interactions for $T_d$ symmetry are
easily obtained from those Eqs. (\ref{KxyD2d})-(\ref{KyzD2d}) by
setting $\theta_1^{D_{2d}}\rightarrow\tan^{-1}(\sqrt{2})$ and
$J_e^{T_d}\rightarrow0$,
\begin{eqnarray}
K_{xy}^{T_d}(n)&=&\frac{(-1)^n}{3}J_a^{T_d},\\
K_{xz}^{T_d}(n)&=&\frac{\sqrt{2}}{3}\cos[(2n-1)\pi/4]J_a^{T_d},\\
K_{yz}^{T_d}(n)&=&\frac{\sqrt{2}}{3}\sin[(2n-1)\pi/4]J_a^{T_d}.
\end{eqnarray}

\subsection{C. Molecular anisotropic exchange interactions}
We first consider the symmetric anisotropic exchange interactions,
letting $q=1,2$ and $p=q+1$.
 For simplicity of presentation, we write
\begin{eqnarray}
J_{q,\pm}^g&=&\frac{1}{2}(J_{c,q}^g\pm J_{f,q}^g).
\end{eqnarray} Then, the isotropic exchange renormalizations may be written as
\begin{eqnarray}
\delta J_{g}&=&\delta J_{g}'=0\>\>{\rm for}\> g=C_{4h}, D_{4h}, C_{4v},T_d\label{deltaJC4h}\nonumber\\
& &\\
\delta J_{D_{2d}}&=&-J_{1,-}^{D_{2d}}\sin^2\theta_{12}^{D_{2d}},\label{deltaJD2d}\\
\delta J_{D_{2d}}'&=&-J_{2,-}^{D_{2d}},\label{deltaJgp}\\
 \delta
J_{S_4},\delta
J_{S_4}'&=&\frac{\sin^2\theta_{1p}^{S_4}}{2}[J_{f,q}^{S_4}+J_{c,q}^{S_4}\cos(2\psi_{1p}^{S_4})].\label{deltaJS4}
\end{eqnarray}
In Eq. (\ref{deltaJS4}),  $q=1,2$ corresponds to $\delta J_{S_4},
\delta J_{s_4}'$, respectively.

 The non-vanishing site-independent symmetric anisotropic exchange interactions in the molecular representation are
\begin{eqnarray}
J_{q,z}^{g}&=&-J_{f,q}^g\>\>{\rm for}\>g=C_{4h}, D_{4h}, C_{4v},\label{J1zC4h}\\
J_{1,z}^{D_{2d}}&=&\frac{J_{f,1}^{D_{2d}}}{2}(1-3\cos^2\theta_{12}^{D_{2d}})\nonumber\\
& &-\frac{3J_{c,1}^{D_{2d}}}{2}\sin^2\theta_{12}^{D_{2d}},\label{J1zD2d}\\
J_{2,z}^{D_{2d}}&=&-J_{2,-}^{D_{2d}}-J_{c,2}^{D_{2d}},\\
J_{q,z}^{S_4}&=&\frac{J_{f,q}^{S_4}}{2}(1-3\cos^2\theta_{1p}^{S_4})\nonumber\\
&
&+\frac{3J_{c,q}^{S_4}}{2}\sin^2\theta_{1p}^{S_4}\cos(2\psi_{1p}^{S_4}).\label{J1zS4}
\end{eqnarray}

For $g=C_{4h}$, the non-vanishing site-dependent symmetric
anisotropic exchange interactions in Eq. (\ref{Haemolecular}) have
strengths
\begin{eqnarray}
K_{q,xy}^{C_{4h}}&=&J_{c,q}^{C_{4h}}\sin(2\chi_{1p}^{C_{4h}}),\\
J_{q,xy}^{C_{4h}}&=&-J_{c,q}^{C_{4h}}\cos(2\chi_{1p}^{C_{4h}}),
\end{eqnarray}
where $\chi_{1p}^g=\phi_{1p}^g+\psi_{1p}^g$.
 For $g=D_{4h}, C_{4v}$, the
non-vanishing site-dependent symmetric anisotropic exchange
interaction strengths are
\begin{eqnarray}
J_{1,xy}^{g}&=&-J_{c,1}^{g},\\
K_{2,xy}^{g}&=&-J_{c,2}^{g}.
\end{eqnarray}

Again, the more interesting group is $g=S_4$.  We find
\begin{eqnarray}
J_{q,xy}^{S_4}&=&-\tilde{J}_1\cos(2\phi_{1p}^{S_4})\nonumber\\
& &\qquad+\tilde{J}_2\sin(2\phi_{1p}^{S_4}),\label{J1xyS4}\\
K_{q,xy}^{S_4}&=&\tilde{J}_1\sin(2\phi_{1p}^{S_4})\nonumber\\
& &\qquad+\tilde{J}_2\cos(2\phi_{1p}^{S_4}),\label{K1xyS4}\\
K_{q,xz}^{S_4}(n)&=&-\tilde{J}_3\cos\Bigl(\phi_{1p}^{S_4}-\frac{n\pi}{2}\Bigr)\nonumber\\
& &\qquad+\tilde{J}_4\sin\Bigl(\phi_{1p}^{S_4}-\frac{n\pi}{2}\Bigr),\label{K1xzS4}\\
K_{q,yz}^{S_4}(n)&=&-\tilde{J}_3\sin\Bigl(\phi_{1p}^{S_4}-\frac{n\pi}{2}\Bigr)\nonumber\\
&
&\qquad-\tilde{J}_4\cos\Bigl(\phi_{1p}^{S_4}-\frac{n\pi}{2}\Bigr),\label{K1xyS4}\\
\tilde{J}_1&=&\frac{1}{2}\Bigl(J_{f,q}^{S_4}\sin^2\theta_{1p}^{S_4}\nonumber\\
& &\>-J_{c,q}^{S_4}(1+\cos^2\theta_{1p}^{S_4})\cos(2\psi_{1p}^{S_4})\Bigr),\label{tildeJ1}\\
 \tilde{J}_2&=&J_{c,q}^{S_4}\cos\theta_{1p}^{S_4}\sin(2\psi_{1p}^{S_4}),\label{tildeJ2}\\
\tilde{J}_3&=&\frac{1}{2}\sin(2\theta_{1p}^{S_4})[J_{f,m}^{S_4}+J_{c,m}^{S_4}\cos(2\psi_{1p}^{S_4})],\label{tildeJ3}\\
\tilde{J}_4&=&J_{c,q}^{S_4}\sin\theta_{1p}^{S_4}\sin(2\psi_{1p}^{S_4}).\label{tildeJ4}
\end{eqnarray}

For $g=D_{2d}$, the non-vanishing site-dependent anisotropic
exchange interaction strengths are
\begin{eqnarray}
J_{1,xy}^{D_{2d}}&=&-\frac{J_{f,1}^{D_{2d}}}{2}\sin^2\theta_{12}^{D_{2d}}\nonumber\\
& &-\frac{J_{c,1}^{D_{2d}}}{2}(1+\cos^2\theta_{12}^{D_{2d}}),\label{J1xyD2d}\\
K_{1,xz}^{D_{2d}}(n)&=&\cos(n\pi/2)J_{1,+}^{D_{2d}}\sin2\theta_{12}^{D_{2d}},\\
K_{1,yz}^{D_{2d}}(n)&=&-\sin(n\pi/2)J_{1,+}^{D_{2d}}\sin2\theta_{12}^{D_{2d}},\\
K_{2,xy}^{D_{2d}}&=&-J_{2,+}^{D_{2d}}.\label{J2xyD2d}
\end{eqnarray}
For $g=T_d$, there are no symmetric or antisymmetric anisotropic
exchange interactions.

The antisymmetric anisotropic exchange interactions in the
molecular representation are given for $g=C_{4h}, D_{4h}, S_4,$
and $D_{2d}$ by

\begin{eqnarray}
d_z^{g}(n)&=&d_z^g\qquad\qquad\qquad{\rm for}\>g=C_{4h}, D_{4h},\\
{\bm d}_q^{g}&=&0\qquad\qquad\qquad\>\>{\rm for}\>g=C_{4h}, D_{4h},\\
d_z^{g}(n)&=&d_z^g(-1)^{n+1}\qquad\>\>{\rm for}\>g=S_4,D_{2d},\\
{\bm d}_1^{D_{2d}}&=&d_{y1}^{D_{2d}}\hat{\bm y},\\
{\bm d}_2^{D_{2d}}&=&d_{x2}^{D_{2d}}(\hat{\bm x}+\hat{\bm y}),\\
{\bm d}_1^{S_4}&=&d_{x1}^{S_4}\hat{\bm x}+d_{y1}^{S_4}\hat{\bm y},\\
{\bm d}_2^{S_4}&=&d_{x2}^{S_4}\hat{\bm x}+d_{y2}^{S_4}\hat{\bm y}.
\end{eqnarray}
 Tetramers with the lowest-symmetry  $S_4$ require
five parameters to describe the full DM interactions, those with
$D_{2d}$  symmetry require three parameters, those with either
$C_{4h}$ or $D_{4h}$ symmetry require just one parameter, and
tetramers with $T_d$ or $C_{4v}$ symmetry have no DM interactions.

\subsection{D. $C_{2v}^{13}$ DM interactions}

For the [2$\times2$] grid compounds with approximate $C_{2v}^{13}$
symmetry, the four spins lie on the corners of a rhombus of side
$a$ with the position vectors relative to the origin given by
\begin{eqnarray}
{\bm r}_n&=&\frac{a}{\sqrt{2}}\Bigl[\hat{\bm
x}\Bigl(\sin[(2n-1)\pi/4]+\cos[(2n-1)\pi/4]\cos\theta_0\Bigr)\nonumber\\
& &+\hat{\bm y}\cos[(2n-1)\pi/4]\sin\theta_0 \Bigr],
\end{eqnarray}
where the acute angle $\theta_0$ satisfies $\pi/2>\theta_0>\pi/3$
for the [2$\times2$] grid
compounds.\cite{WaldmannNi4,Waldmannreview} There are three
$C_{2v}^{13}$ symmetry operations ${\cal
O}_{\lambda}^{C_{2v}^{13}}$.  The first describes rotations about
the $z$ axis by $\pi$, equivalent to ${\cal O}_3^{D_{2d}}$. The
other two are mirror planes containing the $z$ axis and the
diagonals of the rhombus,\cite{KRS} which are described by
\begin{eqnarray}
{\cal O}_{2,3}^{C_{2v}^{13}}&=&\left(\begin{array}{ccc}
\pm\cos\theta_0&\pm\sin\theta_0&0\\
\pm\sin\theta_0&\mp\cos\theta_0&0\\
0&0&1\end{array}\right).
\end{eqnarray}
For the NNN and next-next-nearest-neighbor DM interactions,
corresponding to pairs across the diagonals, vanish due to Moriya
rule (3) and invariance under ${\cal O}_{2,3}^{C_{2v}^{13}}$.
However, as for $C_{4h}$ and $D_{4h}$ symmetries, the DM
interactions between NN spins do not vanish for $C_{2v}^{13}$
symmetry, but are given by
\begin{eqnarray}
{\cal
H}_{DM}^{C_{2v}^{13}}&=&\sum_{n=1}^4\Bigl[d_z(-1)^{n+1}\hat{\bm
z}+{\bm d}\sin(n\pi/2)\nonumber\\
& &-\tilde{\bm d}\cos(n\pi/2)\Bigr]\cdot\Bigl({\bm
S}_n\times{\bm S}_{n+1}\Bigr), \label{HDMC2v}\\
\tilde{\bm d}&=&{\cal O}_2^{C_{2v}^{13}}\cdot{\bm d},
\end{eqnarray}
for a general two-vector ${\bm d}$ in the $xy$ plane. We note that
Eq. (\ref{HDMC2v}) is invariant under all three symmetries of
$C_{2v}^{13}$.

\subsection{E. Compact single-ion matrix elements}

By using the Schwinger boson technique of representing a spin by two
non-interacting bosons, and checking our results using the standard
Clebsch-Gordan algebra with the assistance of symbolic manipulation
software,
 we find  the  single-spin
matrix elements  with  general $\{s_n\}=(s_1,s_2,s_3,s_4)$ to be
\begin{eqnarray}
\langle\nu'|S_{n,\tilde{z}}|\nu\rangle
&=&\delta_{m',m}\biggl(m\delta_{s',s}\Gamma_{s^{}_{13},s_{13}',s^{}_{24},s_{24}'}^{\{s_n\},s,n}\nonumber\\
&
&+\delta_{s',s+1}C_{-s-1}^m\Delta_{s^{}_{13},s_{13}',s^{}_{24},s_{24}'}^{\{s_n\},-s-1,n}\nonumber\\
& &+\delta_{s',s-1}C_s^m\Delta_{s^{}_{13},s_{13}',s^{}_{24},s_{24}'}^{\{s_n\},s,n}\biggr),\label{Mz}\\
\langle\nu'|S_{n,\tilde{\sigma}}|\nu\rangle&=&\delta_{m',m+\tilde{\sigma}}\biggl(A_s^{\tilde{\sigma}
m}\delta_{s',s}
\Gamma_{s^{}_{13},s_{13}',s^{}_{24},s_{24}'}^{\{s_n\},s,n}\nonumber\\
&
&-\delta_{s',s+1}D_{-s-1}^{\tilde{\sigma},m}\Delta_{s^{}_{13},s_{13}',s^{}_{24},s_{24}'}^{\{s_n\},-s-1,n}\nonumber\\
& &+\delta_{s',s-1}D_s^{\tilde{\sigma},m}\Delta_{s^{}_{13},s_{13}',s^{}_{24},s_{24}'}^{\{s_n\},s,n}\biggr),\label{Msigma}\\
C_s^m&=&\sqrt{s^2-m^2},\label{Csm}\\
D_s^{\tilde{\sigma},m}&=&\tilde{\sigma}\sqrt{(s-\tilde{\sigma}m)(s-\tilde{\sigma}m-1)},\label{Dsigmasm}\\
\Gamma_{s_{13},s_{13}',s_{24},s_{24}'}^{\{s_n\},s,n}&=&\delta_{s_{24}',s^{}_{24}}\epsilon_n^{-}\alpha_{s_1,s_3}^{s^{}_{24},s,n}(s^{}_{13},s_{13}')\nonumber\\
&
&+\delta_{s_{13}',s^{}_{13}}\epsilon_n^{+}\alpha_{s_2,s_4}^{s_{13},s,n}(s_{24},s_{24}'),\\
\Delta_{s^{}_{13},s_{13}',s^{}_{24},s_{24}'}^{\{s_n\},s,n}&=&\delta_{s_{24}',s^{}_{24}}\epsilon_n^{-}\beta_{s_1,s_3}^{s_{24},s,n}(s^{}_{13},s_{13}')\nonumber\\
&
&+\delta_{s_{13}',s^{}_{13}}\epsilon_n^{+}\beta_{s_2,s_4}^{s^{}_{13},s,n}(s^{}_{24},s_{24}'),\\
\alpha_{s_1,s_3}^{s^{}_{24},s,n}(s^{}_{13},s_{13}')&=&\frac{1}{4}(1+\xi_{s,s^{}_{13},s^{}_{24}})\delta_{s_{13}',s^{}_{13}}\nonumber\\
&
&-\sqrt{2}\sin[(2n-1)\pi/4]\Bigl(F^{s_{13},s_{24}}_{s_1,s_3,s}\delta_{s_{13}',s^{}_{13}-1}\nonumber\\
& &+F^{s_{13}+1,s_{24}}_{s_1,s_3,s}\delta_{s_{13}',s_{13}+1}\Bigr),\\
\beta_{s_1,s_3}^{s^{}_{24},s,n}(s_{13},s_{13}')&=&-\frac{(-1)^n}{4}\eta_{s,s_{13},s_{24}}\delta_{s_{13}',s^{}_{13}}\nonumber\\
&
&+\sqrt{2}\sin[(2n-1)\pi/4]\Bigl(G_{s_1,s_3,s}^{s_{13},s_{24}}\delta_{s_{13}',s^{}_{13}-1}\nonumber\\
&
&+G_{s_1,s_3,-s}^{s_{13}+1,s_{24}}\delta_{s_{13}',s^{}_{13}+1}\Bigr),\\
F_{s_1,s_3,s}^{s_{13},s_{24}}&=&\frac{\eta_{s_{13},s_1,s_3}A_{s+s_{13}}^{s_{24}}A_{s_{24}}^{s-s_{13}}}{4s(s+1)},\label{A}\nonumber\\
& &\\
G_{s_1,s_3,s}^{s_{13},s_{24}}&=&\frac{\eta_{s_{13},s_1,s_3}A_{s+s_{13}}^{s_{24}}A_{s+s_{13}-1}^{s_{24}}}{4s\sqrt{4s^2-1}},\label{B}\nonumber\\
& &\\
\eta_{z,x,y}&=&\frac{A_{x+z}^yA_y^{x-z}}{\sqrt{z^2(4z^2-1)}},\label{eta}\\
 \xi_{z,x,y}&=&\frac{x(x+1)-y(y+1)}{z(z+1)},\label{xi}\\
 \epsilon_n^{\pm}&=&\frac{1}{2}[1\pm(-1)^n],
\end{eqnarray}
where  $A_s^m$ is given by Eq.  (\ref{Asm}). The prefactors $m$,
$A_s^{\tilde{\sigma} m}$, $C_s^m$, $C_{-s-1}^m$,
$D_s^{\tilde{\sigma},m}$, and $D_{-s-1}^{\tilde{\sigma},m}$ are
consequences of the Wigner-Eckart theorem for a vector
operator.\cite{Tinkham} The challenge was to obtain the
coefficients
$\Gamma_{s^{}_{13},s_{13}',s^{}_{24},s_{24}'}^{\{s_n\},s,n}$ and
$\Delta_{s^{}_{13},s_{13}',s^{}_{24},s_{24}'}^{\{s_n\},s,n}$.
Their hierarchical structure based upon  the unequal-spin dimer
suggests that analogous coefficients  with $n
> 4$ may be obtainable.\cite{ek2} Details will be presented
elsewhere.\cite{ekfuture}

\subsection{F. First-order eigenstate energy constants}

The constants appearing in the first-order eigenstate energies
(\ref{E1}) are
\begin{eqnarray}
c_{\overline{\nu}}^{\pm}&=&\frac{1}{4}\Bigl(1\pm\xi^2_{s,s_{13},s_{24}}-\eta_{s,s_{13},s_{24}}^2-\eta_{s+1,s_{13},s_{24}}^2\Bigr)\nonumber\\
a_{\overline{\nu}}^{\pm}&=&c_{\overline{\nu}}^{+} \pm2\biggl(\sum_{\sigma=\pm1}\Bigl[\Bigl(F_{s_1,s_1,s}^{s_{13}+(\sigma+1)/2,s_{24}}\Bigr)^2\nonumber\\
&
&-\sum_{\sigma'=\pm1}\Bigl(G_{s_1,s_1,\sigma\sigma's+\sigma(1+\sigma')/2}^{s_{13}+(1+\sigma)/2,s_{24}}\Bigr)^2\Bigr]\nonumber\\
&
&+(s_{13}\leftrightarrow s_{24})\biggr),\\
b_{\overline{\nu}}^{\pm}&=&\frac{1}{8}\sum_{\sigma'=\pm1}(2s+1+\sigma')^2\biggl(\eta^2_{s+(1+\sigma')/2,s_{13},s_{24}}\nonumber\\
&
&\pm8\sum_{\sigma=\pm1}\Bigl[\Bigl(G_{s_1,s_1,\sigma\sigma's+\sigma(1+\sigma')/2}^{s_{13}+(1+\sigma)/2,s_{24}}\Bigr)^2\label{b}\nonumber\\
& &+(s_{13}\leftrightarrow s_{24})\Bigr]\biggr),
\end{eqnarray}
where the $F_{s_1,s_3,s}^{s_{13},s_{24}}$,
$G_{s_1,s_3,s}^{s_{13},s_{24}}$, $\eta_{z,x,y}$, and
$\xi_{z,x,y}$, are given by Eqs.  (\ref{A})-(\ref{xi}),
respectively.

 In order to calculate the $\langle\nu|{\cal H}_{{\rm
b},1}^g|\nu\rangle$ we first write it as
\begin{eqnarray}
\langle\nu|{\cal H}_{{\rm b},1}^g|\nu\rangle&=&-\frac{J^g_{{\rm
b},1}}{4}\sum_{n,\nu'}\Bigl|\langle\nu'|{\bm S}_n\cdot{\bm
S}_{n+1}+{\bm S}_{n+1}\cdot{\bm S}_{n}|\nu\rangle\Bigr|^2.\nonumber\\
\end{eqnarray}

We note that for general $n$, ${\bm S}_n\cdot{\bm S}_{n\pm1}$
commutes with the isotropic part of the Hamiltonian, which
includes the Heisenberg, biquadratic, and isotropic three-center
interactions.  Thus, we expect $\langle\nu'|{\bm S}_n\cdot{\bm
S}_{n+1}+ {\bm S}_{n+1}\cdot{\bm S}_n|\nu\rangle$ to vanish unless
$m'=m$ and $s'=s$. It is easy to see that $m'=m$ by inspection.
 The matrix elements for $s'=s\pm1$ and $s'=s\pm2$ can then easily
be shown from their $s,m$ dependencies to vanish.  For example,
the $s'=s-2$ terms are proportional to
$C_s^mC_{s-1}^m+\frac{1}{2}\sum_{\tilde{\sigma}}D_s^{-\tilde{\sigma},m}D_{s-1}^{\tilde{\sigma},m-\tilde{\sigma}}=0$.
   There are three non-vanishing $s'=s$ terms, corresponding to the intermediate states $s''=s,s+1,$ and $s-1$. These
are respectively proportional to
$m^2+\frac{1}{2}\sum_{\tilde{\sigma}}A_s^{-\tilde{\sigma}m}A_s^{\tilde{\sigma}(m+\tilde{\sigma})}=s(s+1)$,
$C_{-s-1}^mC_{s+1}^m-\frac{1}{2}\sum_{\tilde{\sigma}}D_{-s-1}^{-\tilde{\sigma},m}D_{s+1}^{\tilde{\sigma},m-\tilde{\sigma}}=(s+1)(2s+3)$,
and
$C_s^mC_{-s}^m-\frac{1}{2}\sum_{\tilde{\sigma}}D_s^{-\tilde{\sigma},m}D_{-s}^{\tilde{\sigma},m-\tilde{\sigma}}=s(2s-1)$.
 Hence, the Wigner-Eckart theorem guarantees that these are independent of
 $m$. We first performed two checks of our matrix element forms.
 First, we evaluated
 \begin{eqnarray}
 \langle\nu'|{\bm S}_1\cdot{\bm
 S}_3|\nu\rangle&=&\delta_{\nu,\nu'}\biggl(s(s+1)\Bigl[\frac{1}{16}(1+\xi_{s,s_{13},s_{24}})^2\nonumber\\
 & &-\Bigl(F_{s_1,s_1,s}^{s_{13},s_{24}}\Bigr)^2-\Bigl(F_{s_1,s_1,s}^{s_{13}+1,s_{24}}\Bigr)^2\Bigr]\nonumber\\
 &
 &+s(2s-1)\Bigl[\frac{1}{16}\eta^2_{s,s_{13},s_{24}}\nonumber\\
 & &-\Bigl(G_{s_1,s_1,-s}^{s_{13},s_{24}}\Bigr)^2-\Bigl(G_{s_1,s_1,-s}^{s_{13}+1,s_{24}}\Bigr)^2\Bigr]\nonumber\\
 &
 &+(s+1)(2s+3)\Bigl[\frac{1}{16}\eta^2_{s+1,s_{13},s_{24}}\nonumber\\
 & &-\Bigl(G_{s_1,s_1,s+1}^{s_{13},s_{24}}\Bigr)^2-\Bigl(G_{s_1,s_1,s+1}^{s_{13}+1,s_{24}}\Bigr)^2\Bigr]\biggr)\nonumber\\
 &=&\Bigl(\frac{1}{2}s_{13}(s_{13}+1)-s_1(s_1+1)\Bigr)\delta_{\nu,\nu'},\nonumber\\
 & &\\
 \delta_{\nu,\nu'}&=&\delta_{s,s'}\delta_{m,m'}\delta_{s_{13}^{},s_{13}'}\delta_{s_{24}^{},s_{24}'},
 \end{eqnarray}
  as required.  Similarly, $\langle\nu'|{\bm S}_2\cdot{\bm
 S}_4|\nu\rangle$ is found from the above by setting
 $s_{13}\leftrightarrow s_{24}$, as required.  Then, we found
 \begin{eqnarray}
 \sum_{n=1}^4\langle\nu'|{\bm S}_n\cdot{\bm
 S}_{n+1}|\nu\rangle&=&\frac{g_0(\overline{\nu})}{4}\delta_{\nu,\nu'},
 \end{eqnarray}
 where
 \begin{eqnarray}
 g_0(\overline{\nu})&=&s(s+1)(1-\xi^2_{s,s_{13},s_{24}})-s(2s-1)\eta^2_{s,s_{13},s_{24}}\nonumber\\
 & &-(s+1)(2s+3)\eta^2_{s+1,s_{13},s_{24}}\nonumber\\
 &=&2\Bigl[s(s+1)-s_{13}(s_{13}+1)-s_{24}(s_{24}+1)\Bigr],\label{g0}\nonumber\\
 \end{eqnarray}
 as required.
 We then may write
 \begin{eqnarray}
 \langle\nu'|{\bm S}_n\cdot{\bm
 S}_{n+1}+{\bm S}_{n+1}\cdot{\bm
 S}_n|\nu\rangle&=&\delta_{s,s'}\delta_{m,m'}M_{s_{13},s_{13}'}^{s_{24},s_{24}'}(\overline{\nu},n),\nonumber\\
 \end{eqnarray}
 where
 \begin{eqnarray}
M_{s_{13},s_{13}'}^{s_{24},s_{24}'}(\overline{\nu},n)&=&\delta_{s_{13}',s_{13}^{}}\delta_{s_{24}',s_{24}^{}}\frac{g_0(\overline{\nu})}{8}\nonumber\\
&
&+(-1)^{n+1}\sum_{\sigma,\sigma'=\pm1}\delta_{s_{24}',s_{24}^{}+\sigma}\delta_{s_{13}',s_{13}^{}+\sigma'}\nonumber\\
& &\times h_{\sigma,\sigma'}(\overline{\nu})\nonumber\\
&&-\sqrt{2}\sin[(2n-1)\pi/4]\nonumber\\
& &\times\sum_{\sigma=\pm1}\delta_{s_{13}',s_{13}^{}}\delta_{s_{24}',s_{24}^{}+\sigma}\frac{g_{\sigma}(\overline{\nu})}{4}\nonumber\\
&
&-\sqrt{2}\sin[(2n+1)\pi/4]\nonumber\\
&
&\times\sum_{\sigma=\pm1}\delta_{s_{13}',s_{13}^{}+\sigma}\delta_{s_{24}',s_{24}^{}}\frac{\tilde{g}_{\sigma}(\overline{\nu})}{4},
\end{eqnarray}
where
\begin{eqnarray}
g_{\sigma}&=&\sum_{\sigma_1=\pm1}\biggl[s(s+1)(1+\xi_{s,s_{13},s_{24}+\sigma(1+\sigma_1)/2})\nonumber\\
& &\times F_{s_1,s_1,s}^{s_{24}+(1+\sigma)/2,s_{13}}\nonumber\\
&
&+\frac{1}{2}\sum_{\sigma_2=\pm1}(2s+1+\sigma_1)(2s+1+2\sigma_1)\nonumber\\
&
&\times\eta_{s+(\sigma_1+1)/2,s_{13},s_{24}+(\sigma+\sigma_2)/2}\nonumber\\
& &\times G_{s_1,s_1,\sigma_1\sigma_2
s+\sigma_2(1+\sigma_1)/2}^{s_{24}+(1+\sigma)/2,s_{13}},\nonumber\\
& &\\
h_{\sigma,\sigma}&=&s(s+1)F_{s_1,s_1,s}^{s_{13}+(\sigma+1)/2,s_{24}}F_{s_1,s_1,s}^{s_{24}+(1+\sigma)/2,s_{13}+\sigma}\nonumber\\
& &+\frac{1}{2}\sum_{\sigma_1=\pm1}(2s+1+\sigma_1)(2s+1+2\sigma_1)\nonumber\\
& &\times G_{s_1,s_1,\sigma_1\sigma
s+\sigma(1+\sigma_1)/2}^{s_{13}+(1+\sigma)/2,s_{24}}G_{s_1,s_1,-\sigma_1\sigma
s-\sigma(1+\sigma_1)/2}^{s_{24}+(1+\sigma)/2,s_{13}+\sigma}\nonumber\\
& &+(s_{13}\leftrightarrow s_{24}),\\
h_{+,-}&=&\sum_{\sigma=\pm1}s(s+1)F_{s_1,s_1,s}^{s_{13}+1,s_{24}-(1-\sigma)/2}F_{s_1,s_1,s}^{s_{24},s_{13}+(1+\sigma)/2}\nonumber\\
&
&+\frac{1}{2}\sum_{\sigma_1=\pm1}(2s+\sigma_1+1)(2s+2\sigma_1+1)\nonumber\\
& &\times G_{s_1,s_1,\sigma\sigma_1
s+\sigma(1+\sigma_1)/2}^{s_{13}+1,s_{24}-(1-\sigma)/2}G_{s_1,s_1,\sigma\sigma_1
s+\sigma(1+\sigma_1)/2}^{s_{24},s_{13}+(1+\sigma)/2},\nonumber\\
\end{eqnarray}
and where $\tilde{g}_{\sigma}(\overline{\nu})$ and
$h_{-,+}(\overline{\nu})=\tilde{h}_{+,-}(\overline{\nu})$
 are respectively
obtained from $g_{\sigma}(\overline{\nu})$ and the
$h_{+,-}(\overline{\nu})$ by setting $s_{13}\leftrightarrow
s_{24}$.

Letting $x=s_{13}$ and $y=s_{24}$, these expressions may be
simplified to yield
\begin{eqnarray}
h_{\sigma,\sigma}(0,x,x)&=&-\sqrt{(2x+2+\sigma)(2x+\sigma)}\eta^2_{x+(1+\sigma)/2,s_1,s_1}\nonumber\\
& &\times\frac{(2x+1+\sigma)}{2},\\
h_{\sigma,\sigma}(s,x,y)&=&-\frac{(x-y)^2}{s(s+1)}\eta_{x+(1+\sigma)/2,s_1,s_1}\eta_{y+(1+\sigma)/2,s_1,s_1}\nonumber\\
&
&\times\sqrt{(x+y+1+\sigma)^2-s^2}\nonumber\\
& &\times\sqrt{(x+y+1+\sigma)^2-(s+1)^2},\nonumber\\
& &\qquad\>{\rm for}\>s\ge1,\\
h_{+,-}(s,x,y)&=&\frac{(x+y+1)^2}{s(s+1)}\eta_{x,s_1,s_1}\eta_{y+1,s_1,s_1}\nonumber\\
& &\times\sqrt{(s+1)^2-(x-y-1)^2}\nonumber\\
& &\times\sqrt{s^2-(x-y-1)^2},\\
g_{\sigma}(s,x,y)&=&\eta_{x+(1+\sigma)/2,s_1,s_1}\nonumber\\
& &\times\sqrt{[x+(1+\sigma)/2]^2-(y-s)^2}\nonumber\\
& &\times\sqrt{(y+s+1)^2-[x+(1+\sigma)/2]^2}.\nonumber\\
\end{eqnarray}
 The diagonal matrix elements of ${\cal H}^g_{{\rm
b},1}$ are then easily found to be
\begin{eqnarray}
\langle\nu|{\cal H}^g_{{\rm b},1}|\nu\rangle&=&-J^g_{{\rm
b},1}{\cal B}_{\overline{\nu}},\end{eqnarray} where
\begin{eqnarray} {\cal
B}_{\overline{\nu}}&=&\sum_{\sigma=\pm1}\biggl(h^2_{\sigma,\sigma}(\overline{\nu})
+\frac{1}{16}
\Bigl[g^2_{\sigma}(\overline{\nu})+\tilde{g}^2_{\sigma}(\overline{\nu})\Bigr]\biggr)\nonumber\\
& &+h_{+,-}^2(\overline{\nu})+\tilde{h}_{+,-}^2(\overline{\nu})
+\frac{1}{64}g^2_0(\overline{\nu}).
\end{eqnarray}

The three-center isotropic quartic spin-spin interactions proposed
by Kostyuchenko may be written for the six $g$ symmetries as
\begin{eqnarray}
{\cal H}^g_{\rm t}&=&{\cal H}^g_{{\rm t},1}+{\cal H}^g_{{\rm t},2},\\
{\cal H}^g_{{\rm t},1}&=&-\frac{1}{8}J_{{\rm
t},1}^g\sum_{n=1,3}\Bigl({\bm S}_n\cdot{\bm S}_{n+1}+{\bm
S}_{n+1}\cdot{\bm S}_{n}\Bigr)\nonumber\\
& &\times\sum_{n'=2,4}\Bigl({\bm S}_{n'}\cdot{\bm S}_{n'+1}+{\bm
S}_{n'+1}\cdot{\bm S}_{n'}\Bigr)\nonumber\\
& &\qquad +H.c.,\\
{\cal H}^g_{{\rm t},2}&=&-\frac{1}{2}J^g_{{\rm t},2}\Bigl({\bm
S}_{1}\cdot{\bm S}_{3}+{\bm S}_{2}\cdot{\bm
S}_{4}\Bigr)\nonumber\\
& &\times\sum_{n=1}^4\Bigl({\bm S}_n\cdot{\bm S}_{n+1}+{\bm
S}_{n+1}\cdot{\bm S}_{n}\Bigr),
\end{eqnarray}
where for $g=T_d$, we have $J^{T_d}_{{\rm t},1}=J^{T_d}_{{\rm
t},2}=-J_3$ in the notation of  Kostyuchenko.\cite{Kostyuchenko}
From our matrix elements above, it is then easy to see that
\begin{eqnarray}
\langle\nu'|{\cal H}^g_{{\rm
t},2}|\nu\rangle&=&-\frac{g_0(\overline{\nu})\delta_{\nu^{},\nu'}}{8}J^g_{{\rm
t},2}\Bigl(s_{13}(s_{13}+1)\nonumber\\
& &+s_{24}(s_{24}+1)-4s_1(s_1+1)\Bigr),
\end{eqnarray}
where $g_0(\overline{\nu})$ is given by Eq. (\ref{g0}). With
regard to the matrix elements $\langle\nu'|{\cal H}^g_{{\rm
t},1}|\nu\rangle$, we first note that
\begin{eqnarray}
\sum_{n=2,4}\sin[(2n\pm1)\pi/4]&=&\sum_{n=1,3}\sin[(2n\pm1)\pi/4]=0,\nonumber\\
\end{eqnarray}
so that there are no contributions from $g_{\sigma}$ and
$\tilde{g}_{\sigma}$.  The diagonal matrix elements
$\langle\nu|{\cal H}^g_{{\rm t},1}|\nu\rangle$ is then easily
found to be
\begin{eqnarray}
\langle\nu|{\cal H}^g_{{\rm t},1}|\nu\rangle&=&-J^g_{{\rm
t},1}{\cal T}_{\overline{\nu}},\\
{\cal T}_{\overline{\nu}}&=&\frac{g_0^2(\overline{\nu})}{64}-h_{+,-}^2(\overline{\nu})-\tilde{h}_{+,-}^2(\overline{\nu})\nonumber\\
&
&\qquad-\sum_{\sigma=\pm1}h_{\sigma,\sigma}^2(\overline{\nu}).\label{Tnu}
\end{eqnarray}

\begin{table}
\begin{tabular}{ccccccccccc}
$s$&$s_{13}$&$s_{24}$&$a^{+}_{\overline{\nu}}$&$a^{-}_{\overline{\nu}}$&$b^{+}_{\overline{\nu}}$&$b^{-}_{\overline{\nu}}$&$c_{\overline{\nu}}^{+}$&$c_{\overline{\nu}}^{-}$&${\cal B}_{\overline{\nu}}$&${\cal T}_{\overline{\nu}}$\vspace*{3pt}\\
\hline 2&1&1&0&$\frac{1}{3}$&2&$-\frac{2}{3}$&$\frac{1}{6}$&$\frac{1}{6}$&$\frac{1}{4}$&$\frac{1}{4}$\vspace*{3pt}\\
1&1&1&0&-1&2&2&$-\frac{1}{2}$&$-\frac{1}{2}$&$\frac{5}{4}$&$\frac{1}{4}$\vspace*{3pt}\\
1&1&0&0&1&2&-2&$\frac{1}{2}$&0&$\frac{3}{2}$&-1\vspace*{3pt}\\
0&1&1&0&$\frac{11}{3}$&2&$\frac{2}{3}$&$\frac{11}{6}$&$\frac{11}{6}$&$\frac{7}{4}$&$\frac{1}{4}$\vspace*{3pt}\\
0&0&0&2&-1&2&-2&$\frac{1}{2}$&$\frac{1}{2}$&$\frac{3}{4}$&$-\frac{3}{4}$
\vspace*{3pt}\\
\hline
\end{tabular}
\caption{Values of $a_{\overline{\nu}}^{\pm}$,
$b_{\overline{\nu}}^{\pm}$,  $c_{\overline{\nu}}^{\pm}$,  ${\cal
B}_{\overline{\nu}}$, and ${\cal T}_{\overline{\nu}}$ for
$s_1=1/2$.}\label{table4}
\end{table}

\begin{table}
\begin{tabular}{ccccccccccc}
$s$&$s_{13}$&$s_{24}$&$a^{+}_{\overline{\nu}}$&$a^{-}_{\overline{\nu}}$&$b^{+}_{\overline{\nu}}$&$b^{-}_{\overline{\nu}}$&$c_{\overline{\nu}}^{+}$&$c_{\overline{\nu}}^{-}$&${\cal B}_{\overline{\nu}}$&${\cal T}_{\overline{\nu}}$\vspace*{3pt}\\
\hline 4&2&2&$\frac{1}{7}$&$\frac{2}{7}$&$\frac{24}{7}$&$-\frac{8}{7}$&$\frac{3}{14}$&$\frac{3}{14}$&4&4\vspace*{3pt}\\
3&2&2&$\frac{1}{15}$&$\frac{2}{15}$&$\frac{24}{5}$&$\frac{8}{5}$&$\frac{1}{10}$&$\frac{1}{10}$&4&0\vspace*{3pt}\\
3&2&1&$\frac{1}{15}$&$\frac{2}{5}$&$\frac{24}{5}$&$-\frac{16}{5}$&$\frac{7}{30}$&$\frac{8}{45}$&$\frac{43}{9}$&$-\frac{7}{9}$\vspace*{3pt}\\
2&2&2&$-\frac{1}{7}$&$-\frac{2}{7}$&$\frac{124}{21}$&$\frac{80}{21}$&$-\frac{3}{14}$&$-\frac{3}{14}$&$\frac{23}{4}$&$\frac{9}{4}$\vspace*{3pt}\\
2&2&1&$\frac{1}{3}$&0&4&0&$\frac{1}{6}$&$-\frac{1}{18}$&$\frac{307}{36}$&$-\frac{55}{36}$\vspace*{3pt}\\
2&2&0&$\frac{1}{3}$&$\frac{2}{3}$&4&-4&$\frac{1}{2}$&0&7&-3\vspace*{3pt}\\
2&1&1&$-\frac{1}{3}$&$\frac{2}{3}$&$\frac{20}{3}$&$-\frac{16}{3}$&$\frac{1}{6}$&$\frac{1}{6}$&$\frac{31}{4}$&$-\frac{23}{4}$\vspace*{3pt}\\
1&2&2&$-\frac{7}{5}$&$-\frac{14}{5}$&$\frac{36}{5}$&$\frac{32}{5}$&$-\frac{21}{10}$&$-\frac{21}{10}$&$\frac{31}{4}$&$\frac{25}{4}$\vspace*{3pt}\\
1&2&1&$\frac{3}{5}$&$\frac{8}{5}$&$\frac{68}{15}$&$-\frac{32}{15}$&$\frac{11}{10}$&$-\frac{9}{10}$&$\frac{23}{4}$&$-\frac{1}{12}$\vspace*{3pt}\\
1&1&1&1&-2&4&0&$-\frac{1}{2}$&$-\frac{1}{2}$&$\frac{15}{4}$&$\frac{1}{4}$\vspace*{3pt}\\
1&1&0&-1&2&$\frac{20}{3}$&$-\frac{20}{3}$&$\frac{1}{2}$&0&9&$-\frac{23}{3}$\vspace*{3pt}\\
0&2&2&3&6&$\frac{16}{3}$&$\frac{8}{3}$&$\frac{9}{2}$&$\frac{9}{2}$&$\frac{32}{3}$&$\frac{22}{3}$\vspace*{3pt}\\
0&1&1&$-\frac{11}{3}$&$\frac{22}{3}$&$\frac{16}{3}$&$-\frac{8}{3}$&$\frac{11}{6}$&$\frac{11}{6}$&8&-6\vspace*{3pt}\\
0&0&0&$\frac{11}{3}$&$-\frac{8}{3}$&$\frac{16}{3}$&$-\frac{16}{3}$&$\frac{1}{2}$&$\frac{1}{2}$&$\frac{16}{3}$&$-\frac{16}{3}$
\vspace*{3pt}\\
\hline
\end{tabular}
\caption{Values of $a_{\overline{\nu}}^{\pm}$,
$b_{\overline{\nu}}^{\pm}$, $c_{\overline{\nu}}^{\pm}$, ${\cal
B}_{\overline{\nu}}$ and ${\cal T}_{\overline{\nu}}$ for
$s_1=1$.}\label{table5}
\end{table}

For $s_1=1/2$, the matrix $\langle\nu'|{\cal H}^g_{{\rm
b},1}+{\cal H}^g_{{\rm t},1}|\nu\rangle$  is diagonal, so these
interactions can be treated exactly. For $s_1\ge1$, since these
quartic interactions preserve $s$, the resulting matrix is block
diagonal, as noted by Kostyuchenko.\cite{Kostyuchenko} To the
extent that the single-ion and exchange anisotropy interactions
can be neglected or treated in first order only, the matrix of the
resulting Hamiltonian  is block diagonal. Kostyuchenko compiled a
table of the diagonalized eigenstates for $T_d$ symmetry with
$s_1=1$, neglecting the anisotropy interactions. In Table IV, we
calculated the exact eigenstates for $s_1=1$ of the above
Hamiltonian for the six symmetries under consideration.

\begin{table}\begin{tabular}{cc}
$\overline{\nu}=ss_{13}s_{24}$&$E_{\overline{\nu}}$\cr \hline
422&$-4\tilde{J}_g-6\tilde{J}_g'-4J^g_{{\rm b},1}-2J^g_{{\rm
b},2}-4J^g_{{\rm t},1}-8J^g_{{\rm t},2}$\\
322&$-6\tilde{J}_g'-4J^g_{{\rm b},1}-2J^g_{{\rm b},2}$\\
312,321&$-2\tilde{J}_g-4\tilde{J}_g'-\frac{43}{9}J^g_{{\rm
b},1}-2J^g_{{\rm b},2}+\frac{7}{9}J^g_{{\rm t},1}$\cr
212,221&$\tilde{J}_g-4\tilde{J}_g'-\frac{307}{36}J^g_{{\rm
b},1}-2J^g_{{\rm b},2}+\frac{55}{36}J^g_{{\rm t},1}$\cr
211&$-\tilde{J}_g-2\tilde{J}_g'-\frac{31}{4}J^g_{{\rm
b},1}-2J^g_{{\rm b},2}+\frac{23}{4}J^g_{{\rm t},1}+2J^g_{{\rm
t},2}$\cr 222,202,220&$-3\tilde{J}_g'-4J^g_{{\rm b},1}-5J^g_{{\rm
b},2},\> E_{2\pm}$\cr
122&$2\tilde{J}_g-3\tilde{J}_g'-\frac{31}{4}J^g_{{\rm
b},1}-2J^g_{{\rm b},2}-\frac{25}{4}J^g_{{\rm t},1}+10J^g_{{\rm
t},2}$\cr 111&$\tilde{J}_g-2\tilde{J}_g'-\frac{15}{4}J^g_{{\rm
b},1}-2J^g_{{\rm b},2}-\frac{1}{4}J^g_{{\rm t},1}-2J^g_{{\rm t},2}$\\
121,101&$E_{1\pm}$\cr 112,110&$E_{1\pm}$\cr
011&$-2\tilde{J}_g+2\tilde{J}_g'-8J^g_{{\rm b},1}-2J^g_{{\rm
b},2}+6J^g_{{\rm t},1}-4J^g_{{\rm t},2}$\cr 022,000&$E_{0\pm}$
\end{tabular}
\caption{Eigenstate energies $E_{\overline{\nu}}$ for $s_1=1$ as a
function of the quantum numbers $\overline{\nu}=s,s_{13},s_{24}$
in the absence of anisotropy interactions. The two additional
$s=2$ eigenstate energies
$E_{2\pm}=\frac{1}{2}[a\pm\sqrt{b^2+56(J^g_{{\rm b},1})^2}]$,
where $a=3\tilde{J}_g-9\tilde{J}_g'-\frac{63}{4}J^g_{{\rm
b},1}-7J^g_{{\rm b},2}+\frac{15}{4}J^g_{{\rm t},1}+6J^g_{{\rm
t},2}$ and $b=3\tilde{J}_g-3\tilde{J}_g'+\frac{17}{4}J^g_{{\rm
b},1}+3J^g_{{\rm b},2}-\frac{33}{4}J^g_{{\rm t},1}+6J^g_{{\rm
t},2}$. The two sets of doubly-degenerate $s=1$ eigenstates have
$E_{1\pm}=\frac{1}{2}[A+B\pm\sqrt{(A-B)^2+4C^2}]$, where
$A=3\tilde{J}_g-4\tilde{J}_g'-\frac{23}{4}J^g_{{\rm
b},1}-2J^g_{{\rm b},2}+\frac{1}{12}J^g_{{\rm t},1}$,
$B=-\tilde{J}_g'-9J^g_{{\rm b},1}-5J^g_{{\rm
b},2}+\frac{23}{3}J^g_{{\rm t},1}$, and
$C=\frac{\sqrt{5}}{3}(3J^g_{{\rm b},1}-4J^g_{{\rm t},1})$, and
$E_{0\pm}=\frac{1}{2}[\alpha+\beta\pm\sqrt{(\alpha-\beta)^2+4\gamma^2}]$,
where $\alpha=-6\tilde{J}_g+6\tilde{J}_g'-\frac{32}{3}J^g_{{\rm
b},1}-2J^g_{{\rm b},2}-\frac{22}{3}J^g_{{\rm t},1}+12J^g_{{\rm
t},2}$, $\beta=\frac{16}{3}(J^g_{{\rm t},1}-J^g_{{\rm
b},1})-8J^g_{{\rm b},2}$, and
$\gamma=\frac{4\sqrt{5}}{3}(J^g_{{\rm t},1}-J^g_{{\rm b},1})$.}
\end{table}

In Tables IV-VII, and VII-XI, we have listed analytic formulas for
the coefficients $a^{\pm}_{\overline{\nu}}$,
$b^{\pm}_{\overline{\nu}}$, and $c^{\pm}_{\overline{\nu}}$, and
${\cal B}_{\overline{\nu}}$ for the lowest four eigenstate
manifolds of FM and AFM tetramers, respectively. In both cases,
the manifolds are restricted by $0\le s_{13},s_{24}\le 2s_1$ and
$|s_{13}-s_{24}|\le s\le s_{13}+s_{24}$. In addition, the
coefficients are symmetric under $s_{13}\leftrightarrow s_{24}$.
Hence, for $s_1=1/2$, the five distinct allowed
$(s,s_{13},s_{24})$ states are (0,0,0), (0,1,1), (1,1,0), (1,1,1),
and (2,1,1).  For $s_1=1$, the fourteen distinct allowed
$(s,s_{13},s_{24})$ states are (0,0,0), (0,1,1), (0,2,2), (1,1,0),
(1,1,1), (1,2,1), (1,2,2), (2,1,1), (2,2,0), (2,2,1), (2,2,2),
(3,2,1), (3,2,2), and (4,2,2). From these tables, the coefficients
$a^{\pm}_{\overline{\nu}}$, $b^{\pm}_{\overline{\nu}}$, and
$c^{\pm}_{\overline{\nu}}$, and ${\cal B}_{\overline{\nu}}$ for
all of the allowed eigenstates of tetramers with $s_1\le 3/2$ are
given.  For $s_1=2$, the values for the nine states with $s=4$
cannot be obtained from these formulas, but the values for the
other 46 eigenstates with $0\le s\le 3$ and $4\le s\le 8$ are
given.

\begin{table}
\begin{tabular}{ccc}
$s,s_{13},s_{24}$&$a^{+}_{\overline{\nu}}$&$a^{-}_{\overline{\nu}}$\vspace*{3pt}\\
\hline
$4s_1,2s_1,2s_1$&$\frac{2s_1-1}{8s_1-1}$&$\frac{2s_1}{8s_1-1}$\vspace*{3pt}\cr
$4s_1-1,2s_1,2s_1$&$\frac{(2s_1-1)(4s_1-3)}{(4s_1-1)(8s_1-3)}$&$\frac{2s_1(4s_1-3)}{(4s_1-1)(8s_1-3)}$\vspace*{3pt}\cr
$4s_1-1,2s_1,2s_1-1$&$\frac{(2s_1-1)(4s_1-3)}{(4s_1-1)(8s_1-3)}$&$\frac{2s_1}{8s_1-3}$\vspace*{3pt}\cr
$4s_1-2,2s_1,2s_1$&$\frac{(2s_1-1)a_1(s_1)}{(4s_1-1)(8s_1-1)}$&$\frac{2s_1a_1(s_1)}{(4s_1-1)(8s_1-1)}$\vspace*{3pt}\cr
&$\times\frac{1}{(8s_1-5)}$&$\times\frac{1}{(8s_1-5)}$\vspace*{3pt}\cr
$4s_1-2,2s_1,2s_1-1$&$\frac{a_2(s_1)}{(2s_1-1)(4s_1-1)}$&$\frac{4s_1(s_1-1)}{(2s_1-1)(8s_1-5)}$\vspace*{3pt}\cr
&$\times\frac{1}{(8s_1-5)}$&\vspace*{3pt}\cr
$4s_1-2,2s_1,2s_1-2$&$\frac{a_2(s_1)}{(2s_1-1)(4s_1-1)}$&$\frac{2a_3(s_1)}{(2s_1-1)(4s_1-1)}$\vspace*{3pt}\cr
&$\times\frac{1}{(8s_1-5)}$&$\times\frac{1}{(8s_1-5)}$\vspace*{3pt}\cr
$4s_1-2,2s_1-1,2s_1-1$&$\frac{2s_1-3}{8s_1-5}$&$\frac{2s_1}{8s_1-5}$\vspace*{3pt}\cr
$4s_1-3,2s_1,2s_1$&$\frac{(2s_1-1)(8s_1-15)}{(8s_1-3)(8s_1-7)}$&$\frac{2s_1(8s_1-15)}{(8s_1-3)(8s_1-7)}$\vspace*{3pt}\cr
$4s_1-3,2s_1,2s_1-1$&$\frac{a_4(s_1)}{(4s_1-3)^2(2s_1-1)}$&$\frac{4s_1a_5(s_1)}{(4s_1-3)^2(2s_1-1)}$\vspace*{3pt}\cr
&$\times\frac{1}{(4s_1-1)(8s_1-3)}$&$\times\frac{1}{(4s_1-1)(8s_1-3)}$\vspace*{3pt}\cr
&$\times\frac{1}{(8s_1-7)}$&$\times\frac{1}{(8s_1-7)}$\vspace*{3pt}\cr
$4s_1-3,2s_1,2s_1-2$&$\frac{a_6(s_1)}{(2s_1-1)(4s_1-3)}$&$\frac{a_7(s_1)}{(2s_1-1)(4s_1-3)}$\vspace*{3pt}\cr
&$\times\frac{1}{(8s_1-7)}$&$\times\frac{1}{(8s_1-7)}$\vspace*{3pt}\cr
$4s_1-3,2s_1,2s_1-3$&$\frac{(2s_1-1)a_8(s_1)}{(4s_1-3)^2(8s_1-7)}$&$\frac{2a_9(s_1)}{(4s_1-3)^2(8s_1-7)}$\vspace*{3pt}\cr
$4s_1-3,2s_1-1,2s_1-1$&$\frac{(2s_1-3)((4s_1-5)}{(4s_1-3)(8s_1-7)}$&$\frac{2s_1(4s_1-5)}{(4s_1-3)(8s_1-7)}$\vspace*{3pt}\cr
$4s_1-3,2s_1-1,2s_1-2$&$\frac{a_{10}(s_1)}{(4s_1-1)(4s_1-3)}$&$\frac{2a_{11}(s_1)}{(4s_1-1)(4s_1-3)}$\vspace*{3pt}\cr
&$\times\frac{1}{(8s_1-7)}$&$\times\frac{1}{(8s_1-7)}$\vspace*{3pt}\cr
 \hline
\end{tabular}
\caption{Values of $a_{\overline{\nu}}^{\pm}$ for the ground and
first three excited state manifolds for FM tetramers (or the
highest four excited state manifolds of AFM tetramers), where
$a_1(x)=32x^2-44x+3$, $a_2(x)=16x^3-36x^2+26x-3$, and
$a_3(x)=8x^3-6x^2+2x-1$,
$a_4(x)=2048x^6-9472x^5+17344x^4-15440x^3+6924x^2-1530x+135$,
$a_5(x)=512x^5-1856x^4+2416x^3-1420x^2+399x-45$
$a_6(x)=16x^3-52x^2+58x-9$, $a_7(x)=4x^3-7x^2+3x-1$,
$a_8(x)=16x^2-48x+45$, $a_9(x)=16x^3-24x^2+15x-9$,
$a_{10}(x)=32x^3-96x^2+70x-9$, and
$a_{11}(x)=16x^3-16x^2+5x-2$.}\label{table4}
\end{table}

\begin{table}
\begin{tabular}{ccc}
$s,s_{13},s_{24}$&$b^{+}_{\overline{\nu}}$&$b^{-}_{\overline{\nu}}$\vspace*{3pt}\\
\hline
$4s_1,2s_1,2s_1$&$\frac{24s_1^2}{8s_1-1}$&$-\frac{8s_1^2}{8s_1-1}$\vspace*{3pt}\cr
$4s_1-1,2s_1,2s_1$&$\frac{8s_1(5s_1-2)}{8s_1-3}$&$\frac{8s_1^2}{8s_1-3}$\vspace*{3pt}\cr
$4s_1-1,2s_1,2s_1-1$&$\frac{8s_1(5s_1-2)}{8s_1-3}$&$-\frac{8s_1(3s_1-1)}{8s_1-3}$\vspace*{3pt}\cr
$4s_1-2,2s_1,2s_1$&$\frac{4(112s_1^3-102s_1^2+22s_1-1)}{(8s_1-1)(8s_1-5)}$&$\frac{8s_1(24s_1^2-15s_1+1)}{(8s_1-1)(8s_1-5)}$\vspace*{3pt}\cr
$4s_1-2,2s_1,2s_1-1$&$\frac{4(14s_1^2-12s_1+1)}{8s_1-5}$&$-\frac{8s_1(s_1-1)}{8s_1-5}$\vspace*{3pt}\cr
$4s_1-2,2s_1,2s_1-2$&$\frac{4(14s_1^2-12s_1+1)}{8s_1-5}$&$-\frac{4(10s_1^2-8s_1+1)}{8s_1-5}$\vspace*{3pt}\cr
$4s_1-2,2s_1-1,2s_1-1$&$\frac{4(14s_1^2-10s_1+1)}{8s_1-5}$&$-\frac{8s_1(5s_1-3)}{8s_1-5}$\vspace*{3pt}\cr
$4s_1-3,2s_1,2s_1$&$\frac{12(48s_1^3-74s_1^2+34s_1-5)}{(8s_1-3)(8s_1-7)}$&$\frac{8s_1(40s_1^2-51s_1+15)}{(8s_1-3)(8s_1-7)}$\vspace*{3pt}\cr
$4s_1-3,2s_1,2s_1-1$&$\frac{4b_1(s_1)}{(4s_1-1)(4s_1-3)}$&$\frac{8s_1b_2(s_1)}{(4s_1-1)(4s_1-3)}$\vspace*{3pt}\cr
&$\times\frac{1}{(8s_1-3)(8s_1-7)}$&$\times\frac{1}{(8s_1-3)(8s_1-7)}$\vspace*{3pt}\cr
$4s_1-3,2s_1,2s_1-2$&$\frac{12(6s_1^2-8s_1+1)}{8s_1-7}$&$-\frac{4(6s_1^2-8s_1+1)}{8s_1-7}$\vspace*{3pt}\cr
$4s_1-3,2s_1,2s_1-3$&$\frac{12(24s_1^2-50s_1^2+30s_1-5)}{(4s_1-3)(8s_1-7)}$&$-\frac{4b_3(s_1)}{(4s_1-3)(8s_1-7)}$\vspace*{3pt}\cr
$4s_1-3,2s_1-1,2s_1-1$&$\frac{4(18s_162-22s_1+5)}{8s_1-7}$&$-\frac{24s_1(s_1-1)}{8s_1-7}$\vspace*{3pt}\cr
$4s_1-3,2s_1-1,2s_1-2$&$\frac{4(72s_1^3-98s_1^2+34s_1-3)}{(4s_1-1)(8s_1-7)}$&$-\frac{4b_4(s_1)}{(4s_1-1)(8s_1-7)}$\vspace*{3pt}\cr
 \hline
\end{tabular}
\caption{Values of $b_{\overline{\nu}}^{\pm}$ for the ground and
first three excited state manifolds for FM tetramers (or the
highest four excited state manifolds of AFM tetramers), where
$b_1(x)=2304x^5-6112x^4+5968x^3-2650x^2+552x-45$,
$b_2(x)=1024x^4-2304x^3+1808x^2-576x+63$,
$b_3(x)=56x^3-114x^2+72x-15$,
$b_4(x)=56x^3-70x^2+20x-1$.}\label{table5}
\end{table}

\begin{table}
\begin{tabular}{ccc}
$s,s_{13},s_{24}$&$c^{+}_{\overline{\nu}}$&$c^{-}_{\overline{\nu}}$\vspace*{3pt}\\
\hline
$4s_1,2s_1,2s_1$&$\frac{4s_1-1}{2(8s_1-1)}$&$\frac{4s_1-1}{2(8s_1-1)}$\vspace*{3pt}\cr
$4s_1-1,2s_1,2s_1$&$\frac{4s_1-3}{2(8s_1-3)}$&$\frac{4s_1-3}{2(8s_1-3)}$\vspace*{3pt}\cr
$4s_1-1,2s_1,2s_1-1$&$\frac{16s_1^2-12s_1+3}{2(4s_1-1)(8s_1-3)}$&$\frac{8s_1(2s_1-1)^2}{(4s_1-1)^2(8s_1-3)}$\vspace*{3pt}\cr
$4s_1-2,2s_1,2s_1$&$\frac{32s_1^2-44s_1+3}{2(8s_1-1)(8s_1-5)}$&$\frac{32s_1^2-44s_1+3}{2(8s_1-1)(8s_1-5)}$\vspace*{3pt}\cr
$4s_1-2,2s_1,2s_1-1$&$\frac{c_1(s_1)}{2(2s_1-1)(4s_1-1)}$&$\frac{c_2(s_1)}{2(2s_1-1)^2(4s_1-1)^2}$\vspace*{3pt}\cr
&$\times\frac{1}{(8s_1-5)}$&$\times\frac{1}{(8s_1-5)}$\vspace*{3pt}\cr
$4s_1-2,2s_1,2s_1-2$&$\frac{8s_1^2-10s_1+5}{2(2s_1-1)(8s_1-5)}$&$\frac{2s_1(4s_1^2-7s_1+3)}{(2s_1-1)^2(8s_1-5)}$\vspace*{3pt}\cr
$4s_1-2,2s_1-1,2s_1-1$&$\frac{4s_1-3}{2(8s_1-5)}$&$\frac{4s_1-3}{2(8s_1-5)}$\vspace*{3pt}\cr
$4s_1-3,2s_1,2s_1$&$\frac{(4s_1-1)(8s_1-15)}{2(8s_1-3)(8s_1-7)}$&$\frac{(4s_1-1)(8s_1-15)}{2(8s_1-3)(8s_1-7)}$\vspace*{3pt}\cr
$4s_1-3,2s_1,2s_1-1$&$\frac{c_3(s_1)}{2(2s_1-1)(4s_1-3)}$&$\frac{c_4(s_1)}{2(2s_1-1)^2(4s_1-3)^2}$\vspace*{3pt}\cr
&$\times\frac{1}{(8s_1-3)(8s_1-7)}$&$\times\frac{1}{(8s_1-3)(8s_1-7)}$\vspace*{3pt}\cr
$4s_1-3,2s_1,2s_1-2$&$\frac{(4s_1-1)}{2(2s_1-1)(4s_1-3)}$&$\frac{2(s_1-1)(4s_1-1)}{(2s_1-1)^2(4s_1-3)^2}$\vspace*{3pt}\cr
&$\times\frac{(8s_1^2-18s_1+3)}{(8s_1-7)}$&$\times\frac{(16s_1^3-40s_1^2+29s_1-8)}{(8s_1-7)}$\vspace*{3pt}\cr
$4s_1-3,2s_1,2s_1-3$&$\frac{16s_1^2-28s_1+21}{2(4s_1-3)(8s_1-7)}$&$\frac{16s_1(s_1-1)(2s_1-3)}{(4s_1-3)^2(8s_1-7)}$\vspace*{3pt}\cr
$4s_1-3,2s_1-1,2s_1-1$&$\frac{4s_1-5}{2(8s_1-7)}$&$\frac{4s_1-5}{2(8s_1-7)}$\vspace*{3pt}\cr
$4s_1-3,2s_1-1,2s_1-2$&$\frac{16s_1^2-28s_1+13}{2(4s_1-3)(8s_1-7)}$&$\frac{16(s_1-1)^2(2s_1-1)}{(4s_1-3)^2(8s_1-7)}$\vspace*{3pt}\cr
 \hline
\end{tabular}
\caption{Values of $c_{\overline{\nu}}^{\pm}$ for the ground and
first three excited state manifolds for FM tetramers (or the
highest four excited state manifolds of AFM tetramers), where
$c_1(x)=32x^3-56x^2+30x-3$ and
$c_2(x)=256x^5-640x^4+576x^3-240x^2+48x-3$,
$c_3(x)=256x^4-800x^3+840x^2-330x+45$,
$c_4(x)=2048x^6-8960x^5+15232x^4-13120x^3+6096x^2-1440x+135$.}\label{table6}
\end{table}

\begin{table}
\begin{tabular}{ccc}
$s,s_{13},s_{24}$&$a^{+}_{\overline{\nu}}$&$a^{-}_{\overline{\nu}}$\vspace*{3pt}\\
\hline \vspace*{3pt}\\
$0,x,x$&$\frac{\tilde{a}_1(x)[\tilde{a}_2(x)-4s_1(s_1+1)]}{3(2x+3)(2x-1)}$&$\frac{\tilde{a}_1(x)[x+x^2+4s_1(s_1+1)]}{3(2x+3)(2x-1)}$\vspace*{3pt}\\
$1,x,x$&$\frac{-\tilde{a}_2(x)+4s_1(s_1+1)}{5}$&$-\frac{x+x^2+4s_1(s_1+1)}{5}$\vspace*{3pt}\\
$1,x,x-1$&$\frac{\tilde{a}_3(x)-4s_1(s_1+1)(1+2x^2)}{5(4x^2-1)}$&$\frac{4x^2+2x^4+4s_1(s_1+1)(1+2x^2)}{5(4x^2-1)}$\vspace*{3pt}\\
$2,x,x$&$\frac{\tilde{a}_4(x)[\tilde{a}_5(x)+4s_1(s_1+1)]}{21(2x+3)(2x-1)}$&$-\frac{\tilde{a}_4(x)[x^2+x+4s_1(s_1+1)]}{21(2x+3)(2x-1)}$\vspace*{3pt}\\
$2,x,x-1$&$\frac{\tilde{a}_6(x)-4s_1(s_1+1)(5-2x^2)}{21(4x^2-1)}$&$\frac{\tilde{a}_7(x)+4s_1(s_1+1)(5-2x^2)}{21(4x^2-1)}$\vspace*{3pt}\\
$2,x,x-2$&$-\frac{3(-3+x+x^2)+4s_1(s_1+1)}{21}$&$\frac{4-x+x^2+4s_1(s_1+1)}{21}$\vspace*{3pt}\\
$3,x,x$&$\frac{\tilde{a}_8(x)[-3\tilde{a}_2(x)+4s_1(s_1+1)]}{45(2x+3)(2x-1)}$&$-\frac{\tilde{a}_8(x)[x+x^2+4s_1(s_1+1)]}{45(2x+3)(2x-1)}$\vspace*{3pt}\\
$3,x,x-1$&$\frac{\tilde{a}_9(x)+8s_1(s_1+1)\tilde{a}_{10}(x)}{30(4x^2-1)(4x^2-9)}$&$-\frac{\tilde{a}_{11}(x)+8s_1(s_1+1)\tilde{a}_{10}(x)}{30(4x^2-1)(4x^2-9)}$\vspace*{3pt}\\
$3,x,x-2$&$\frac{1}{2}$&$\frac{1}{6}$\vspace*{3pt}\\
$3,x,x-3$&$\frac{\tilde{a}_{12}(x)-8s_1(s_1+1)x(x-2)}{18(2x-1)(2x-3)}$&$\frac{\tilde{a}_{13}(x)+8s_1(s_1+1)x(x-2)}{18(2x-1)(2x-3)}$\vspace*{3pt}\\
 \hline
\end{tabular}
\caption{Values of $a_{\overline{\nu}}^{\pm}$ for the ground and
first three excited state manifolds for AFM tetramers (or the
highest four excited state manifolds of FM tetramers), where $x$
represents any value of $s_{13}$ that satisfies $0\le
s_{13},s_{24}\le 2s_1$ and $|s_{13}-s_{24}|\le s\le
s_{13}+s_{24}$, and $\tilde{a}_1(x)=3+4x+4x^2$,
$\tilde{a}_2(x)=3(-1+x+x^2)$, $\tilde{a}_3(x)=-3+6x^2+6x^4$,
$\tilde{a}_4(x)= (2x+5)(2x-3)$, $\tilde{a}_5(x)=3(1-x-x^2)$,
$\tilde{a}_6(x)=3(5-16x^2+2x^4)$, $\tilde{a}_7(x)=2x^2(7-x^2)$,
$\tilde{a}_8(x)=-33+4x+4x^2$,
$\tilde{a}_9(x)=198-729x^2+330x^4-24x^6$,
$\tilde{a}_{10}(x)=33-37x^2+4x^4$,
$\tilde{a}_{11}(x)=169x^2-102x^4$,
$\tilde{a}_{12}(x)=27-78x+63x^2-24x^3+6x^4$, and
$\tilde{a}_{13}(x)=9-30x+23x^2-8x^3+2x^4$.}\label{table8}
\end{table}

\begin{table}
\begin{tabular}{ccc}
$s,s_{13},s_{24}$&$b^{+}_{\overline{\nu}}$&$b^{-}_{\overline{\nu}}$\vspace*{3pt}\\
\hline
$0,x,x$&$\frac{8s_1(s_1+1)}{3}$&$\frac{4x(x+1)-8s_1(s_1+1)}{3}$\vspace*{3pt}\\
$1,x,x$&$\frac{4[-1+x+x^2+2s_1(s_1+1)]}{5}$&$\frac{8[x(x+1)-s_1(s_1+1)]}{5}$\vspace*{3pt}\\
$1,x,x-1$&$\frac{4[\tilde{b}_1(x)-2s_1(s_1+1)(1-8x^2)]}{5(4x^2-1)}$&$\frac{4[\tilde{b}_2(x)+2s_1(s_1+1)(1-8x^2)]}{5(4x^2-1)}$\vspace*{3pt}\\
$2,x,x$&$\frac{\tilde{b}_3(x)+8s_1(s_1+1)\tilde{b}_4(x)}{21(2x+3)(2x-1)}$&$\frac{\tilde{b}_5(x)-8s_1(s_1+1)\tilde{b}_4(x)}{21(2x+3)(2x-1)}$\vspace*{3pt}\\
$2,x,x-1$&$\frac{4[\tilde{b}_6(x)+2s_1(s_1+1)(1+8x^2)]}{7(4x^2-1)}$&$-\frac{4[\tilde{b}_7(x)+2s_1(s_1+1)(1+8x^2)]}{7(4x^2-1)}$\vspace*{3pt}\\
$2,x,x-2$&$\frac{4[-3+x-x^2+6s_1(s_1+1)]}{7}$&$\frac{4[1-2x+2x^2-6s_1(s_1+1)]}{7}$\vspace*{3pt}\\
$3,x,x$&$\frac{8[\tilde{b}_8(x)+s_1(s_1+1)\tilde{b}_9(x)]}{45(2x+3)(2x-1)}$&$-\frac{4[\tilde{b}_{10}(x)+2s_1(s_1+1)\tilde{b}_9(x)]}{45(2x+3)(2x-1)}$\vspace*{3pt}\\
$3,x,x-1$&$\frac{4[\tilde{b}_{11}(x)-2s_1(s_1+1)\tilde{b}_{12}(x)]}{15(4x^2-1)(4x^2-9)}$&$\frac{8[\tilde{b}_{13}(x)+s_1(s_1+1)\tilde{b}_{12}(x)]}{15(4x^2-1)(4x^2-9)}$\vspace*{3pt}\\
$3,x,x-2$&$\frac{4[-3+2s_1(s_1+1)]}{3}$&$-\frac{4[x(1-x)+2s_1(s_1+1)]}{3}$\vspace*{3pt}\\
$3,x,x-3$&$\frac{4[\tilde{b}_{14}(x)+2s_1(s_1+1)\tilde{b}_{15}(x)]}{9(2x-3)(2x-1)}$&$-\frac{8[\tilde{b}_{16}(x)+s_1(s_1+1)\tilde{b}_{15}(x)]}{9(2x-3)(2x-1)}$\vspace*{3pt}\\
 \hline
\end{tabular}
\caption{Values of $b_{\overline{\nu}}^{\pm}$ for the ground and
first three excited state manifolds for AFM tetramers (or the
highest four excited state manifolds of FM tetramers), where
$\tilde{b}_1(x)=1-2x^2-2x^4$, $\tilde{b}_2(x)=-3x^2(1-2x^2)$,
$\tilde{b}_3(x)12(15-19x-15x^2+8x^3+4x^4)$,
$\tilde{b}_4(x)=9+20x+20x^2$,
$\tilde{b}_5(x)=-16x(9+x-16x^2-8x^3)$,
$\tilde{b}_6(x)=5-16x^2+2x^4$, $\tilde{b}_7(x)=7x^2-10x^4$,
$\tilde{b}_8(x)=3(33-37x-33x^2+8x^3+4x^4)$,
$\tilde{b}_9(x)=87+44x+44x^2$,
$\tilde{b}_{10}(x)=x(111+43x-136x^2-68x^3)$,
$\tilde{b}_{11}(x)=3(-66+243x^2-110x^4+8x^6)$,
$\tilde{b}_{12}(x)=87+52x^2-64x^4$,
$\tilde{b}_{13}(x)=x^2(107-151x^2+44x^4)$,
$\tilde{b}_{14}(x)=-3(9-26x+21x^2-8x^3+2x^4)$,
$\tilde{b}_{15}(x)=9-32x+16x^2$, and
$\tilde{b}_{16}(x)=-9+30x-35x^2+20x^3-5x^4$.}\label{table9}
\end{table}

\begin{table}
\begin{tabular}{ccc}
$s,s_{13},s_{24}$&$c^{+}_{\overline{\nu}}$&$c^{-}_{\overline{\nu}}$\vspace*{3pt}\\
\hline
$0,x,x$&$\frac{3+4x+4x^2}{6}$&$\frac{3+4x+4x^2}{6}$\vspace*{3pt}\\
$1,x,x$&$\frac{3-4x-4x^2}{6}$&$\frac{3-4x-4x^2}{6}$\vspace*{3pt}\\
$1,x,x-1$&$\frac{3+2x^2}{10}$&$\frac{3(1-x^2)}{10}$\vspace*{3pt}\\
$2,x,x$&$\frac{15-4x-4x^2}{42}$&$\frac{15-4x-4x^2}{42}$\vspace*{3pt}\\
$2,x,x-1$&$\frac{15-2x^2}{42}$&$\frac{15+x^2}{42}$\vspace*{3pt}\\
$2,x,x-2$&$\frac{13-4x+4x^2}{42}$&$\frac{8(2+x-x^2)}{63}$\vspace*{3pt}\\
$3,x,x$&$\frac{33-4x-4x^2}{90}$&$\frac{33-4x-4x^2}{90}$\vspace*{3pt}\\
$3,x,x-1$&$\frac{11-x^2}{30}$&$\frac{132-17x^2}{360}$\vspace*{3pt}\\
$3,x,x-2$&$\frac{1}{3}$&$\frac{23+4x-4x^2}{72}$\vspace*{3pt}\\
$3,x,x-3$&$\frac{6-2x+x^2}{18}$&$\frac{5(3+2x-x^2)}{72}$\vspace*{3pt}\\
 \hline
\end{tabular}
\caption{Values of $c_{\overline{\nu}}^{\pm}$ for the ground and
first three excited state manifolds for AFM tetramers (or the
highest four excited state manifolds of FM
tetramers).}\label{table10}
\end{table}

For Types I and II tetramers, the $s$th AFM level-crossing
induction in the first-order approximation may be written as
\begin{eqnarray}
\gamma B_{s_1,s}^{g,\rm lc
(1)}(\theta)&=&-\tilde{J}_gs-\Theta(\tilde{J}_g'-\tilde{J}_g)2ss_1^2J^g_{{\rm
t},2}\nonumber\\
& &+\Theta(\tilde{J}_g-\tilde{J}_g')
\biggl(\Bigl(\tilde{J}_g-\tilde{J}_g'\nonumber\\
& &+2s_1(s_1+1)J_{\rm b,2}^g\Bigr)E\Bigl(\frac{s+1}{2}\Bigr)\nonumber\\
& &-J_{\rm b,2}^g\Bigl[E\Bigl(\frac{s+1}{2}\Bigr)\Bigr]^3\nonumber\\
& &-J^g_{{\rm t},2}E\Bigl(\frac{s}{2}\Bigr)\Bigl[sE\Bigl(\frac{s+1}{2}\Bigr)-2s_1(s_1+1)\Bigr]\biggr)\nonumber\\
&&-\frac{J_z^g}{2}(a_2^{+}+2b^{+}+a_1^{+}\cos^2\theta)\nonumber\\
&
&+\frac{J_{1,z}^g}{2}[c_2^{-}+\frac{1}{2}(b^{+}+b^{-})+c_1^{-}\cos^2\theta)]\nonumber\\
&
&+\frac{J_{2,z}^g}{4}(a_2^{-}+b^{-}+a_1^{-}\cos^2\theta)\nonumber\\
& &-J_{\rm b,1}^gd-J_{\rm t,1}^ge, \label{levelcrossing}
\end{eqnarray}
 $\Theta(s)$ is the
standard Heaviside step function, $E(x)$ is the largest integer in
$x$ and the level-crossing parameters
 $a_{j}^{\pm}$, $b^{\pm}$, and $c_{j}^{-}$ for $j=1,2$ and
 $d$
are functions of $s,s_1$ and the tetramer type.  For Type II, the
functions are different for even and odd $s$.

 The AFM level-crossing parameters are defined according to
\begin{eqnarray}
a_{1}^{\pm}&=&s(2s-1)a_{s,s_{13},s_{24}}^{s_1,\,\pm}\nonumber\\
& &-(s-1)(2s-3)a_{s-1,s_{13}',s_{24}'}^{s_1,\,\pm},\label{am}\\
a_{2}^{\pm}&=&sa_{s,s_{13},s_{24}}^{s_1,\,\pm}-(s-1)a^{s_1,\,\pm}_{s-1,s_{13}',s_{24}'},\\
b^{\pm}&=&b_{s,s_{13},s_{24}}^{s_1,\,\pm}-b_{s-1,s_{13}',s_{24}'}^{s_1,\,\pm},\\
c_{1}^{-}&=&s(2s-1)c^{s_1,\,-}_{s,s_{13},s_{24}}\nonumber\\
& &-(s-1)(2s-3)c_{s-1,s_{13}',s_{24}'}^{s_1,\,-},\\
c_{2}^{-}&=&sc_{s,s_{13},s_{24}}^{s_1,\,-}-(s-1)c_{s-1,s_{13}',s_{24}'}^{s_1,\,-},\label{cm}\\
d&= &{\cal B}_{\overline{\nu}}-{\cal B}_{\overline{\nu}'}, \label{d}\\
e&=&{\cal T}_{\overline{\nu}}-{\cal T}_{\overline{\nu}'},
\label{e}
\end{eqnarray}
where \begin{eqnarray}
\overline{\nu}'&=&\{s-1,s_{13}',s_{24}',s_1\}
\end{eqnarray}
and the $s_{13}',s_{24}'$ values depend upon the tetramer type.
 In the next two sections, we evaluate the $a_j^{\pm}$, $b^{\pm}$
 and $c_j^{\pm}$ for Type I and Type II AFM tetramers.  The NN biquadratic
exchange level-crossing parameter $d$ has differently complicated
forms for Type I and Type II AFM tetramers, the general forms of
which are not given for brevity.

For $s_1=1/2$, the Types I and II first-order level-crossing
inductions $\gamma B_{1/2,s}^{g,{\rm lc}(1)}(\theta)$ are given in
the text.  In that simple example, there are no effects of
single-ion anisotropy.  Hence to illustrate the full dependencies
on all of the microscopic parameters, we list the $s_1=1$
first-order level-crossing inductions.  For Type I, we have
\begin{eqnarray}
\gamma B_{1,1}^{g,{\rm
lc}(1)}(\theta)&=&-\tilde{J}_g+\frac{1}{12}(35J_{{\rm
b},1}^g+13J^g_{{\rm t},1})-2J^g_{{\rm t},2}\nonumber\\
&
&-J_z^g\Bigl(\frac{7}{6}-\frac{7}{10}\cos^2\theta\Bigr)\nonumber\\&
&+\frac{7}{10}J_{\rm
eff}^g(1-3\cos^2\theta),\\
\gamma B_{1,2}^{g,{\rm lc}(1)}(\theta)&=&-2\tilde{J}_g+2J_{{\rm
b},1}^g+J_z^g\Bigl(\frac{31}{42}-\frac{19}{70}\cos^2\theta\Bigr)\nonumber\\
& &+4J^g_{{\rm t},1}-4J^g_{{\rm t},2}-\frac{19}{70}J_{\rm
eff}^g(1-3\cos^2\theta),\nonumber\\
& &\\ \gamma B_{1,3}^{g,{\rm
lc}(1)}(\theta)&=&-3\tilde{J}_g+\frac{1}{4}(7J_{{\rm
b},1}^g-9J^g_{{\rm t},1})-6J^g_{{\rm t},2}\nonumber\\
&
&+J_z^g\Bigl(\frac{181}{210}-\frac{13}{14}\cos^2\theta\Bigr)\nonumber\\&
&-\frac{13}{14}J_{\rm
eff}^g(1-3\cos^2\theta),\\
\gamma B_{1,4}^{g,{\rm
lc}(1)}(\theta)&=&-4\tilde{J}_g-4J^g_{{\rm t},1}+J_z^g\Bigl(\frac{83}{70}-\frac{3}{2}\cos^2\theta\Bigr)\nonumber\\
& &-8J^g_{{\rm t},2}-\frac{3}{2}J_{\rm eff}^g(1-3\cos^2\theta).
\end{eqnarray}
The Type II first-order level-crossing inductions for $s_1=1$ are
\begin{eqnarray}
\gamma B_{1,1}^{g,{\rm
lc}(1)}(\theta)&=&-\tilde{J}_g'-\frac{1}{3}(11J_{{\rm
b},1}^g+9J^g_{{\rm t},1})+3J_{{\rm
b},2}^g\nonumber\\
& &-\frac{J_z^g}{6}(5-3\cos^2\theta)+\frac{J_{2,z}^g}{6}(1+3\cos^2\theta),\nonumber\\
& &\\
 \gamma B_{1,2}^{g,{\rm
lc}(1)}(\theta)&=&-\tilde{J}_g-\tilde{J}_g'+\frac{1}{12}(15J_{{\rm
b},1}^g+23J^g_{{\rm t},1})\nonumber\\
& &+\frac{J_z^g}{2}(5+\cos^2\theta)+\frac{J_{1,z}^g}{2}(1+\cos^2\theta)\nonumber\\
& &+3J_{{\rm
b},2}^g+2J^g_{{\rm t},2}+\frac{J_{2,z}^g}{6}(1+3\cos^2\theta),\\
\gamma B_{1,3}^{g,{\rm
lc}(1)}(\theta)&=&-\tilde{J}_g-2\tilde{J}_g'+\frac{1}{36}(107J_{{\rm
b},1}^g+179J^g_{{\rm t},1})\nonumber\\& &+J_z^g\Bigl(\frac{43}{30}-\frac{3}{2}\cos^2\theta\Bigr)+\frac{J_{1,z}^g}{6}(1+5\cos^2\theta)\nonumber\\
& &-2J^g_{{\rm t},2}+\frac{J_{2,z}^g}{2}(1+\cos^2\theta),\\
\gamma B_{1,4}^{g,{\rm
lc}(1)}(\theta)&=&-2\tilde{J}_g-2\tilde{J}_g'+\frac{1}{9}(7J_{{\rm
b},1}^g}{-29J_{{\rm t},1})-8J^g_{{\rm t},2}\nonumber\\& &+J_z^g\Bigl(\frac{83}{70}-\frac{3}{2}\cos^2\theta\Bigr)+\frac{J_{1,z}^g}{3}(1+5\cos^2\theta)\nonumber\\
& &+\frac{J_{2,z}^g}{2}(1+\cos^2\theta).
\end{eqnarray}
\subsection{G. Type I First-order AFM level-crossing
constants}

The Type I constants are relevant  for both
$\tilde{J}_g'-\tilde{J}_g>0$
 AFM level-crossing inductions and for
 some low energy FM manifold states (with
$\tilde{J}_g'>\tilde{J}_g>0$).  The low-energy states within an
arbitrary $s$ manifold have $s_{13}=s_{24}=2s_1$. We let
$a_{s,s_{13},s_{24}}^{s_1,\pm}\equiv a_{\overline{\nu}}^{\pm}$,
etc. For general $s, s_1$, we have
\begin{eqnarray}
a_{s,2s_1,2s_1}^{s_1,\,\pm}&=&c_{s,2s_1,2s_1}^{s_1,\,-}\Bigl(1\mp\frac{1}{4s_1-1}\Bigr),\label{apm}\\
 b_{s,2s_1,2s_1}^{s_1,\,\pm}&=&\frac{1}{2(2s+3)(2s-1)}\biggl[s(s+1)
\nonumber\\
& &+[2s(s+1)-1][8s_1(2s_1+1)-s(s+1)]\nonumber\\
 & &\pm\frac{1}{4s_1-1}
\Bigl(2[16s_1^2+s(s+1)][s(s+1)-1]\nonumber\\
& &-8s_1[2s(s+1)-1]\Bigr)
\biggr],\label{bpm}\\
c_{s,2s_1,2s_1}^{s_1,\,\pm}&=&\frac{3[s(s+1)-1]-8s_1(2s_1+1)}{2(2s-1)(2s+3)},\label{cpm}\\
{\cal
B}_{s,2s_1,2s_1}^{s_1}&=&\frac{1}{16}[s(s+1)-4s_1(2s_1+1)]^2\nonumber\\
&
&+\frac{s(s+1)(4s_1-s)(s+4s_1+1)}{8(4s_1-1)}\nonumber\\
& &+\frac{s_1^2(4s_1+1)}{(4s_1-1)}\delta_{s,0},\label{calB}\\
{\cal T}_{s,2s_1,2s_1}^{s_1}&=&\frac{1}{16}[s(s+1)-4s_1(2s_1+1)]^2\nonumber\\
& &-\frac{s_1^2(4s_1+1)}{(4s_1-1)}\delta_{s,0},
\end{eqnarray}

From these expressions, we may evaluate the Type I first-order
level-crossing inductions for AFM tetramers. From the definitions of
the level-crossing constants in Eqs. (\ref{am})-(\ref{cm}), we
rewrite them to explicitly indicate the $s_1, s$ and type
dependencies, and for Type I, it is easy to show that
\begin{eqnarray}
a_{I,j}^{s_1,\,\pm}(s)&=&c_{I,j}^{s_1,\,-}(s)\Bigl(1\mp\frac{1}{4s_1-1}\Bigr),\\
b_{I}^{s_1,\,\pm}(s)&=&-\frac{2s[8s_1(2s_1+1)+4s^4-10s^2+3]}{(4s^2-1)(4s^2-9)}\nonumber\\
& &\times\Bigl(1\mp\frac{1}{4s_1-1}\Bigr),\\
c_{I,1}^{s_1,\,-}(s)&=&\frac{3[4s^3+5s^2-3s-3-8s_1(2s_1+1)]}{2(2s+1)(2s+3)},\nonumber\\
& &\label{c1mofs}\\
c_{I,2}^{s_1,\,-}(s)&=&\frac{c_{20}(s)+(4s^2-4s+3)8s_1(2s_1+1)}{2(4s^2-1)(4s^2-9)},\nonumber\\
& &\\ c_{20}(s)&=&3(4s^4-9s^2-s+3),\label{c20ofs}\\
d_{I}^{s_1}(s)&=&\frac{s[s^2(4s_1-3)+8s_1(s_1+1-4s_1^2)]}{4(4s_1-1)}\nonumber\\
& &-\frac{s_1^2(4s_1+1)}{(4s_1-1)}\delta_{s,1},\label{dI}\\
e_{I}^{s_1}(s)&=&\frac{s}{4}[s^2-4s_1(2s_1+1)]\nonumber\\
& &+\frac{s_1^2(4s_1+1)}{(4s_1-1)}\delta_{s,1},\label{eI}
\end{eqnarray}
for $j=1,2$.  For $s_1=1/2$, it is easy to see that
$a_{I,j}^{1/2,+}(s)=b_{I}^{1/2,+}(s)=0$ for $s=1,2$, as expected.

Letting $a_{I,j}^{s_1,\pm}(s)=a_j^{\pm}$,
$b_I^{s_1,\pm}(s)=b^{\pm}$, and $c_{I,j}^{s_1,\pm}(s)=c_j^{\pm}$,
it is easy to show that for Type I tetramers,
\begin{eqnarray}
c_2^{-}+\frac{1}{2}(b^{+}+b^{-})&=&-\frac{1}{3}c_1^{-},\\
a_2^{-}+b^{-}&=&-\frac{1}{3}a_1^{-}=-\frac{4s_1}{3(4s_1-1)}c_1^{-},
\end{eqnarray}
where $c_1^{-}=c_{I,1}^{s_1,\,-}$.

This implies that for Type I, the axial NN and NNN axial
anisotropic exchange interactions may be combined to yield an
effective axial anisotropic exchange interaction given by Eq.
(\ref{Jeff}).

For the single-ion contributions to the level-crossing inductions,
no such simple relation can be found.  We note that
\begin{eqnarray}
a_2^{+}+b^{+}&=&-\frac{1}{3}a_1^{+},
\end{eqnarray}
but the overall quantity $a_2^{+}+2b^{+}+a_1^{+}\cos^2\theta$ in Eq.
(\ref{levelcrossing}) contains the extra quantity $b^{+}$, which
depends upon $s,s_1$.

\subsection{H. Type II First-order AFM level-crossing constants}

For AFM tetramers with $\tilde{J}_g<0, \tilde{J}_g'-\tilde{J}_g<0$,
Type II, there are two classes of minimum energy configurations for
each $s$ manifold, depending upon whether $s$ is even or odd.  For
even $s$, the relevant states have $s_{13}=s_{24}=s/2$, and for $s$
odd, they are $s_{13},s_{24}=(s\pm1)/2, (s\mp1)/2$. This type is
also relevant for FM tetramers with $\tilde{J}_g>\tilde{J}_g'>0$,
especially for the first excited manifold of states with $s=4s_1-1$.
For even $s$ the relevant parameters are
\begin{eqnarray}
a_{s,s/2,s/2}^{s_1,\,\pm}&=&\frac{1}{2(2s-1)}\Bigl[s-1\mp\frac{f_1(s,s_1)}{2(s+3)}\Bigr],\\
f_1(s,s_1)&=&16s_1(s_1+1)-s^2-2s+6,\\
b_{s,s/2,s/2}^{s_1,\,\pm}&=&\frac{1}{2(2s-1)}\Bigl[s^2\pm\frac{f_2(s,s_1)}{s+3}\Bigr],\\
f_2(s,s_1)&=&16s_1(s_1+1)(s^2+2s-1)\nonumber\\
& &-s(s^3+4s^2+s-4),\\
c_{s,s/2,s/2}^{\pm}&=&\frac{s-1}{2(2s-1)},\\
{\cal
B}^{s_1}_{s,s/2,s/2}&=&\frac{s^4}{64}+\frac{s(s+1)[16s_1(s_1+1)-s(s+4)]}{16(s+3)}\nonumber\\
&
&+(1-\delta_{s,0})\frac{(s+1)[16s_1(s_1+1)-s^2+4]}{128s(s+3)}\nonumber\\
& &\times(s+2)[16s_1(s_1+1)-s(s+4)]\nonumber\\
& &+\delta_{s,0}\frac{4s_1^2(s_1+1)^2}{3},\\
{\cal T}^{s_1}_{s,s/2,s/2}&=&\frac{s^4}{64}-\delta_{s,0}\frac{4s_1^2(s_1+1)^2}{3}\nonumber\\
& &-(1-\delta_{s,0})\frac{(s+1)[16s_1(s_1+1)-s^2+4]}{128s(s+3)}\nonumber\\
& &\times(s+2)[16s_1(s_1+1)-s(s+4)].
\end{eqnarray}

For odd $s$, the relevant parameters are
\begin{eqnarray}
a^{s_1,\,\pm}_{s,(s+1)/2,(s-1)/2}&=&\frac{1}{2s(2s-1)}\Bigl[s^2-s+1\nonumber\\
& &\qquad\mp\frac{f_3(s,s_1)}{2(s+2)(s+4)}\Bigr],\\
f_3(s,s_1)&=&16s_1(s_1+1)(s^2+3s-1)\nonumber\\
& &-s^4-5s^3+11s-11,\\
b^{s_1,\,\pm}_{s,(s+1)/2,(s-1)/2}&=&\frac{1}{2(2s-1)}\Bigl[s^2-1\nonumber\\
& &\qquad\pm\frac{f_4(s,s_1)}{(s+2)(s+4)}\Bigr],\\
f_4(s,s_1)&=&16s_1(s_1+1)(s^3+5s^2+4s-3)\nonumber\\
& &-(s+1)(s^4+6s^3+7s^2-3s+1),\nonumber\\
c^{+}_{s,(s+1)/2,(s-1)/2}&=&\frac{s^2-s+1}{2s(2s-1)},\\
c^{-}_{s,(s+1)/2,(s-1)/2}&=&\frac{(s+1)(s-1)^2}{2s^2(2s-1)},\\
{\cal
B}^{s_1}_{s,(s+1)/2,(s-1)/2}&=&\frac{(s^2-1)^2}{64}+\frac{(s+1)^2}{32(s+2)}\nonumber\\
& &\times[16s_1(s_1+1)-(s+3)(s-1)]\nonumber\\
& &+[16s_1(s_1+1)-(s+3)(s-1)]^2\nonumber\\
& &\times\frac{(s+1)^4}{256s^2(s+2)^2}\nonumber\\ &
&+[16s_1(s_1+1)-(s+3)(s-1)]\nonumber\\
& &\times[16s_1(s_1+1)-(s+5)(s+1)]\nonumber\\
& &\times\frac{(2s+3)(s^2-2s-1)^2}{256s^3(s+1)(s+4)(2s+1)^2},\nonumber\\
& &\\
{\cal T}^{s_1}_{s,(s+1)/2,(s-1)/2}&=&\frac{(s^2-1)^2}{64}-\frac{(s+1)^4}{256s^2(s+2)^2}\nonumber\\
& &\times[16s_1(s_1+1)-(s+3)(s-1)]^2\nonumber\\
& &-[16s_1(s_1+1)-(s+3)(s-1)]\nonumber\\
& &\times[16s_1(s_1+1)-(s+5)(s+1)]\nonumber\\
&
&\times\frac{(2s+3)(s^2-2s-1)^2}{256s^3(s+1)(s+4)(2s+1)^2}.\nonumber\\
\end{eqnarray}
From these expressions, we obtain the level-crossing inductions for
the AFM type $\tilde{J}_g<0$ and $\tilde{J}_g'-\tilde{J}_g<0$.

For $s$ even, we have
\begin{eqnarray}
a_{II{\rm e},1}^{s_1,\pm}(s)&=&s(2s-1)a_{s,s/2,s/2}^{s_1,\pm}\nonumber\\
& &-(s-1)(2s-3)a_{s-1,s/2,(s-2)/2}^{s_1,\pm},\\
a_{II{\rm e},2}^{s_1,\pm}(s)&=&sa_{s,s/2,s/2}^{s_1,\pm}\nonumber\\
& &-(s-1)a_{s-1,s/2,(s-2)/2}^{s_1,\pm},\\
b_{II{\rm e}}^{s_1,\pm}(s)&=&b_{s,s/2,s/2}^{s_1,\pm}-b_{s-1,s/2,(s-2)/2}^{s_1,\pm},\\
c_{II{\rm e},1}^{s_1,\pm}(s)&=&s(2s-1)c_{s,s/2,s/2}^{\pm}\nonumber\\
& &-(s-1)(2s-3)c_{s-1,s/2,(s-2)/2}^{\pm},\\
c_{II{\rm e},2}^{\pm}(s)&=&sc_{s,s/2,s/2}^{\pm}\nonumber\\
& &-(s-1)c_{s-1,s/2,(s-2)/2}^{\pm},\\
d_{II{\rm e}}^{s_1}(s)&=&{\cal B}^{s_1}_{s,s/2,s/2}-{\cal B}^{s_1}_{s-1,s/2,(s-2)/s},\label{dIIeven}\\
e_{II{\rm e}}^{s_1}(s)&=&{\cal T}^{s_1}_{s,s/2,s/2}-{\cal
T}^{s_1}_{s-1,s/2,(s-2)/s}.\label{eIIeven}
\end{eqnarray}

For $s$ odd, we have
\begin{eqnarray}
a_{II{\rm o},1}^{s_1,\pm}(s)&=&s(2s-1)a_{s,(s+1)/2,(s-1)/2}^{s_1,\pm}\nonumber\\
& &-(s-1)(2s-3)a_{s-1,(s-1)/2,(s-1)/2}^{s_1,\pm},\nonumber\\
& &\\
a_{II{\rm o},2}^{s_1,\pm}(s)&=&sa_{s,(s+1)/2,(s-1)/2}^{s_1,\pm}\nonumber\\
& &-(s-1)a_{s-1,(s-1)/2,(s-1)/2}^{s_1,\pm},\\
b_{II{\rm o}}^{s_1,\pm}(s)&=&b_{s,(s+1)/2,(s-1)/2}^{s_1,\pm}-b_{s-1,(s-1)/2,(s-1)/2}^{s_1,\pm},\nonumber\\
& &\\
c_{II{\rm o},1}^{s_1,\pm}(s)&=&s(2s-1)c_{s,(s+1)/2,(s-1)/2}^{\pm}\nonumber\\
& &-(s-1)(2s-3)c_{s-1,(s-1)/2,(s-1)/2}^{\pm},\nonumber\\
& &\\
c_{II{\rm o},2}^{s_1,\pm}(s)&=&sc_{s,(s+1)/2,(s-1)/2}^{\pm}\nonumber\\
& &-(s-1)c_{s-1,(s-1)/2,(s-1)/2}^{\pm},\\
d_{II{\rm o}}^{s_1}(s)&=&{\cal B}^{s_1}_{s,(s+1)/2,(s-1)/2}\nonumber\\
& &-{\cal B}^{s_1}_{s-1,(s-1)/2,(s-1)/2},\label{dIIodd}\\
e_{II{\rm o}}^{s_1}(s)&=&{\cal T}^{s_1}_{s,(s+1)/2,(s-1)/2}\nonumber\\
& &-{\cal T}^{s_1}_{s-1,(s-1)/2,(s-1)/2}.\label{eIIodd}
\end{eqnarray}

From these expressions, we may obtain the Type II AFM level-crossing
induction parameters.  For even $s$, we find
\begin{eqnarray}
a_{II{\rm e},1}^{s_1,\,\pm}(s)&=&\frac{2s-3}{2}\mp\frac{48s_1(s_1+1)+a_{10}^{\rm e}(s)}{4(s+1)(s+3)},\label{aII1pme}\\
a_{10}^{\rm e}(s)&=&-2s^3-5s^2+6s+18,\\
a_{II{\rm e},2}^{s_1,\,\pm}(s)&=&\frac{1}{2(2s-1)(2s-3)}\biggl[2s^2-6s+3\nonumber\\
&
&\pm\frac{16s_1(s_1+1)(2s^2-4s+3)+a_{20}^{\rm e}(s)}{2(s+1)(s+3)}\biggr],\label{aII2pme}\nonumber\\
& &\\
a_{20}^{\rm e}(s)&=&2s^4+2s^3-9s^2-18s+18,\\
b_{II{\rm e}}^{s_1,\,\pm}(s)&=&\frac{s}{(2s-1)(2s-3)}\biggl[s-1\nonumber\\
& &\pm\frac{16s_1(s_1+1)(s-2)+b_0^{\rm e}(s)}{2(s+1)(s+3)}\biggr],\label{bIIpme}\\
b_0^{\rm e}(s)&=&-4s^4-7s^3+20s^2+14s-18,\\
c_{II{\rm e},1}^{-}(s)&=&\frac{s(2s-3)}{2(s-1)},\label{cII1me}\\
c_{II{\rm
e},2}^{-}(s)&=&\frac{s(2s^2-4s+1)}{2(s-1)(2s-1)(2s-3)},\label{cII2me}\\
\end{eqnarray}
and $d_{II{\rm e}}^{s_1}(s)$ and $e_{II{\rm e}}^{s_1}(s)$ are
given by Eqs. (\ref{dIIeven}) and (\ref{eIIeven}).   Combining
$a_{II{\rm e},2}^{s_1,\,-}(s)$ and $b_{II{\rm e}}^{s_1,\,-}(s)$,
we find
\begin{eqnarray}
a_{II{\rm e},2}^{s_1,\,-}+b_{II{\rm
e}}^{s_1,\,-}&=&\frac{1}{2}-\frac{16s_1(s_1+1)-d_0^{\rm
e}(s)}{4(s+1)(s+3)},\\
d_0^{\rm e}(s)&=&2s^3+7s^2+2s-6.
\end{eqnarray}
We note that  this expression differs substantially from that for
$a_{II{\rm e},1}^{s_1,\,-}(s)$, except for $s_1=1/2$ and $s=2$.
Similarly, it is elementary to combine $c_{II{\rm e},2}^{-}(s)$ and
$[b_{II{\rm e}}^{s_1,\,+}(s)+b_{II{\rm e}}^{s_1,\,-}(s)]/2$.  We
find
\begin{eqnarray}
c_{II{\rm e},2}^{-}(s)+\frac{1}{2}\Bigl(b_{II{\rm
e}}^{s_1,\,+}(s)+b_{II{\rm
e}}^{s_1,\,-}(s)\Bigr)&=&\frac{s}{2(s-1)}.
\end{eqnarray}
This simple expression differs from that for $c_{II{\rm
e},1}^{-}(s)$ by the factor $2s-3$.  However, at $s=2$, the only
even $s$ value for $s_1=1/2$,  they are equivalent. In addition, as
for Type I, there is no simple relation between the single-ion
parameters $a_{II{\rm e},2}^{s_1,\,+}(s)+2b_{II{\rm
e}}^{s_1,\,+}(s)$ and $a_{II{\rm e},1}^{s_1,\,+}(s)$.

For odd $s$, the Type II level-crossing induction parameters are
\begin{eqnarray}
a_{II{\rm o},1}^{s_1,\,\pm}(s)&=&\frac{2s-1}{2}\mp\frac{48s_1(s_1+1)+a_{10}^{\rm o}(s)}{4(s+2)(s+4)},\label{aIIpmo}\\
a_{10}^{\rm o}(s)&=&-2s^3-11s^2-10s+17,\\
a_{II{\rm o},2}^{s_1,\,\pm}(s)&=&\frac{1}{2(2s-1)(2s-3)}\biggl[2s^2-2s-1\nonumber\\
& &\pm\frac{16s_1(s_1+1)(2s^2+1)+a_{20}^{\rm
o}(s)}{2(s+2)(s+4)}\biggr],\label{aII2pmo}\\
a_{20}^{\rm o}(s)&=&2s^4+10s^3+9s^2-22s-5,\\
b_{II{\rm
o}}^{s_1,\,\pm}(s)&=&\frac{1}{(2s-1)(2s-3)}\biggl[(s-1)(s-2)\nonumber\\
& &\pm\frac{16s_1(s_1+1)(s^2-4s+1)+b_0^{\rm
o}(s)}{2(s+2)(s+4)}\biggr],\label{bIIpmo}\nonumber\\
& &\\
b_0^{\rm o}(s)&=&-4s^5-19s^4+54s^2-2s-5,\\
c_{II{\rm o},1}^{-}(s)&=&\frac{(s-1)(2s-1)}{2s},\label{cII1o}\\
c_{II{\rm
o},2}^{-}(s)&=&\frac{(s-1)(2s^2-4s+3)}{2s(2s-1)(2s-3)},\label{cII2o}\\
d_{II{\rm o}}^{s_1}(1)&=&\frac{s_1}{6}(4s_1^3+8s_1^2+7s_1+3),\\
e_{II{\rm o}}^{s_1}(1)&=&-\frac{s_1}{6}(4s_1^3+8s_1^2+3s_1-1),
\end{eqnarray}
and the other $d_{II{\rm o}}^{s_1}(s)$ and $e_{II{\rm
o}}^{s_1}(s)$ values are found from Eqs. (\ref{dIIodd}) and
(\ref{eIIodd}).

 We note that $a_{II{\rm e},j}^{1/2,\,+}(2)=a_{II{\rm
o},j}^{1/2,\,+}(1)=b_{II{\rm e}}^{1/2,\,+}(2)=b_{II{\rm
o}}^{1/2,\,+}(1)=0$ for $j=1,2$, as expected.  However, by
combining $a_{II{\rm o},2}^{s_1,\,-}(s)$ and $b_{II{\rm
o}}^{s_1,\,-}(s)$, we have
\begin{eqnarray}
a_{II{\rm o},2}^{s_1,\,-}+b_{II{\rm
o}}^{s_1,\,-}&=&\frac{1}{2}-\frac{16s_1(s_1+1)-d_0^{\rm
o}(s)}{4(s+2)(s+4)},\\
d_o^{\rm o}(s)&=&2s^3+13s^2+22s+5,
\end{eqnarray}
which differs substantially from the version with $s_1=1/2$.  In
addition, it is elementary to combine $c_{II{\rm
o},2}^{-}(s)+[b_{II{\rm o}}^{s_1,\,+}(s)+b_{II{\rm
o}}^{s_1,\,-}(s)]/2$.  We find
\begin{eqnarray}
c_{II{\rm o},2}^{-}(s)+\frac{1}{2}\Bigl(b_{II{\rm
o}}^{s_1,\,+}(s)+b_{II{\rm
o}}^{s_1,\,-}(s)\Bigr)&=&\frac{(s-1)}{2s},\end{eqnarray} which
differs from $c_{II{\rm o},1}^{-}(s)$ by the factor $2s-1$.  At
$s=1$, the only relevant odd $s$ value for $s_1=1/2$, these are
equivalent.  In addition, as for Type I and the even crossings of
Type II, there is no simple relation between the single-ion
parameters $a_{II{\rm o},2}^{s_1,\,+}(s)+2b_{II{\rm
o}}^{s_1,\,+}(s)$ and $a_{II{\rm o},1}^{s_1,\,+}(s)$.

\subsection{I. Hartree INS functions}

The functions $L_{\nu,\nu'}({\bm q})$ and $M_{\nu,\nu'}({\bm q})$ in
the self-consistent Hartree  INS $S_g^{(1)}({\bm q},\omega)$ in the
induction representation are given by

\begin{eqnarray}
L_{\nu,\nu'}({\bm
q})&=&\delta_{m',m}\delta_{s_{24}',s_{24}^{}}\biggl[m^2\delta_{s',s}\nonumber\\
& &\times\Bigl(\delta_{s_{13}',s_{13}^{}}f_{\overline{\nu},0}({\bm
q})+\sum_{\sigma''=\pm1}\delta_{s_{13}',s_{13}^{}+\sigma''}f^{\sigma''}_{\overline{\nu},1}({\bm
q})\Bigr)\nonumber\\
&
&+\sum_{\sigma'=\pm1}\delta_{s',s+\sigma'}\Bigl(C^m_{-\sigma's-(\sigma'+1)/2}\Bigr)^2\nonumber\\
& &\times\Bigl(\delta_{s_{13}',s_{13}^{}}
f^{\sigma'}_{\overline{\nu},2}({\bm
q})\nonumber\\
&
&\qquad+\sum_{\sigma''=\pm1}\delta_{s_{13}',s_{13}^{}+\sigma''}f^{\sigma',\sigma''}_{\overline{\nu},3}({\bm
q})\Bigr)\biggr]\nonumber\\
& &+\left(\begin{array}{c} s_{13}\leftrightarrow s_{24}\\
s_{13}'\leftrightarrow s_{24}'\\
q_y\rightarrow -q_y\end{array}\right),\\
M_{\nu,\nu'}({\bm
q})&=&\sum_{\sigma=\pm1}\delta_{m',m+\sigma}\delta_{s_{24}',s_{24}^{}}\biggl[\Bigl(A_s^{\sigma m}\Bigr)^2\delta_{s',s}\nonumber\\
& &\times\Bigl(\delta_{s_{13}',s_{13}^{}}f_{\overline{\nu},0}({\bm
q})+\sum_{\sigma''=\pm1}\delta_{s_{13}',s_{13}^{}+\sigma''}f^{\sigma''}_{\overline{\nu},1}({\bm
q})\Bigr)\nonumber\\
&
&+\sum_{\sigma'=\pm1}\delta_{s',s+\sigma'}\Bigl(D^{\sigma,m}_{-\sigma's-(\sigma'+1)/2}\Bigr)^2\nonumber\\
& &\times\Bigl(\delta_{s_{13}',s_{13}^{}}
f^{\sigma'}_{\overline{\nu},2}({\bm
q})\nonumber\\
&
&\qquad+\sum_{\sigma''=\pm1}\delta_{s_{13}',s_{13}^{}+\sigma''}f^{\sigma',\sigma''}_{\overline{\nu},3}({\bm
q})\Bigr)\biggr]\nonumber\\
& &+\left(\begin{array}{c} s_{13}\leftrightarrow s_{24}\\
s_{13}'\leftrightarrow s_{24}'\\
q_y\rightarrow -q_y\end{array}\right),\\
f_{\overline{\nu},0}({\bm q})&=&\frac{1}{8}\Bigl(f_{+}({\bm
q})+\xi^2_{s,s_{13},s_{24}}f_{-}({\bm q})\nonumber\\
& &-2\xi_{s,s_{13},s_{24}}\sin(q_xa)\sin(q_ya)\Bigr),\\
f_{\overline{\nu},1}^{\sigma''}({\bm
q})&=&2\Bigl(1-\cos[a(q_x+q_y)]\Bigr)\nonumber\\
&
&\times\Bigl(F_{s_1,s_1,s}^{s_{13}+(\sigma''+1)/2,s_{24}}\Bigr)^2\\
f_{\overline{\nu},2}^{\sigma'}({\bm q})&=&\frac{1}{8}f_{-}({\bm
q})\eta^2_{s+(\sigma'+1)/2,s_{13},s_{24}},\\
f_{\overline{\nu},3}^{\sigma',\sigma''}({\bm
q})&=&2\Bigl(1-\cos[a(q_x+q_y)]\Bigr)\nonumber\\
&
&\times\Bigl(G_{s_1,s_1,\sigma'\sigma''s+\sigma''(\sigma'+1)/2}^{s_{13}+(\sigma''+1)/2,s_{24}}\Bigr)^2,\\
f_{\pm}({\bm
q})&=&1+\cos(q_xa)\cos(q_ya)\nonumber\\
& &\pm\cos(q_zc)[\cos(q_xa)+\cos(q_ya)],
\end{eqnarray}
where the $A_s^m$, $C_s^m$, $D_s^{\tilde{\sigma},m}$,
$F_{s_1,s_3,s}^{s_{13},s_{24}}$,  $G_{s_1,s_3,s}^{s_{13},s_{24}}$,
$\eta_{z,x,y}$, $\xi_{z,x,y}$,
 are given by Eqs. (\ref{Asm}),
(\ref{Csm}), (\ref{Dsigmasm}),  and (\ref{A})-(\ref{xi}),
respectively.


\begin{thebibliography}{99}
\bibitem{general}
R. Sessoli, D. Gatteschi, A. Caneschi, and M. Novak, Nature
(London), {\bf 365}, 141 (1993); W. Wernsdorfer and R. Sessoli,
Science {\bf 284}, 133 (1999).

\bibitem{moregeneral}
M. N. Leuenberger and D. Loss, Nature (London), {\bf 410}, 789
(2001).

\bibitem{ek}
D. V. Efremov and R. A. Klemm, Phys. Rev. B {\bf 66}, 174427
(2002).
\bibitem{ekshort}
R. A. Klemm and D. V. Efremov, AIP Conf. Proc. {\bf 850}, 1151
(2006).
\bibitem{ek2}
D. V. Efremov and R. A. Klemm, Phys. Rev. B {\bf 74}, 064408
(2006)(cond-mat/0601591).



\bibitem{Shapira}
Y. Shapira, M. T. Liu, S. Foner, C. E. Dub{\'e}, and P. J.
Bonitatebus, Jr., Phys. Rev. B {\bf 59}, 1046 (1999).

\bibitem{Mennerich}
C. Mennerich, H.-H. Klauss, M. Broekelmann, F. J. Litterst, C.
Golze, R. Klingeler, V. Kataev, B. B{\"u}chner, S.-N. Grossjohann,
W. Brenig, M. Goiran, H. Rakoto, J.-M. Broto, O. Kataeva, and D.
J. Price, Phys. Rev. B {\bf 73}, 174415 (2006).


\bibitem{Dalal}
D. Zipse, J. M. North, N. S. Dalal, S. Hill, and R. S. Edwards,
Phys. Rev. B {\bf 68}, 184408 (2003).

\bibitem{Fe8}
S. Carretta, E. Liviotti, N. Magnani, P. Santini, and G. Amoretti,
Phys. Rev. Lett. {\bf 92}, 207205 (2004).

\bibitem{Hill1}
D. Zipse, N. S. Dalal, R. M. Achey, J. M. North, S. Hill, and R.
S. Edwards, Appl. Magn. Reson. {\bf 27}, 151 (2004).

\bibitem{Hill2}
S. Hill, R. S. Edwards, J. M. North, S. Maccagnano, and N. S.
dalal, Polyhedron {\bf 22}, 1897 (2003).

\bibitem{Black1}
R. S. Rubins, T. D. Black, and J. Barak, J. Chem. Phys. {\bf 85},
3770 (1985).

\bibitem{Black2}
T. D. Black. R. S. Rubins, D. K. De, R. C. Dickinson, and W. A.
Baker, Jr., J. Chem. Phys. {\bf 80-}, 4620 (2984).

\bibitem{Black3}
R. C. Dickinson, W. A. Baker, Jr., T. D. Black, and R. S. Rubins,
J. Chem. Phys. {\bf 79}, 2609 (1983).

\bibitem{Buluggiu}
 E. Buluggiu, J. Chem. Phys. {\bf 84}, 1243 (1986)

\bibitem{Cr4}
A. Bino, D. C. Johnston, D. P. Goshorn, T. R. Halbert, and E. I.
Stiefel, Science {\bf 241}, 1479 (1988).

\bibitem{Co4}
E.-C. Yang, D. N. Hendrickson, W. Wernsdorfer, M. Nakano, L. N.
Zakharov, R. D. Sommer, A. L. Rheingold, M. Ledezma-Gairaud, and
G. Christou, J. Appl. Phys. {\bf 91}, 7382 (2002).

\bibitem{Ni4}
A. Sieber, C. Boskovic, R. Bircher, O. Waldmann, W. T. Ochsenbein,
G. Chaboussant, H. U. G{\"u}del, N. Kirchner, J. van Slageren, W.
Wernsdorfer, A. Neels, H. Stoeckli-Evans, S. Jannsen, F. Juranyi,
and H. Mutka, Inor. Chem. {\bf 44}, 4315 (2005), and references
therein.

\bibitem{Ni4Maria}
M. Moragues-C{\'a}novas, M. Helliwell, L. Ricard, {\'E}ric
Rivi{\`e}re. W. Wernsdorfer, E. Brechin, and T. Mallah, Eur. J.
Inorg. Chem. {\bf 2004}, 2219.



\bibitem{Edwards}
R. S. Edwards, S. Maccagnano, E.-C. Yang, S. Hill, W. Wernsdorfer,
D. Hendrickson, and G. Christou, J. Appl. Phys. {\bf 93}, 7807
(2003).

\bibitem{Ni4S4}
E. del Barco, A. D. Kent, E.-C. Yang, and D. N. Hendrickson,
Polyhedron {\bf 24}, 2695 (2005).

\bibitem{Hendrickson}
D. N. Hendrickson, E.-C. Yang, R. M. Isidro, C. Kirman, J.
Lawrence, R. S. Edwards, S. Hill, A. Yamaguchi, H. Ishimoto, W.
Wernsdorfer, C. Ramsey, N. Dalal, and M. M. Olmstead, Polyhedron
{\bf 24}, 2280 (2005).

\bibitem{Boskovic}
C. Boskovic, R. Bircher, P. L. W. Tregenna-Piggott, H. U.
G{\"u}del, C. Paulsen, W. Wernsdorfer, A.-L. Barra, E. Khatsko, A.
Neels, and H. Stoeckli-Evans, J. Am. Chem. Soc. {\bf 125}, 14046
(2003).

\bibitem{Parkpublished}
K. Park, M. R. Pederson, and C. S. Hellberg, Phys. Rev. B {\bf
69}, 014416 (2004).

\bibitem{Pedersonpublished}
J. Ribas-Arino, T. Baruah, and M. R. Pederson, J. Chem. Phys. {\bf
123}, 044303 (2005).

\bibitem{Pedersonreview}
A. V. Postnikov, J. Kortus, and M. R. Pederson, Phys. Stat. Solidi
B:  Basic Research {\bf 243}, 2533 (2006).

\bibitem{Stolbov}
S. Stolbov, R. A. Klemm, and T. S. Rahman, cond-mat/0501178
(unpublished).

\bibitem{Park}
K. Park, M. R. Pederson, and N. Bernstein, J. Phys. Chem. Solids
{\bf 65}, 805 (2004).

\bibitem{NRL} M. R. Pederson, private communications.

\bibitem{Harmon}
D. W. Boukhvalov, M. Al-Sager, E. Z. Kurmaev, A. Moewes, V. R.
Galakhov, L. D. Finkelstein, S. Chiuzbaian, M. Neumann, V. V.
Dobrovitskii, M. I. Katsnelson, A. I. Lichtenstein, B. N. Harmon,
K. Endo, J. M. North, and N. S. Dalal, Phys. Rev. B {\bf 75},
014419 (2007).



\bibitem{Bocabook}
R. Bo{\v c}a, {\it Theoretical Foundations of Molecular Magnetism},
(Elsevier, Amsterdam, 1999).


\bibitem{WG}
O. Waldmann and H. U. G{\" u}del, Phys. Rev. B
 {\bf 72}, 094422 (2005).


\bibitem{Tinkham}
M. Tinkham, {\it Group Theory and Quantum Mechanics}, (McGraw-Hill,
New York, 1964).


\bibitem{ka}
R. A. Klemm and M. Ameduri, Phys. Rev. B {\bf 66}, 012403  (2002).

\bibitem{Nd4}
D. M. Barnhart, D. L. Clark, J. C. Gordon, J. C. Huffman, J. G.
Watkin, and B. D. Zwick, J. Am. Chem. Soc. {\bf 115}, 8461 (1993).

\bibitem{KlemmLuban}
R. A. Klemm and M. Luban, Phys. Rev. B {\bf 64}, 104424 (2001).

\bibitem{BenciniGatteschi}
A. Bencini and D. Gatteschi, {\it Electron Paramagnetic
Resonanceof Exchange Coupled Systems}, (Springer, Berlin, 1990).

\bibitem{Moriya}
T. Moriya, Phys. Rev. {\bf 120}, 91 (1960).

\bibitem{Dzyaloshinskii} I. Dzyaloshinskii, J. Phys. Chem. Solids {\bf 4}, 241 (1958).



\bibitem{Jackson}
J. D. Jackson, {\it Classical Electrodynamics, 3rd Edition} (Wiley
\& Sons, Hoboken, NJ, 1999), p. 186.

\bibitem{Goldstein}
H. Goldstein, {\it Classical Mechanics}, (Addison-Wesley, Reading,
MA, 1965), p. 109.

\bibitem{Marisol}
M. Alc{\'a}ntara Ortigoza, R. A. Klemm, and T. S. Rahman, Phys.
Rev. B {\bf 72}, 174416 (2005).

\bibitem{NaV2O5} R. Valenti, C. Gros, and W. Brenig, Phys. Rev. B {\bf
62}, 14164 (2000).

\bibitem{Katsura}
H. Katsura, N. Nagaosa, and A. V. Balatsky, Phys. Rev. Lett. {\bf
95}, 057205 (2005).

\bibitem{Mostovoy}
S.-W. Cheong and M. Mostovoy, Nature Materials {\bf 6}, 13 (2007).

\bibitem{Schnack}
J. Schnack, M. Br{\"u}ger, M. Luban, P. K{\"o}gerler, E. Morosan,
R. Fuchs, R. Modler, H. Nojiri, R. C. Rai, J. Cao, J. L. Musfeldt,
and X. Wei, Phys. Rev. B {\bf 73}, 094401 (2006).

\bibitem{WaldmannNi4}
O. Waldmann, J. Hassmann, P. M{\"u}ller, D. Volkmer, U. S.
Schubert, and J.-M. Lehn, Phys. Rev. B {\bf 58}, 3277 (1998).

\bibitem{Waldmannreview}
O. Waldmann, Coord. Chem. Rev. {\bf 249}, 2550 (2005).

\bibitem{KRS}
R. A. Klemm, C. T. Rieck, and K. Scharnberg, Phys. Rev. B {\bf
61}, 5913 (2000).

\bibitem{Kostyuchenko}
V. V. Kostyuchenko, Phys. Rev. B {\bf 76}, 21204 (2007).

\bibitem{ekfuture}
D. V. Efremov  and R. A. Klemm, unpublished.





\bibitem{Fe4}
S. Carretta, P. Santini, G. Amoretti, T. Guidi, R. Caciuffo, A.
Candini, A. Cornia, D. Gatteschi, M. Plazanet, and J. A. Stride,
Phys. Rev. B {\bf 70}, 214403 (2004).

\bibitem{CorniaFe4}
A. Cornia, A. C. Fabretti, P. Garrisi, C. Mortal{\`o}, D.
Bonacchi, D. Gatteschi, R. Sessoli, L. Sorace, W. Wernsdorfer, and
A.-L. Barra, Angew. Chem. Int. Ed. {\bf 43}, 1136 (2004).

\bibitem{Rastelli}
E. Rastelli and A. Tossi, Phys. Rev. B {\bf 75}, 13414 (2007).

\bibitem{Fe4spin5}
G. Amoretti, S. Carretta, R. Caciuffo, H. Casalta, A. Cornia, M.
Affronte, and D. Gatteschi, Phys. Rev. B {\bf 64}, 104403 (2001).

\bibitem{Ni4C2v}
J. M. Clemente-Juan, H. Andres, J. J. Borr{\'a}s-Almenar, E.
Coronado, H. U. G{\"u}del, M. Aebersold, G. Kearly, H.
B{\"u}ttner, and M. Zolliker, J. Am. Chem. Sic. {\bf 121}, 10021
(1999).

\bibitem{Lecren}
L. Lecren, W. Wernsdorfer, Y.-G. Li, O. Roubeau, H. Miyasaka, and
R. Cl{\'e}rac, J. Am. Chem. Soc. {\bf 127}, 11311 (2005).

\bibitem{Ni2Mn2}
M. Koikawa, M. Ohba, and T. Tokii, Polyhedron {\bf 24}, 2257
(2005).

\bibitem{CrNi3}
J.-N. Rebilly, L. Catala, E. Rivi{\`e}re, R. Guillot, W.
Wernsdorfer, and T. Mallah, Inorg. Chem. {\bf 44}, 8194 (2005).

\bibitem{ak}
M. Ameduri and R. A. Klemm, J. Phys. A: Math. Gen. {\bf 37}, 1095
(2004).

\bibitem{Janprivate}
J. L. Musfeldt, private comuunication.



\end{thebibliography}
\end{document}